\documentclass[apj]{emulateapj}

\usepackage{apjfonts}
\usepackage{amsmath}
\usepackage{epstopdf}

\usepackage{subfigure}

\usepackage{longtable}

\def\wig#1{\mathrel{\hbox{\hbox to 0pt{%
          \lower.5ex\hbox{$\sim$}\hss}\raise.4ex\hbox{$#1$}}}}

\shorttitle{Neglected Clouds in Cool Brown Dwarfs}
\shortauthors{Morley et al.}

\newcommand{\mj}{$M_{\mathrm{J}}$}

\newcommand{\teff}{$T_{\rm eff}$}

\newcommand{\cp}{\citep}
\newcommand{\ct}{\citet}

\newcommand{\icarus}{Icarus} 
\newcommand{\fsed}{$f_{\textrm{sed}}$} 
\newcommand{\nas}{Na$_2$S}

\begin{document}

\title{Neglected Clouds in T and Y Dwarf Atmospheres}

\author{Caroline V. Morley\altaffilmark{1}, Jonathan J. Fortney\altaffilmark{1, 2}, Mark S. Marley\altaffilmark{3}, Channon Visscher\altaffilmark{4}, Didier Saumon\altaffilmark{5}, S. K. Leggett\altaffilmark{6}}

\altaffiltext{1}{Department of Astronomy and Astrophysics, University of California, Santa Cruz, CA 95064, USA; cmorley@ucolick.org}
\altaffiltext{2}{Alfred P. Sloan Research Fellow}
\altaffiltext{3}{NASA Ames Research Center, Moffett Field, CA 94035, USA} 
\altaffiltext{4}{Southwest Research Institute, Boulder, CO 80302, USA} 
\altaffiltext{5}{Los Alamos National Laboratory, Los Alamos, NM 87545, USA} 
\altaffiltext{6}{Gemini Observatory, Northern Operations Center, Hilo, HI 96720, USA}

\begin{abstract}
As brown dwarfs cool, a variety of species condense in their atmospheres, forming clouds. Iron and silicate clouds shape the emergent spectra of L dwarfs, but these clouds dissipate at the L/T transition.  A variety of other condensates are expected to form in cooler T dwarf atmospheres.  These include Cr, MnS, \nas, ZnS, and KCl, but the opacity of these optically thinner clouds has not been included in previous atmosphere models. Here, we examine their effect on model T and Y dwarf atmospheres.  The cloud structures and opacities are calculated using the \ct{AM01} cloud model, which is coupled to an atmosphere model to produce atmospheric pressure-temperature profiles in radiative-convective equilibrium. We generate a suite of models between \teff\ = 400 and 1300 K, log $g$=4.0 and 5.5, and condensate sedimentation efficiencies from \fsed=2 to 5. Model spectra are compared to two red T dwarfs, Ross 458C and UGPS 0722--05; models that include clouds are found to match observed spectra significantly better than cloudless models.  The emergence of sulfide clouds in cool atmospheres, particularly \nas, may be a more natural explanation for the ``cloudy'' spectra of these objects, rather than the re-emergence of silicate clouds that wane at the L-to-T transition.  We find that sulfide clouds provide a mechanism to match the near- and mid-infrared colors of observed T dwarfs. Our results indicate that including the opacity of condensates in T dwarf atmospheres is necessary to accurately determine the physical characteristics of many of the observed objects. 

\end{abstract}

 
\section{Introduction}

Since the first brown dwarfs were discovered two decades ago \cp{Becklin88, Nakajima95}, hundreds more brown dwarfs have been discovered using wide field infrared surveys. These substellar objects, too low in mass to fuse hydrogen in their cores, range in mass from $\sim$ 13 to 75 \mj\ and are classified by their spectra into L, T, and most recently Y dwarfs \cp{Kirkpatrick05, Cushing11}. Without hydrogen fusion as an internal energy source, brown dwarfs never reach a main-sequence state of constant luminosity; instead, they cool over time and will transition through the brown dwarf spectral sequence as different molecules and condensates form in their atmospheres. To model their atmospheres accurately requires an understanding of both the chemistry and physics of the materials that will condense into clouds.  

\subsection{Modeling L and T Dwarfs}

\subsubsection{L dwarfs}

Grain or condensate formation has been predicted to play an important role in L dwarf atmospheres since before the first brown dwarfs were discovered \cp{Lunine86, Lunine89}. Modern equilibrium thermochemical models predict that a variety of different condensates will form in L dwarf atmospheres \cp{Fegley94, Lodders99}; by comparing models to observations, it is now well-established that a variety of refractory materials condense in L dwarfs \cp[see, e.g.][]{Tsuji96, Allard01,Marley02,Burrows06,Cushing08}. The condensates that appear to dominate, based on the abundances of elements available to condense, are corundum (Al$_2$O$_3$), iron (Fe), enstatite (MgSiO$_3$), and forsterite (Mg$_2$SiO$_4$), and these species form cloud layers, removing atoms found within the clouds from the lower pressure atmosphere above \cp{Fegley96, Lodders02, Lodders03, Lodders06, Visscher10}. Within windows between major molecular absorption bands, there is little gas opacity so, in models without clouds, the emergent flux comes from hotter layers deep within the atmosphere. Cloud opacity tends to suppress the flux in the near-infrared within these windows; a thick cloud layer limits the depth from which the planet can radiate, removing some of the flux at these wavelengths, and forcing it to other wavelengths \cp{AM01}. 

When the opacity of these clouds is included in radiative-convective equilibrium models of brown dwarf atmospheres, the resulting model spectra match those of observed L dwarfs \cp{Cushing06, Cushing08, Saumon08, Stephens09}. Observations show that there is a range of colors for a given spectral type, which are believed to be associated with cloud variations or metallicity, but the details of this are not fully understood. Regardless, observed colors and spectra of L dwarfs cannot be well-matched without a significant cloud layer \cp{Burrows06}. 

\subsubsection{T Dwarfs}

As a brown dwarf continues to cool, it undergoes a significant transformation in its observed spectrum when it reaches an effective temperature of approximately 1400 K. Objects cooler than this transitional effective temperature begin to show methane absorption features in their near-infrared spectra and, when these features appear, are classified as T dwarfs \cp{Burgasser02, Kirkpatrick05}. Within a small range of effective temperature, the iron and silicate clouds become dramatically less important. \ct{Marley10} show that this transition could potentially be explained by the breaking up of these cloud layers into patchy clouds, but the details of the transition are still very much unknown. However, the recent discovery of highly photometrically variable early T dwarfs suggests that cloud patchiness may indeed play a role \cp{Radigan12, Artigau09}.  Regardless, as the clouds dissipate, the atmospheric windows in the near-infrared clear. Flux emerges from deeper, hotter atmospheric layers, and the brown dwarf becomes much bluer in $J-K$ color (see Figure \ref{intro-colormag}). 

\subsubsection{History of Modeling T Dwarfs}

\begin{figure}[]
\begin{minipage}[b]{\linewidth}
 \includegraphics[width=3.6in]{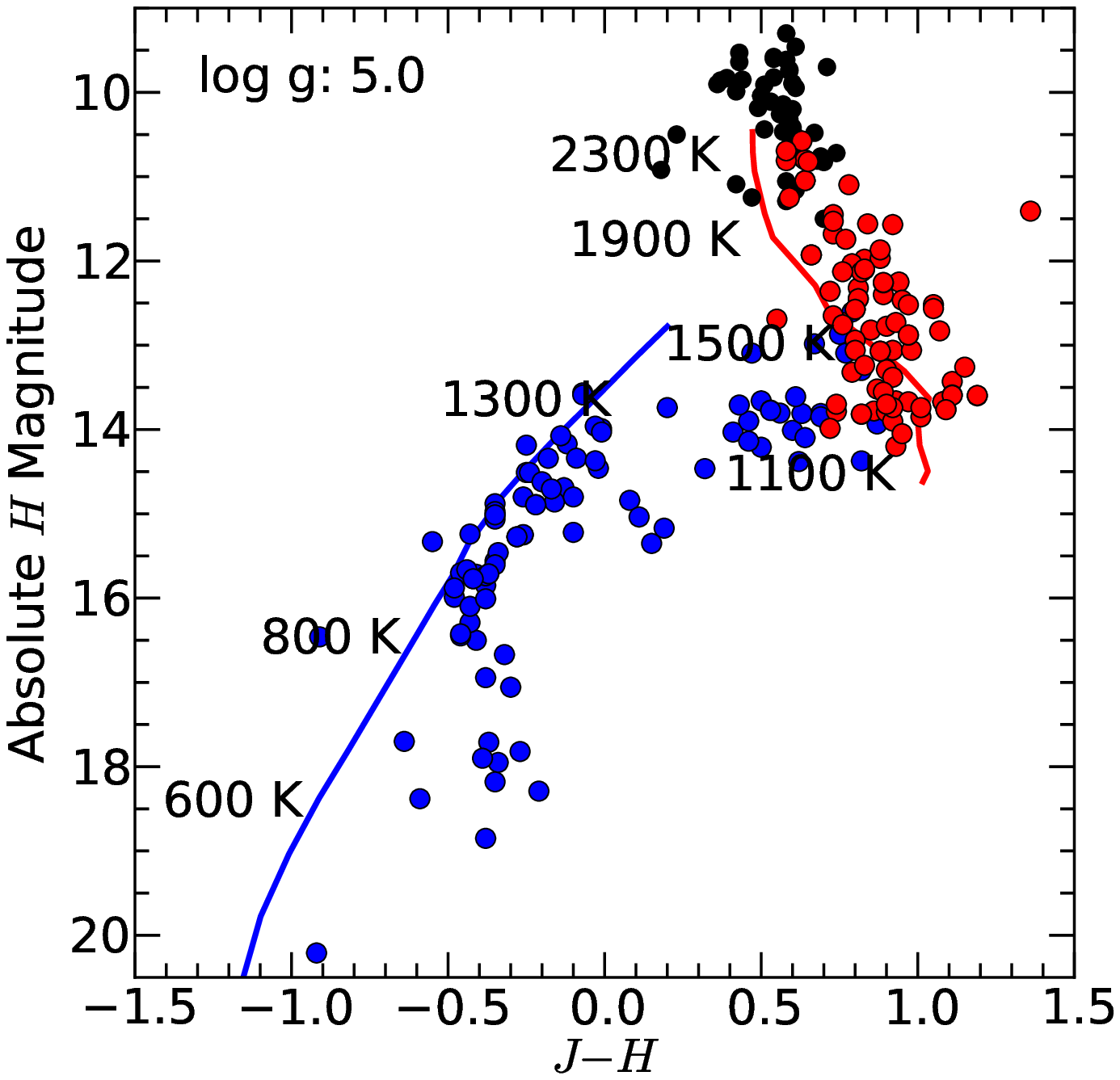}
 \vspace{-4mm}
\end{minipage}
\begin{minipage}[b]{\linewidth}
 \includegraphics[width=3.6in]{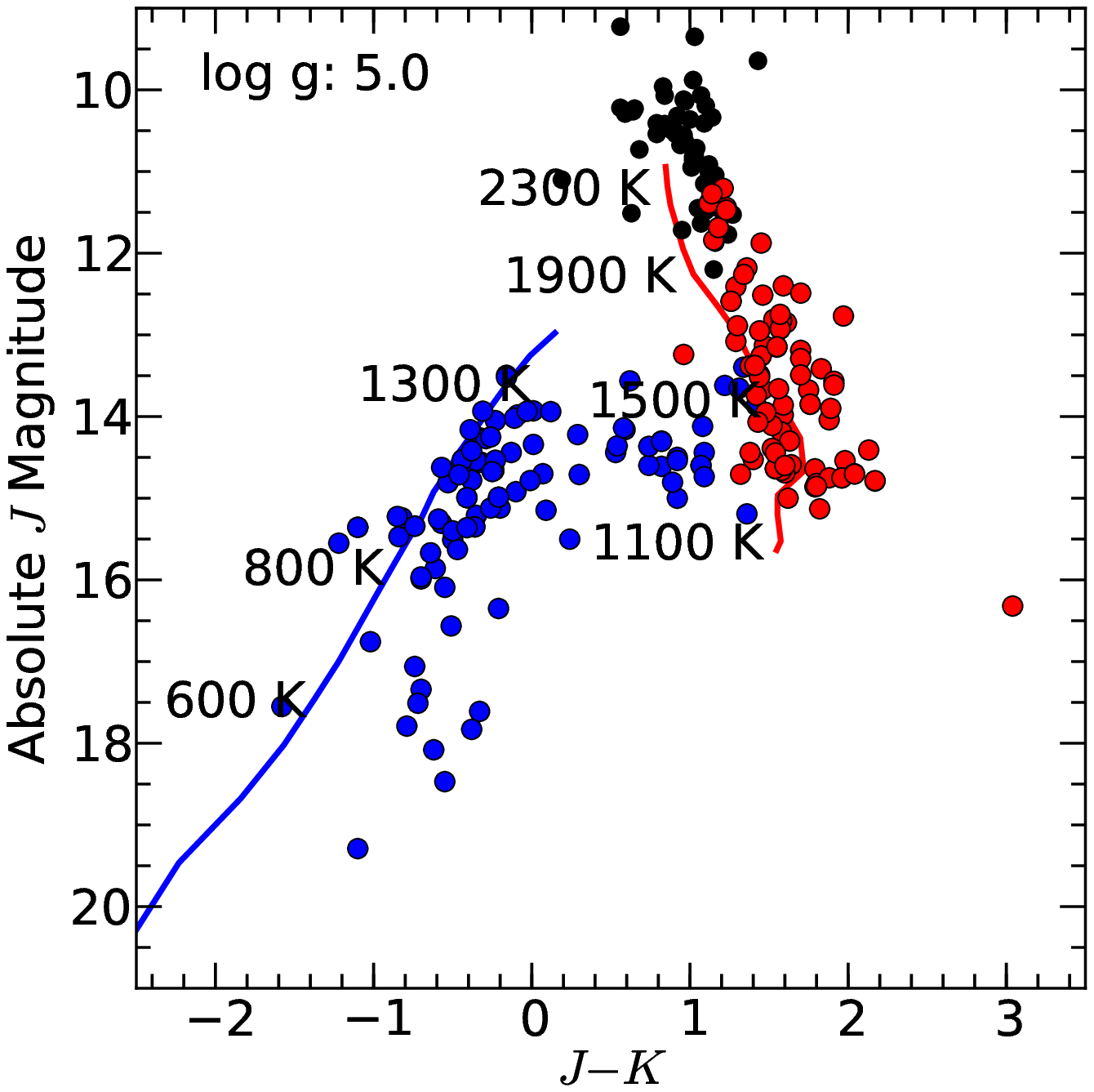}
  \vspace{-4mm}
\end{minipage}
 \caption{Color-magnitude diagrams of L and T dwarfs. \emph{Top:} Observed brown dwarf $J-H$ color is plotted against the absolute $H$ magnitude for all known brown dwarfs with measured parallax. M dwarfs are plotted as black circles, L dwarfs as red circles, and T dwarfs as blue circles. Observational data are from \ct{Dupuy12}. Models are plotted as solid lines. Blue lines are cloudless models and red lines are cloudy (\fsed=2) models that include iron, silicate, and corundum clouds. Each labeled temperature marks the approximate location of the model with that effective temperature. The surface gravity of all models is log $g$ =5.0 (1000m/s$^2$).  \emph{Bottom: } Same as above, but $J-K$ color is plotted against the absolute $K$ magnitude. }
   \label{intro-colormag} 
\end{figure}

The first T dwarf discovered, Gl229B \cp{Nakajima95, Oppenheimer95}, was modeled by \ct{Marley96}, \ct{Allard96}, \ct{Fegley96}, and \ct{Tsuji96} using cloud-free models. These models assume that the condensate-forming materials have been removed from the gas phase, but do not contribute to the cloud opacity. Early T dwarfs are generally quite well-modeled using cloudless atmospheric models. However, recent observations of cooler T dwarfs suggest that T dwarfs of type T8 or later (\teff $\lesssim$ 800 K) appear to be systematically redder in $J-K$ and $J-H$ colors than the cloudless models predict (see Figure \ref{intro-colormag}). 

One of the challenges of modeling brown dwarf spectra is the uncertainties in the absorption bands of major gas species such as methane and ammonia, as well as absorption due to collisional processes. Recent work by \ct{Saumon12} has modeled a range of brown dwarfs using improved line lists for ammonia from \ct{Yurchenko11} and an improved treatment of the pressure-induced opacity of H$_2$ collisions from \ct{Richard12}. This work improves the accuracy of model near-infrared spectra and reddens the $J-K$ colors of the model spectra with effective temperatures between 500 and 1500 K. The color shift is due to decreased opacity in $K$ band from collision-induced absorption and, for \teff $\lesssim$ 500 K model only, increased ammonia opacity in $J$ band. However, these improvements do not change the colors enough to match the colors of the coolest T dwarfs. 

Clouds are a natural way to redden near-infrared spectra. Cloud opacity limits the emergent flux most prominently in $J$ band, so it reddens the $J-K$ and $J-H$ colors of the models.  \ct{Burgasser10} suggest that the remnants of the iron and silicate clouds could redden these cool T dwarfs, but here we suggest instead that the formation of other condensates, which naturally arise from equilibrium chemistry calculations, may play an important role. 

\subsubsection{Y Dwarfs}

The proposed spectral class Y encompasses brown dwarfs that have cooled below \teff$\sim$500 K; a handful of these cool objects have recently been discovered \cp{Cushing11, Kirkpatrick12}. At these temperatures, NH$_3$ begins to play a more significant role in shaping the near-infrared spectra, and sodium and potassium wane in importance in the optical because they condense into clouds. Appreciable amounts of H$_2$O and NH$_3$ will condense into clouds at \teff$\sim$350 K and $\sim$200 K, respectively, and will further alter Y dwarf spectra. As we discover and characterize more of these cold objects, the study of clouds will be crucial to understand their spectral characteristics. 

\subsection{Secondary Cloud Layers}

Silicate, iron, and corundum, which are the condensates that dominate the cloud opacity in our L dwarf models, are not the only condensates that thermochemical models predict will form in substellar atmospheres as they cool. Other condensates will form at lower temperatures and add to the cloud opacity via the same physical processes that formed the iron and silicate cloud layers. In cool substellar atmospheres, \nas\ (sodium sulfide) has been predicted to play a potentially significant role \cp{Lodders99, Lodders06, Visscher06}. Other species expected to condense at these lower temperatures (roughly 600 to 1400 K) include Cr, MnS, ZnS, and KCl. 

To our knowledge, none of these five condensates have been included in a brown dwarf atmosphere model before now.  \ct{Marley00} estimated column optical depths for several of these species and recognized that \nas\ could be important at low $T_{\rm eff}$ but did not include this species in subsequent modeling because of lack of adequate optical constant data. \ct{Burrows01, Burrows02} noted that \nas\ and KCl will condense in cool T dwarfs, but also noted that the indices of refraction are difficult to find. Helling and collaborators \cp{Helling06} also recognized that some of these species will form condensates in some cases but also did not compute model atmospheres that included this opacity source.  \ct{Fortney05c} noted that some of these species might be detectable in extrasolar planet transit spectra which probe a longer path length through the atmosphere.

\section{Methods}

\subsection{Cloud Model}

To model cloudy T dwarf atmospheres, we modify the \ct{AM01} cloud model. This model has successfully been used to model the effects of the iron, silicate, and corundum clouds on the spectra of L dwarfs \cp{Saumon08,Stephens09}. Here, we do not include the opacity of iron, silicate, and corundum clouds; based on observed trends, we assume that the opacity of these clouds becomes negligible for the early T dwarfs.  We instead include Cr, MnS, \nas, ZnS, and KCl. 

The \ct{AM01} approach avoids treating the highly uncertain microphysical processes that create clouds in brown dwarf and planetary atmospheres. Instead, it aims to balance the advection and diffusion of each species' vapor and condensate at each layer of the atmosphere. It balances the upward transport of vapor and condensate by turbulent mixing in the atmosphere with the downward transport of condensate by sedimentation. This balance is achieved using the equation 

\begin{equation}
-K_{zz} \dfrac{\partial q_t}{ \partial z} - f_{\textrm{sed}} w_*q_c =0, 
\end{equation}
where $K_{zz}$ is the vertical eddy diffusion coefficient, $q_t$ is the mixing ratio of condensate and vapor, $q_c$ is the mixing ratio of condensate, $w_*$ is the convective velocity scale, and \fsed\ is a parameter that describes the efficiency of sedimentation in the atmosphere. 

This calculation provides the total amount of condensate at each layer of the atmosphere. The distribution of particle sizes at each level of the atmosphere is represented by a log-normal distribution in which the modal particle size is calculated using the sedimentation flux. A high sedimentation efficiency parameter \fsed\ results in vertically thinner clouds with larger particle sizes, whereas a lower \fsed\ results in more vertically extended clouds with smaller particles sizes. As a result, a higher \fsed\ corresponds to optically thinner clouds and a lower \fsed\ corresponds to optically thicker clouds. 

The \ct{AM01} cloud model code computes the available quantity of condensible gas above the cloud base by comparing the local gas abundance (accounting for upwards transport by mixing via $K_{\rm zz}$) to the local condensate vapor pressure $p_{\rm vap}$. In cases where the formation of condensates does not proceed by homogeneous condensation we nevertheless compute an equivalent vapor pressure curve as described in Section \ref{cloudcondchem}.

\subsection{Atmosphere Model}

The cloud code is coupled to a 1D atmosphere model that calculates the pressure-temperature profile of an atmosphere in radiative-convective equilibrium. The atmosphere models are described in \ct{Mckay89, Marley96, Burrows97, MM99, Marley02, Saumon08, Fortney08b}.  This methodology has been successfully applied to modeling brown dwarfs with both cloudy and clear atmospheres \cp{Marley96, Marley02, Burrows97, Saumon06, Saumon07, Leggett07a, Leggett07b, Mainzer07, Blake07, Cushing08, Geballe09, Stephens09}. 

In the atmosphere model, the thermal radiative transfer is determined using the ``source function technique'' presented in \ct{Toon89}. Within this method, it is possible to include Mie scattering of particles as an opacity source in each layer. Our opacity database for gases, described extensively in \ct{Freedman08}, includes all the important absorbers in the atmosphere. This opacity database includes two significant updates since \ct{Freedman08}, which are described in \ct{Saumon12}: a new molecular line list for ammonia \cp{Yurchenko11} and an improved treatment of the pressure-induced opacity of H$_2$ collisions \cp{Richard12}.  

Both the cloud model and the chemical equilibrium calculations (see Section \ref{cc}) are coupled with the radiative transfer calculations and the pressure-temperature profile of the atmosphere; this means that a converged model will have a temperature structure that is self-consistent with the clouds and chemistry. 

\subsection{Mie Scattering by Cloud Particles}
\begin{figure}[]
\begin{center}

 \begin{minipage}[b]{\linewidth}
    \includegraphics[width=3.5in]{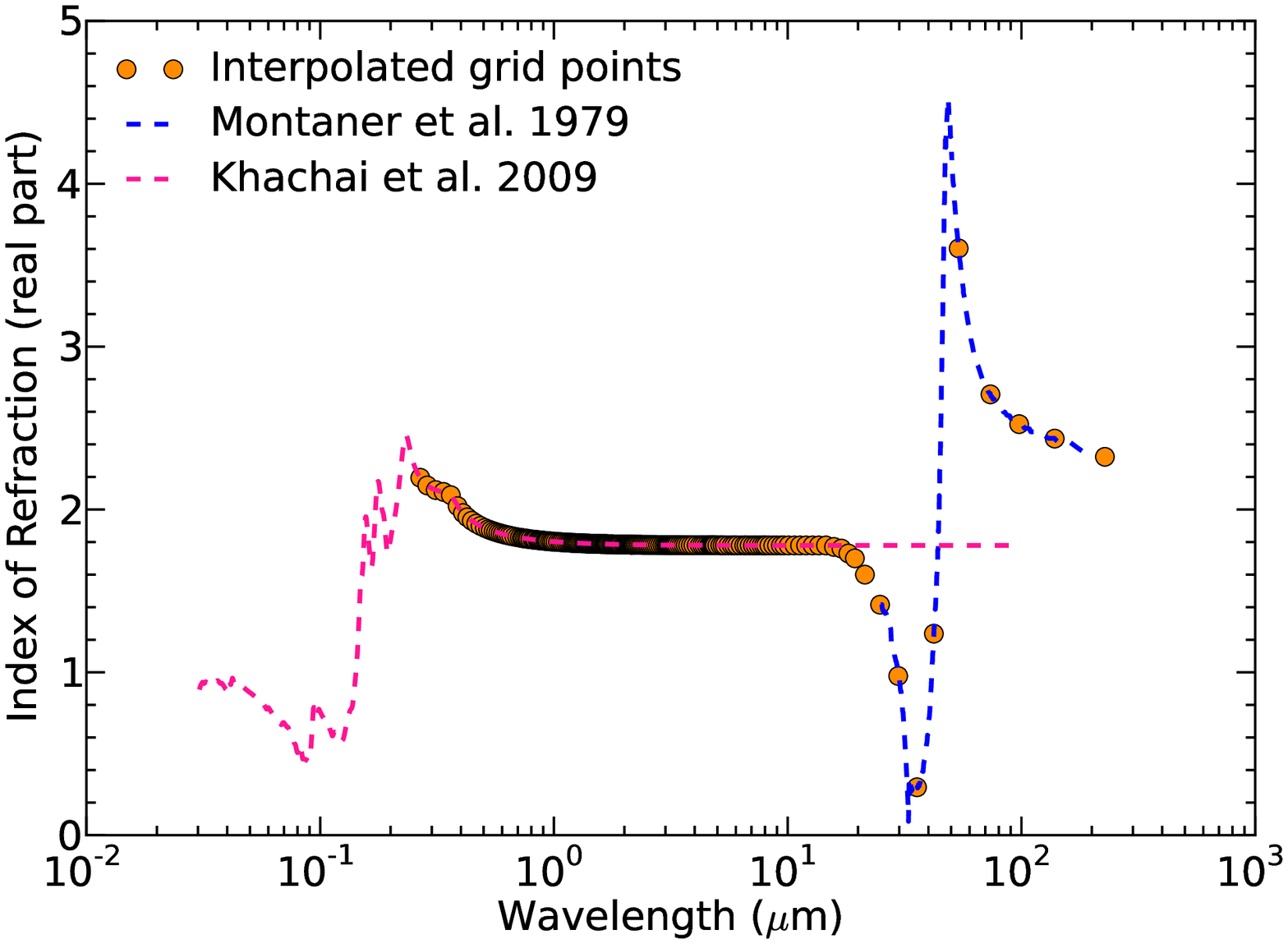}
    \vspace{-5mm}
  \end{minipage}
 \begin{minipage}[b]{\linewidth}
    \includegraphics[width=3.5in]{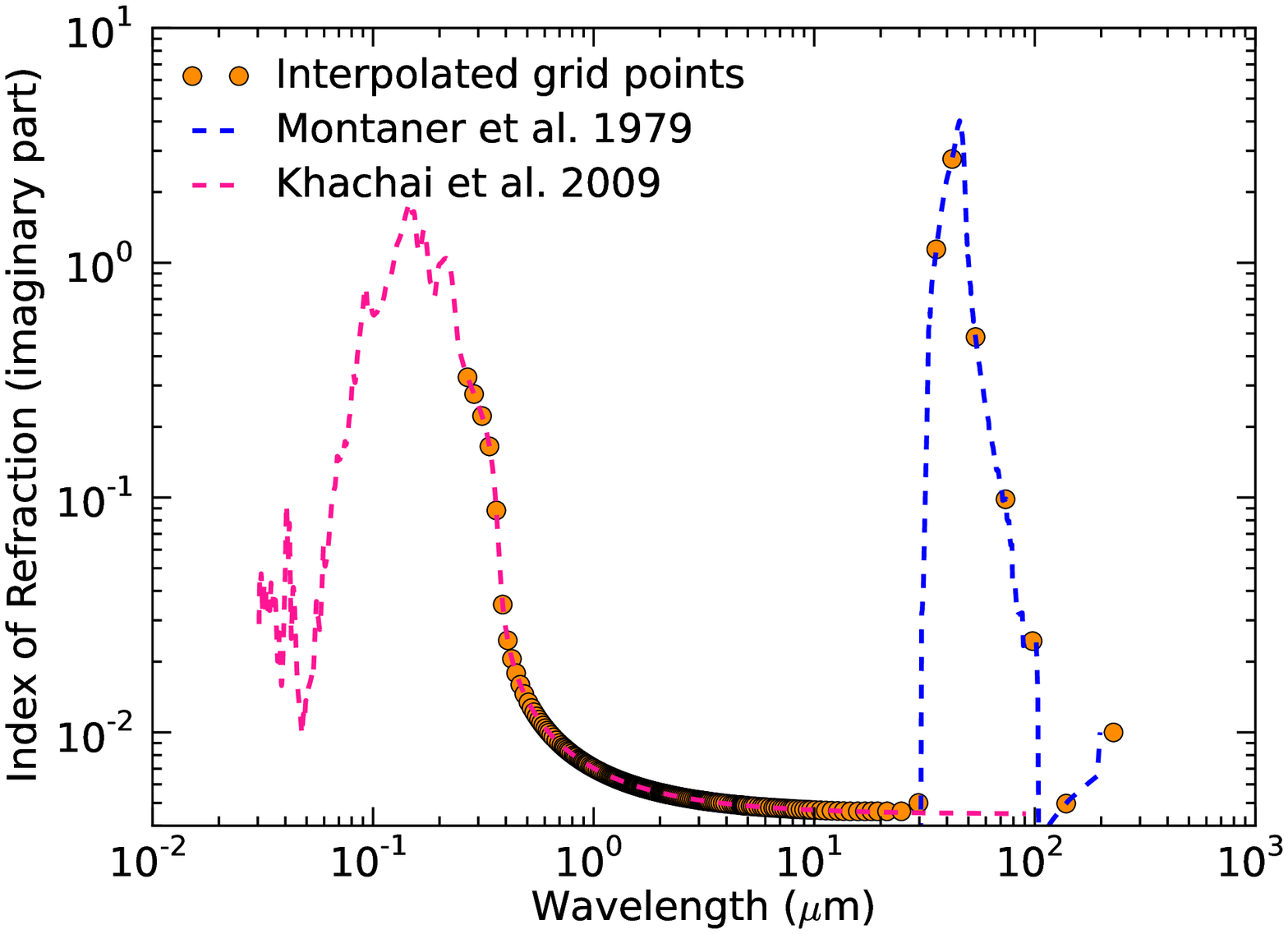}
    \vspace{-5mm}
  \end{minipage}
  \caption{\nas\ index of refraction. The real and imaginary parts of the sodium sulfide index of refraction from the two sources used are plotted as a function of wavelength. \ct{Montaner79} observational data are shown as a blue dashed line. \ct{Khachai09} calculations are shown as a pink dashed line. The interpolated values used for the Mie scattering calculation are shown as pink circles. }
   \label{na2s_n} 
\end{center}
\end{figure}

We calculate the effect of the model cloud distribution on the flux using Mie scattering theory to describe the cloud opacity.  Assuming that particles are spherical and homogeneous, we calculate the scattering and absorption coefficients of each species for each of the particle sizes within the model. In order to make these scattering calculations, we need to understand the optical properties (the real and imaginary parts of the index of refraction) of each material.

The optical properties were found from a variety of diverse sources, summarized in Table \ref{indexofref}. To calculate Mie scattering within the model atmosphere, we use a grid of optical properties at wavelengths from 0.268 to 227 \micron. Where data were not available, we extrapolated the available data, following trends for similar known molecules. 

The molecules with the most complete published optical properties are ZnS and KCl, both of which are obtained from \ct{Querry87}, who tabulates the optical constants for 24 different minerals. 

Optical properties for Cr are published in \ct{Stashchuk84} from 0.26 to 15 \micron. The optical properties from 15 \micron\ to 227 \micron\ were linearly extrapolated from these experimental data following the trend of the optical properties of Fe. Various extrapolations were tested; the choice of optical properties beyond 15 \micron\ does not change the results of the calculations in any meaningful way. 

Optical properties for MnS are published in \ct{Huffman67}, from 0.09 to 13 \micron. Optical properties from 15 \micron\ to 227 \micron\ are extrapolated, following the trends of the other two studied sulfide condensates ZnS and \nas. 

The optical properties for \nas, the clouds with the largest optical depth, are combined from two different sources. \ct{Montaner79} provides experimental data in the infrared, from 25 to 198 \micron. \ct{Khachai09} provides first principles calculations of the optical properties from 0.03 to 91 \micron. In the region of overlap, the \ct{Montaner79} laboratory values are used.  The real and imaginary parts of the index of refraction are plotted in Figure \ref{na2s_n}. 

\begin{deluxetable}{lll}
\small
\tablecolumns{3}
\tablewidth{0pc}
\tablecaption{Sources of Optical Properties}
\tablehead{Species & Source &  Wavelength Range}
\startdata
  KCl   & \ct{Querry87} & 0.22-167 \micron \\
  ZnS  & \ct{Querry87} & 0.22-167 \micron \\
  MnS &  \ct{Huffman67} & 0.09-13 \micron \\
  Cr     &  \ct{Stashchuk84} & 0.26-15 \micron \\
  Na$_2$S & \ct{Montaner79} & 25-198 \micron \\
             & \ct{Khachai09} & 0.03-91 \micron \\
\enddata
  \label{indexofref}
\end{deluxetable}

\subsection{Chemistry Models} \label{cc}

\subsubsection{Gas Phase Chemistry} \label{gaschem}
The abundances of molecular, atomic, and ionic species are calculated using thermochemical equilibrium following the models of \ct{Fegley94, Fegley96, Lodders99, Lodders02,Lodders02b, Lodders03, Lodders06, Lodders09}. We adopt solar-composition elemental abundances from \ct{Lodders03}. The differences between \ct{Lodders03} and newer abundance measurements \cp[e.g.][]{Asplund09} are not large enough to significantly alter the condensation temperatures considered in the paper. \ct{Lodders03} abundances were therefore selected for consistency with previous modeling efforts by our groups.   The abundances of condensate-forming elements are listed in Table {2}.  We assume uniform heavy element abundance ratios over a range of metallicities from [Fe/H] = -0.5 to [Fe/H] = +0.5 in order to explore the metallicity dependence of the condensation temperature expressions. 

\subsubsection{Cloud Condensation Chemistry} \label{cloudcondchem}

A simplified equilibrium condensation approach is used to calculate saturation vapor pressures and condensation curves (see Figure 3) for Cr, MnS, Na$_{2}$S, ZnS, and KCl as a function of pressure, temperature, and metallicity, based upon the more comprehensive thermochemical models of Lodders \& Fegley (see \ref{gaschem}) and \ct{Visscher06,Visscher10}.  In each case, we consider condensation from the most abundant Cr-, Mn-, Na-, Zn-, and K-bearing gas phases at the cloud base as predicted by the chemical models.  The relative mass of each cloud (relative to Na$_{2}$S) is listed in Table 3, assuming complete removal of available condensate material from the gas phase.

\begin{deluxetable}{lll}
\small
\tablecolumns{3}
\tablewidth{0pc}
\tablecaption{Abundances of condensate-forming elements}
\tablehead{Element & $A(\rm El)$$^a$ &  Major condensate}
\startdata
  Fe   & $7.54 \pm 0.03$ & Fe \\
  Si  &  $7.61 \pm 0.02$& Mg$_2$SiO$_4$, MgSiO$_3$ \\
  Mg  &  $7.62 \pm 0.02$& Mg$_2$SiO$_4$, MgSiO$_3$ \\
  O  &  $8.76 \pm 0.05$& Mg$_2$SiO$_4$, MgSiO$_3$, Al$_2$O$_3$, H$_2$O \\
  Al  &  $6.54 \pm 0.02$& Al$_2$O$_3$, CaAl$_{12}$O$_{19}$, \\
        &                               & CaAl$_2$O$_4$, Ca$_2$Al$_2$SiO$_7$\\
  Na  &  $6.37 \pm 0.03$& \nas \\
  Zn  &  $4.70 \pm 0.04$& ZnS \\
  Mn  &  $5.58 \pm 0.03$& MnS \\
  S  &  $7.26 \pm 0.03$& \nas, ZnS, MnS \\  
  Cr  &  $5.72 \pm 0.05 $& Cr \\
  K  &  $5.18 \pm 0.05 $& KCl \\
  Cl  &  $5.33 \pm 0.06 $& KCl \\
\enddata
\tablecomments{$^{a}$ Where $A(\rm El) = log[n(\rm{El)}/n(H)] + 12$ }

  \label{abundances}
\end{deluxetable}

Chromium metal is the most refractory of the clouds considered here and condenses from monatomic Cr gas via the reaction
\begin{equation}
\textrm{Cr} = \textrm{Cr(s)}.
\end{equation}
where `(s)' indicates a solid phase.  The condensation condition for Cr-metal is defined by
\begin{equation}\label{rxn:Cr}
p_{\textrm{Cr}}^* \ge p_{\textrm{Cr}}',
\end{equation}
where $p_{\textrm{Cr}}'$ is the saturation vapor pressure of Cr gas in equilibrium with Cr-metal and $p_{\textrm{Cr}}^*$ is the partial pressure of Cr below the cloud for a solar-composition gas ($p_{\textrm{Cr}}^*=q_{\textrm{Cr}}^* p_t$, where $q_{\textrm{Cr}}^*$ is the mole fraction abundance of Cr and $p_t$ is the total atmospheric pressure).  Upon condensation, the thermodynamic activity of Cr-metal is unity and the equilibrium constant ($K_{p}$) expression for reaction (\ref{rxn:Cr}) can be written as
\begin{equation}
p_{\textrm{Cr}}'= K_{p}^{-1}.
\end{equation}
Substituting for the temperature-dependent value of $K_p$, the saturation vapor pressure of Cr above the cloud base can be estimated using the expression
\begin{equation}
\log p_{\textrm{Cr}}' \approx 7.490 - 20592/T,
\end{equation}
for $T$ in Kelvin and $p$ in bars.  Below the cloud, we assume that Cr gas is approximately representative of the elemental Cr abundance in solar composition gas (see Table \ref{cloudstable}):
\begin{equation}
\log p_{\textrm{Cr}}^* \approx -6.052 + \log p_t + [\textrm{Fe/H}].
\end{equation}
The condensation temperature as a function of the total atmospheric pressure ($p_t$) and metallicity can therefore be approximated by setting $p_{\textrm{Cr}}^*=p_{\textrm{Cr}}'$ and rearranging to give
\begin{equation}
10^{4}/T_{\textrm{cond}}\textrm{(Cr)} \approx 6.576 - 0.486 \log p_t - 0.486 [\textrm{Fe/H}].
\end{equation}
This expression yields a condensation temperature near $\sim1520$ K at 1 bar and solar metallicity \cp[cf.][]{Lodders06}, and shows that greater total pressures and/or metallicities will lead to higher condensation temperatures.  Condensation of Cr-metal effectively removes gas-phase chromium from the atmosphere, and the abundances of Cr-bearing gases rapidly decrease with altitude above the cloud.

Our modeling of sulfide condensation chemistry follows that of \ct{Visscher06}, and the condensation reactions and temperature-dependent expressions presented here are taken from that study.  The deepest sulfide cloud expected in brown dwarf atmospheres is MnS, which forms via the reaction
\begin{equation}  \label{rxn:MnS}
\textrm{H}_{2}\textrm{S} + \textrm{Mn} = \textrm{MnS(s)} + \textrm{H}_{2},
\end{equation}
The formation of the MnS cloud is limited by the total manganese abundance, which is 2\% of the sulfur abundance in a solar-composition gas.  The condensation curve for MnS is thus derived by exploring the chemistry of monatomic Mn, which is the dominant Mn-bearing gas near the cloud base.  Using results from \ct{Visscher06}, the saturation vapor pressure of Mn above the cloud is given by
\begin{equation}
\log p_{\textrm{Mn}}' \approx 11.532 - 23810/T - [\textrm{Fe/H}],
\end{equation}
where the metallicity dependence comes from H$_{2}$S (the dominant S-bearing gas) remaining in the gas phase above the MnS cloud base.  By setting $p_{\textrm{Mn}}^*=p_{\textrm{Mn}}'$, the MnS condensation curve  is approximated by \ct{Visscher06}:
\begin{equation}
10^{4}/T_{\textrm{cond}}\textrm{(MnS)} \approx 7.447 - 0.42 \log p_t - 0.84[\textrm{Fe/H}],
\end{equation}
giving a condensation temperature near $\sim1340$ K at 1 bar in a solar-metallicity gas.  

The Na$_{2}$S cloud is the most massive of the metal sulfide clouds expected to form in brown dwarf atmospheres  because Na is more abundant than either Mn or Zn in a solar-composition gas (see Table \ref{cloudstable}).  Sodium sulfide condenses via the net thermochemical reaction
\begin{equation}
\textrm{H}_{2}\textrm{S} + 2\textrm{Na} = \textrm{Na}_{2}\textrm{S(s)} + \textrm{H}_{2}.
\end{equation}
The mass of the Na$_{2}$S cloud is limited by the elemental abundance of sodium, which is 13\% of the abundance of sulfur in a solar composition gas.  Using results from \ct{Visscher06}, the saturation vapor pressure of Na above the cloud base is given by
\begin{equation}
\log p_{\textrm{Na}}' \approx 8.550 - 13889/T - 0.50 [\textrm{Fe/H}],
\end{equation}
where the metallicity dependence results from H$_{2}$S remaining in the gas phase above the Na$_{2}$S cloud and from the stoichiometry of Na and H$_{2}$S in the condensation reaction.
The condensation temperature (where $p_{\textrm{Na}}^*=p_{\textrm{Na}}'$) is given by \ct{Visscher06}:
\begin{equation}
10^{4}/T_{\textrm{cond}}(\textrm{Na}_{2}\textrm{S}) \approx 10.045 - 0.72 \log p_t - 1.08 [\textrm{Fe/H}],
\end{equation}
indicating condensation near $\sim1000$ K at 1 bar in a solar-metallicity gas.  

The ZnS cloud layer forms via the reaction
\begin{equation}
\textrm{H}_{2}\textrm{S} + \textrm{Zn} = \textrm{ZnS(s)} + \textrm{H}_{2},
\end{equation}
The formation of the ZnS cloud is limited by the total Zn abundance, which is 0.3\% of the S abundance in a solar-composition gas.  Using results from \ct{Visscher06}, the saturation vapor pressure of Zn over condensed ZnS is given by
\begin{equation}
\log p_{\textrm{Zn}}' \approx 12.812 - 15873/T -  [\textrm{Fe/H}]
\end{equation}
The condensation curve (where $p_{\textrm{Zn}}^*=p_{\textrm{Zn}}'$) is approximated by \ct{Visscher06}:
\begin{equation}
10^{4}/T_{\textrm{cond}}\textrm{(ZnS)} \approx 12.527 - 0.63 \log p_t - 1.26 [\textrm{Fe/H}],
\end{equation}
giving a condensation temperature of $\sim800$ K at 1 bar in a  solar-metallicity gas.

Our treatment of KCl condensation chemistry is similar to that for Cr-metal and the metal sulfides.  With decreasing temperatures, KCl replaces neutral K as the dominant K-bearing gas in brown dwarf atmospheres \cp{Lodders99,Lodders06}.  The KCl cloud layer is thus expected to form via the net thermochemical reaction
\begin{equation}\label{rxn:KCl}
\textrm{KCl} = \textrm{KCl(s)},
\end{equation}
and condenses as a solid over the range of conditions considered here.  The vapor pressure of KCl above condensed KCl(s) is given by
\begin{equation}
\log p_{\textrm{KCl}}' \approx 7.611 - 11382/T,
\end{equation}
derived from the equilibrium constant expression for the condensation reaction.
The mass of the KCl cloud is limited by the total potassium abundance, which is 70\% of the chlorine abundance in a solar-composition gas \cp{Lodders03}.  Note that other K-bearing species may remain relatively abundant near cloud condensation temperatures, particularly at higher pressures (e.g., see \citealt{Fegley94} and \citealt{Lodders99} for a more detailed discussion of chemical speciation).  However, KCl is the dominant K-bearing gas near the cloud base for the relevant $P-T$ conditions expected in cool brown dwarf atmospheres (see Figure 3) over the range of metallicities (-0.5 to +0.5 dex) considered here.   For simplicity we therefore assume that KCl is approximately representative of the elemental K abundance below the cloud, given by
\begin{equation}
\log p_{\textrm{KCl}}^* \approx -6.593 + \log p_t + [\textrm{Fe/H}].
\end{equation}
\noindent The condensation temperature as a function of pressure and metallicity is  estimated by setting $p_{\textrm{KCl}}^*=p_{\textrm{KCl}}'$ and rearranging to give
\begin{equation}
10^{4}/T_{\textrm{cond}}\textrm{(KCl)} \approx 12.479 - 0.879 \log p_t - 0.879 [\textrm{Fe/H}],
\end{equation}
yielding a condensation temperature near $\sim 800$ K at 1 bar in a solar-metallicity gas \cp[cf.][]{Lodders99,Lodders06}.  In general, the condensation curve expressions demonstrate that condensation temperatures increase with total pressure, as illustrated in Figure 4.  Furthermore, higher metallicities are expected to result in higher condensation temperatures and more massive cloud layers in brown dwarf atmospheres.  In each case, the saturation vapor pressures of cloud-forming species rapidly decrease with altitude above the cloud layers.

\begin{deluxetable}{llc}
\small
\tablecolumns{3}
\tablewidth{0pc}
\tablecaption{Abundances of Condensate-Forming Species}
\tablehead{condensate & $p_x^*$ below cloud base$^a$ &  cloud mass$^b$}
\startdata
Cr & $p_{\textrm{Cr}}^* \approx 8.87\times10^{-7} p_{t} m$ & 0.30\\
MnS & $p_{\textrm{Mn}}^* \approx 6.32\times10^{-7}p_{t} m$ & 0.36\\
Na$_{2}$S & $p_{\textrm{Na}}^* \approx 3.97\times10^{-6} p_{t} m$ & $1.00$ \\
ZnS & $p_{\textrm{Zn}}^* \approx 8.45\times10^{-8} p_{t} m$ & 0.05\\
KCl & $p_{\textrm{KCl}}^* \approx 2.55\times10^{-7} p_{t} m$ & 0.12\\
Fe & $p_{\textrm{Fe}}^* \approx 5.78\times10^{-5} p_{t} m$ & 20.85\\
\enddata
\tablecomments{$^{a}$Where $p_x^*$ is the partial pressure (in bars) of each gas phase species $x$ below the predicted cloud base using solar-composition abundances from \ct{Lodders03}, $p_t$ is the total atmospheric pressure, and the metallicity factor $m$ is defined by $\log m = [\textrm{Fe/H}]$. $^{b}$Total condensate mass relative to the Na$_{2}$S cloud.  Values for Fe shown for comparison.}
\label{cloudstable}
\end{deluxetable}

\subsection{Comparison to Other Cloud Models}

The \ct{AM01} model is one method of several that have been applied to 
cloudy brown dwarf atmospheres. \ct{Helling08} review various cloud modeling techniques
and compare model predictions for various cases.  The most important conceptual differences
between these approaches lies in the assumptions of how condensed phases interact with 
the gas.  

In the chemical equilibrium
approach \cp[e.g.][]{Allard01}, condensed phases remain in contact with the gas phase and can continue to react
with the gas even at temperatures well below the condensation temperature.  As an example, when following this approach,
Fe grains which first condense at temperatures of over 2000 K react with atmospheric $\rm H_2S$ to form FeS below 1000 K.  In the condensation
chemistry approach we employ here, the condensed phases are assumed to sediment out of the atmosphere
and are not available to interact with gas phases at temperatures below the condensation temperature.
Thus Fe grains form a discrete cloud layer and do not react to form FeS.  $\rm H_2S$ consequently
remains in the gas phase and reacts to form condensates as outlined in Section 2.4.2.  Jupiter is an
excellent example of the applicability of this framework, as the presence of H$_2$S in the observable atmosphere is only possible because Fe is sequestered in a deep cloud layer, which prevents the formation of FeS which otherwise deplete other gas phase S species \cp{Fegley94}. The presence of alkali absorption in T dwarfs likewise demonstrates the applicability of condensation chemistry \citep{Marley02}.  A detailed comparison of true equilibrium condensation
and cloud condensate removal from equilibrium can be found in \ct{Fegley94, Lodders06} and references therein.

A different approach is taken by Helling \& Woitke \cp{Helling06} who follow the trajectory of
tiny seed particles of TiO$_2$ that are assumed to be emplaced high in the atmosphere and sink downwards.
As the seeds fall through the atmosphere they collect condensate material.  In \ct{Helling06} and
numerous follow on papers \cp{Helling08, Witte09, Witte11, deKok11} this group
models the microphysics of grain growth given these conditions. Because the background atmosphere
is not depleted of gaseous species until the grains fall through the atmosphere, a compositionally very different set of
grains are formed.  In particular they predict `dirty' grains composed of layers of varying condensates.

A direct comparison between the predictions of the various cloud modeling schools is often difficult because of differing
assumptions of elemental abundances and the background thermal profile.  Modeling
tests in which predictions of the various groups are compared to data would be illuminating, but this
is far beyond the scope of the work reported here.

\subsection{Evolution Model}

In order to calculate absolute magnitudes of the modeled brown dwarfs, we use the results of evolution models which determine the radius of a brown dwarf as it cools and contracts over its lifetime. We use the evolution models of \ct{Saumon08} with the surface boundary condition from cloudless atmospheres. Using a cloudless boundary condition instead of one consistent with these clouds changes the calculated magnitudes of the models very slightly, but does not change the overall trends or results. 

\subsection{Model Grid}

To analyze the effect of these clouds, we generate a grid of 182 model atmospheres at effective temperatures and surface gravities spanning the full range of T dwarfs. We calculate pressure-temperature profiles and synthetic spectra for atmospheres from 400 to 1300 K (50 to 100 K increments), with log($g$) (cgs) of 4.0, 4.5, 5.0, and 5.5 and cloud sedimentation efficiency parameter \fsed=2, 3, 4, and 5. For this study, we use only solar metallicity composition. We then compare these, both photometrically and spectroscopically, to observed T dwarfs.

\section{Results}

\subsection{Model Pressure-Temperature Profiles}

\begin{figure}[]
\includegraphics[width=3.5in]{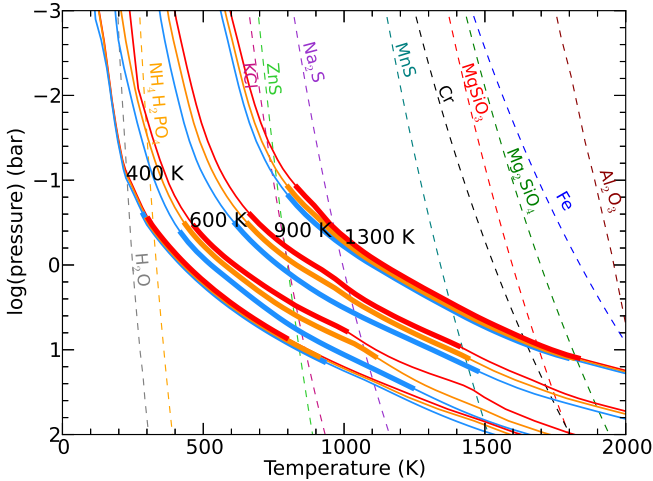}
\caption{The pressure-temperature profiles of model atmospheres are plotted. Models at 400, 600, 900, and 1300 K are shown, and the effective temperature of the model is labeled on the plot. The surface gravity of the 400 K model is log $g$=4.5; for the hotter models, log $g$=5.0. We show cloudless models in blue, and cloudy models with \fsed=2 (red) and 4 (orange). Condensation curves for each condensate species are plotted. The cloudy models include the condensates Cr, MnS, \nas, ZnS, and KCl. Note that for each case, increasing the cloud thickness increases the temperature at a given atmospheric pressure. The 1-6 \micron\ photosphere of each model is shown as a thicker line. } 
\label{PT-results}
\end{figure}

In Figure \ref{PT-results}, we show the pressure-temperature profiles of models with effective temperatures of 400 K, 600 K, 900 K, and 1300 K. The surface gravity of the 400 K models is log $g$=4.5; for the hotter models, log $g$=5.0. We plot models with two different cloud sedimentation efficiencies which include only the \nas, MnS, ZnS, Cr, and KCl clouds. Because \nas\ and MnS are by far the most dominant cloud species (see Section \ref{cloudstructures}), as a shorthand we refer to this collection of clouds as `sulfide clouds.' 

The condensation curves of major and minor species predicted to form by equilibrium chemistry are also plotted. The location of a given cloud base is expected to be where the pressure-temperature profile of the model atmosphere crosses the condensation curve. Each model crosses the condensation curve of each species at very different pressures and, to a lesser extent, temperatures, so we expect that the significance of the clouds will be strongly controlled by the effective temperature of the model. 

The cloudy 400 K and 600 K models have two convection zones. All 900 and 1300 K models have a single deep convection zone. 

For all models, it is clear that, as in previous cloudy models of L dwarfs, adding cloud opacity has a ``blanketing'' effect on the model, increasing the temperature of the atmosphere for a given atmospheric pressure. As the cloud becomes optically thicker, the entire pressure-temperature profile becomes hotter; thus, on a plot of pressure-temperature profiles such as Figure \ref{PT-results}, increasing the cloud opacity moves the whole profile to the right. 

\subsection{Model Spectra} \label{modelspecs}

\begin{figure*}[]
  \begin{minipage}[b]{0.5\linewidth}
    \includegraphics[width=3.7in]{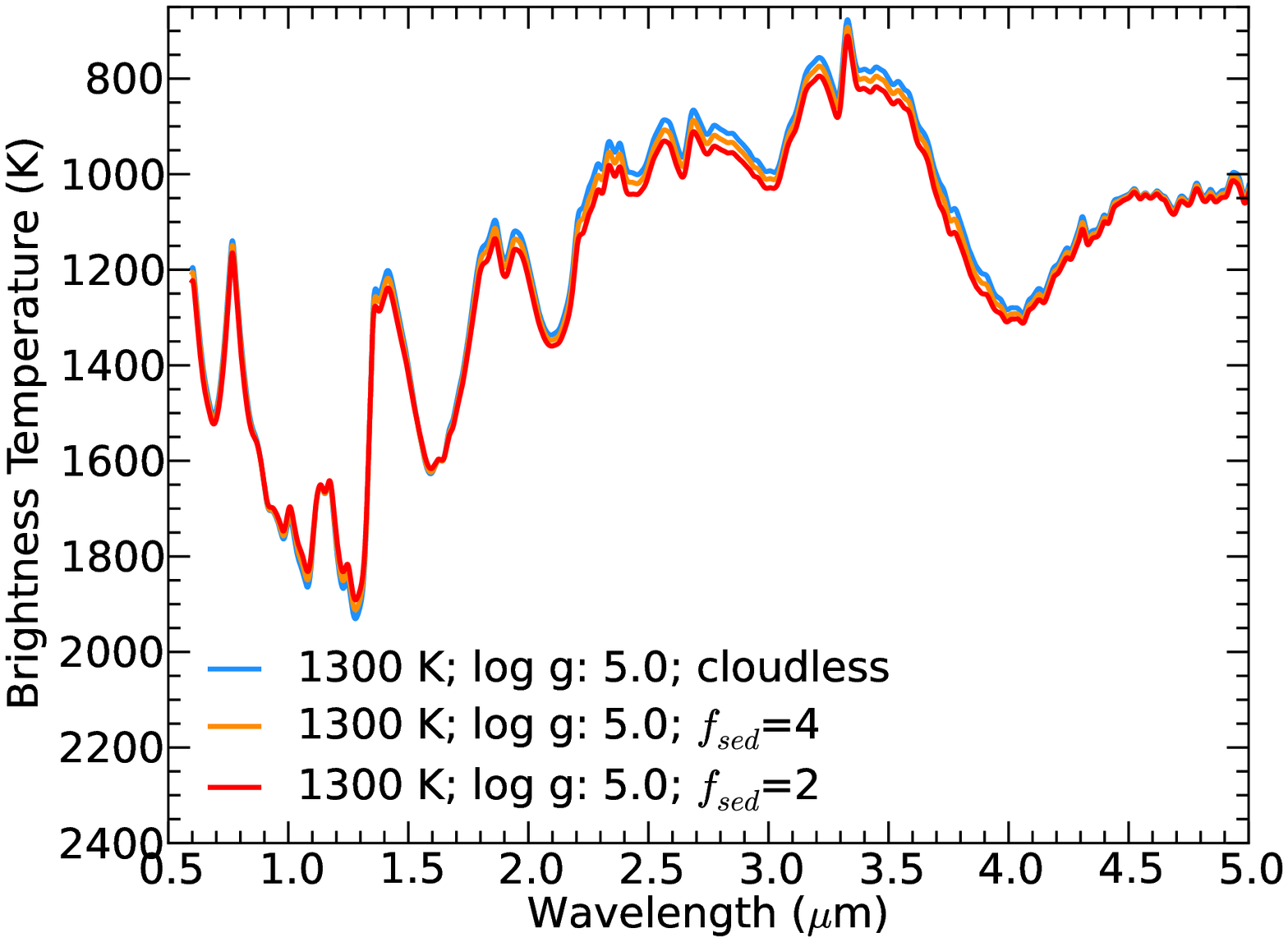}
    \vspace{-2mm}
  \end{minipage}
    \begin{minipage}[b]{0.5\linewidth}
    \includegraphics[width=3.7in]{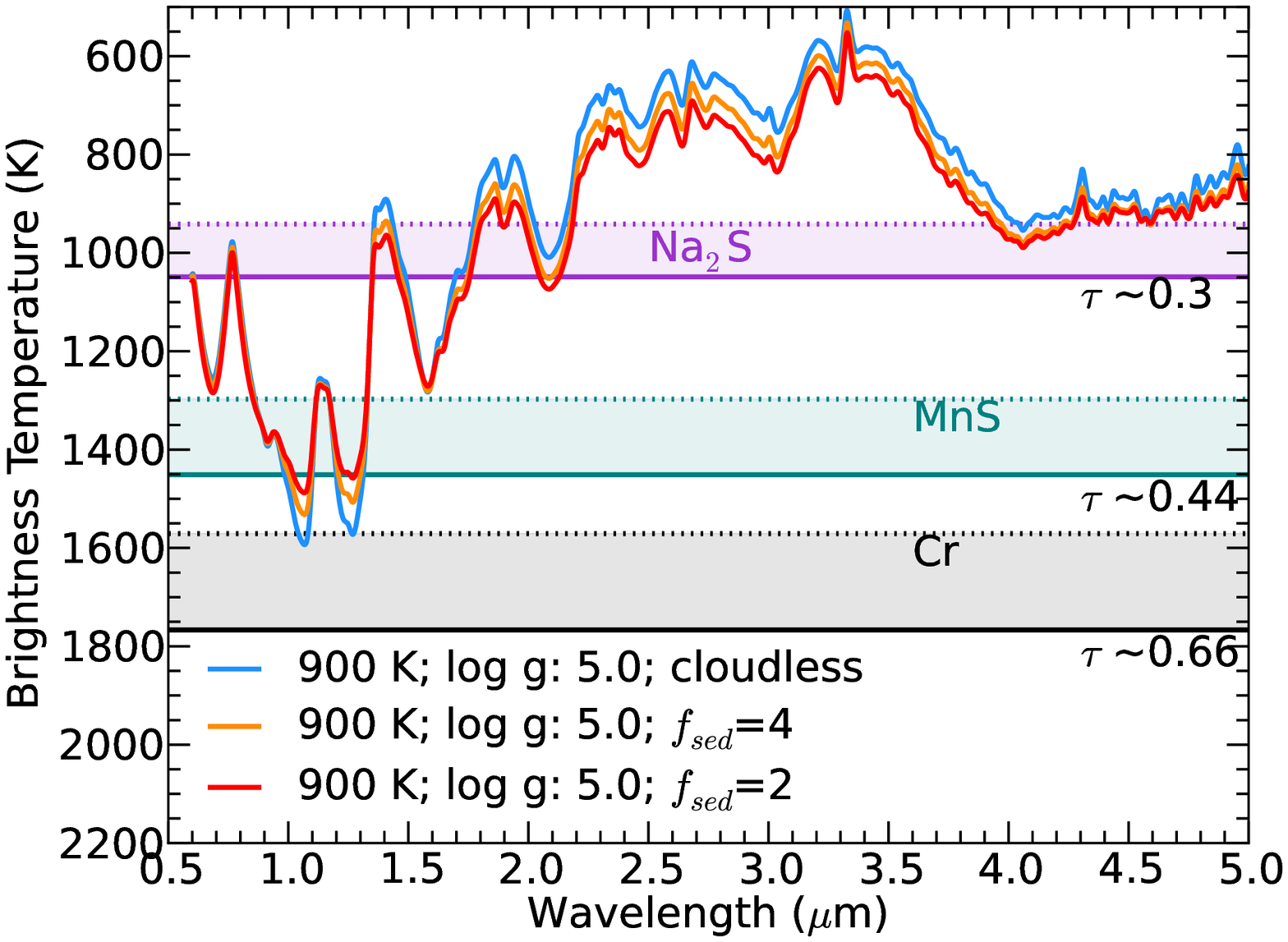}
    \vspace{-2mm}
  \end{minipage}
    \begin{minipage}[b]{0.5\linewidth}
    \includegraphics[width=3.7in]{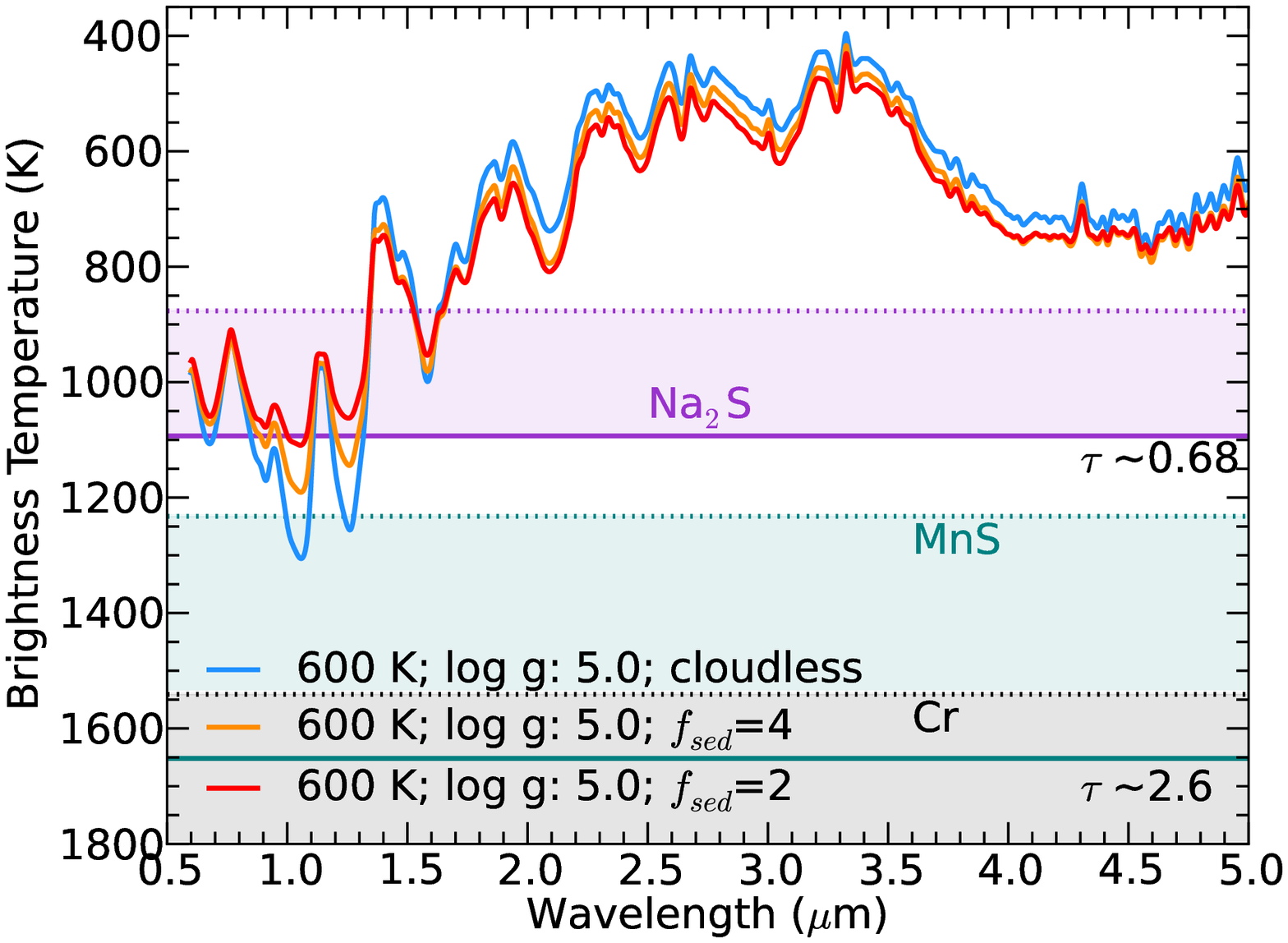}
    \vspace{-2mm}
  \end{minipage}
    \begin{minipage}[b]{0.5\linewidth}
    \includegraphics[width=3.7in]{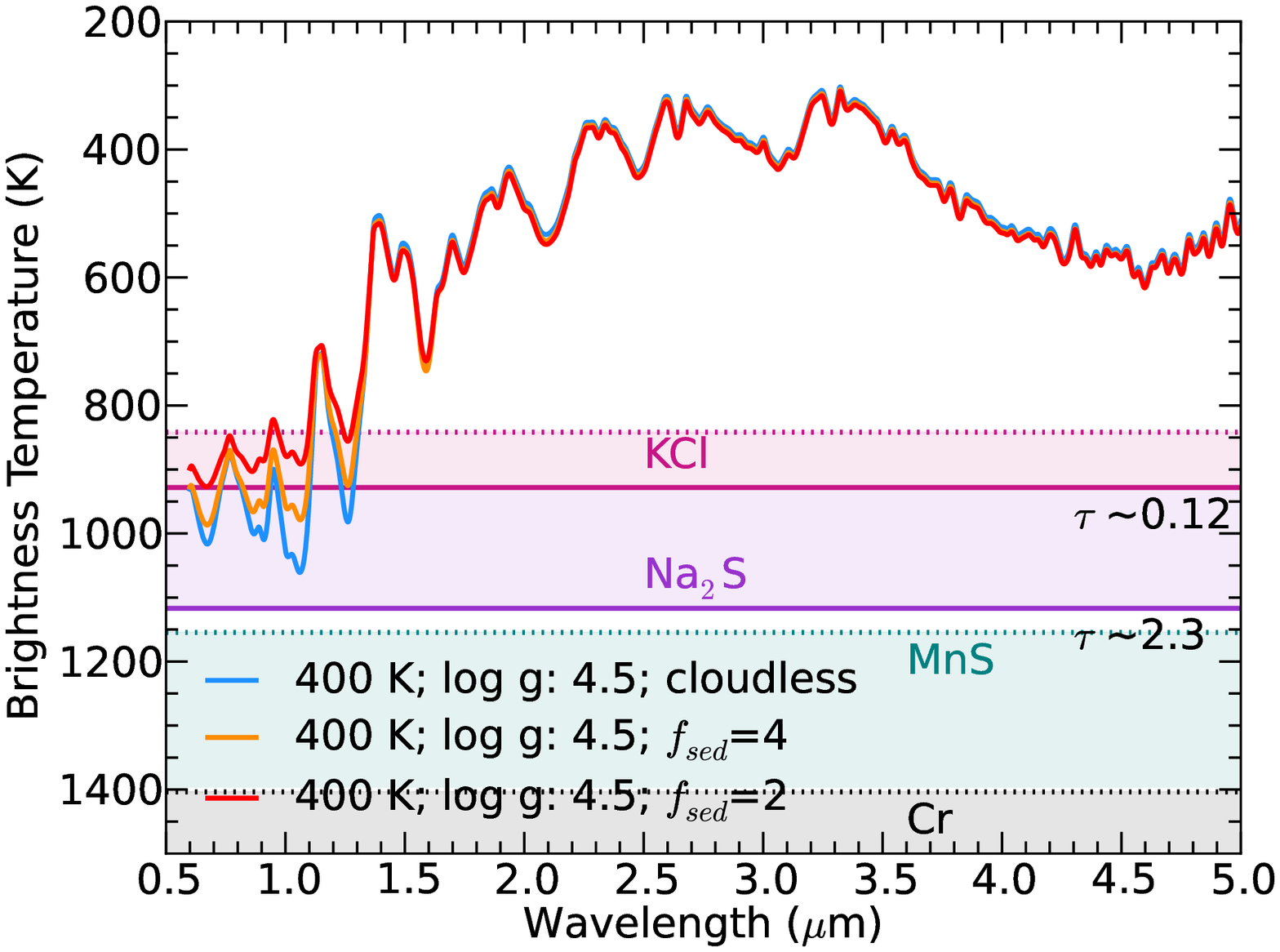}
    \vspace{-2mm}
  \end{minipage}
\caption{The model spectra are plotted as brightness temperature vs.~wavelength. cloudless, \fsed=2, and \fsed=4 models are shown. The solid horizontal line indicates the temperature at the base of the each cloud, and the dashed horizontal line denotes the temperature of the layer in which column extinction optical depth of the cloud reaches 0.1. Note that for all clouds in the \teff\ 1300 K model, the column optical depth model never exceeds 0.1. The maximum column optical depth of the \nas\ clouds ($\tau$ at the cloud base) is calculated using the \fsed=4 models and labeled on each plot. 
}
\label{BT}
\end{figure*}

\begin{figure}[]
\begin{center}
\begin{minipage}[b]{\linewidth}
\includegraphics[width=3.5in]{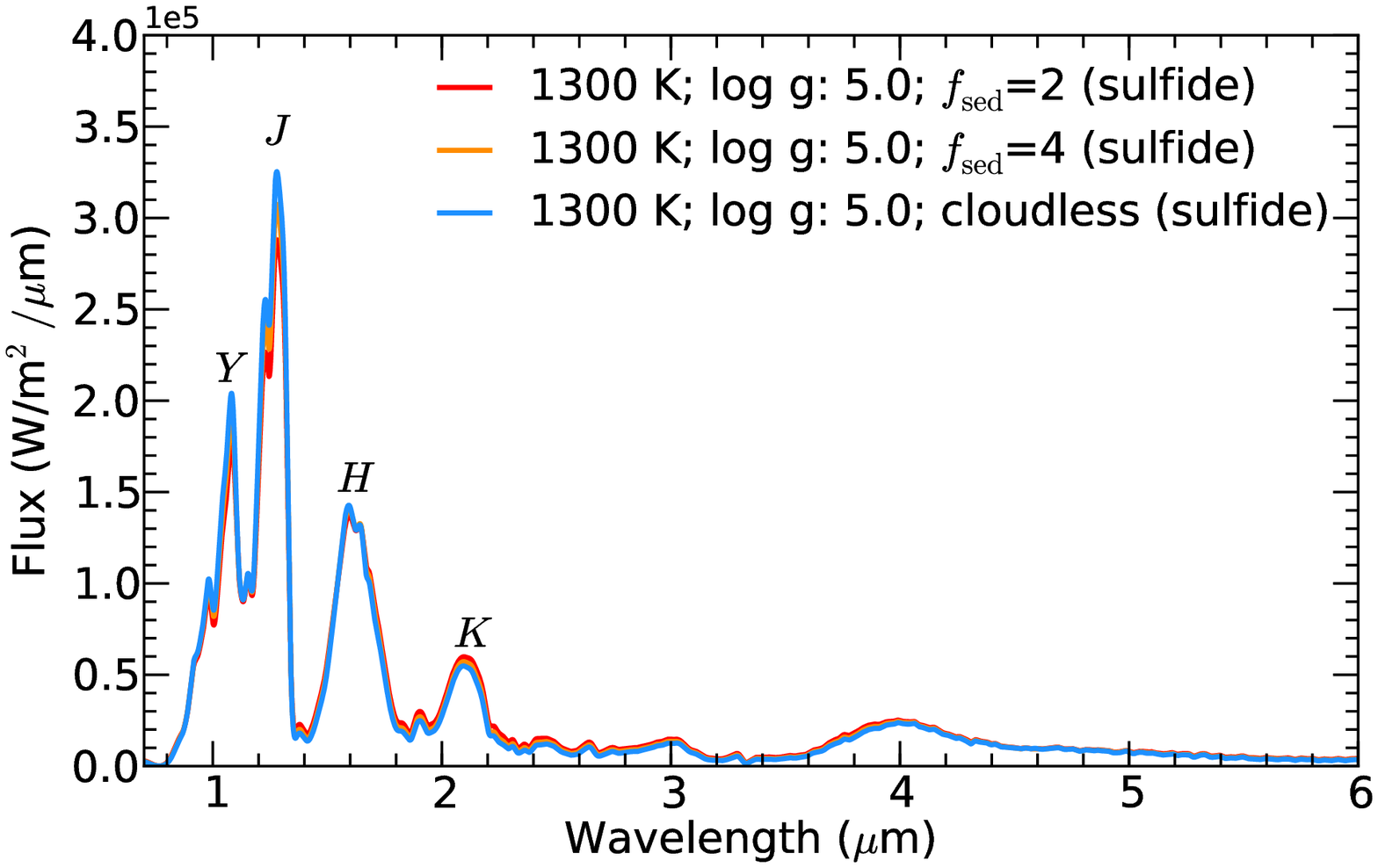}
\vspace{-6mm}
\end{minipage}
\begin{minipage}[b]{\linewidth}
\includegraphics[width=3.5in]{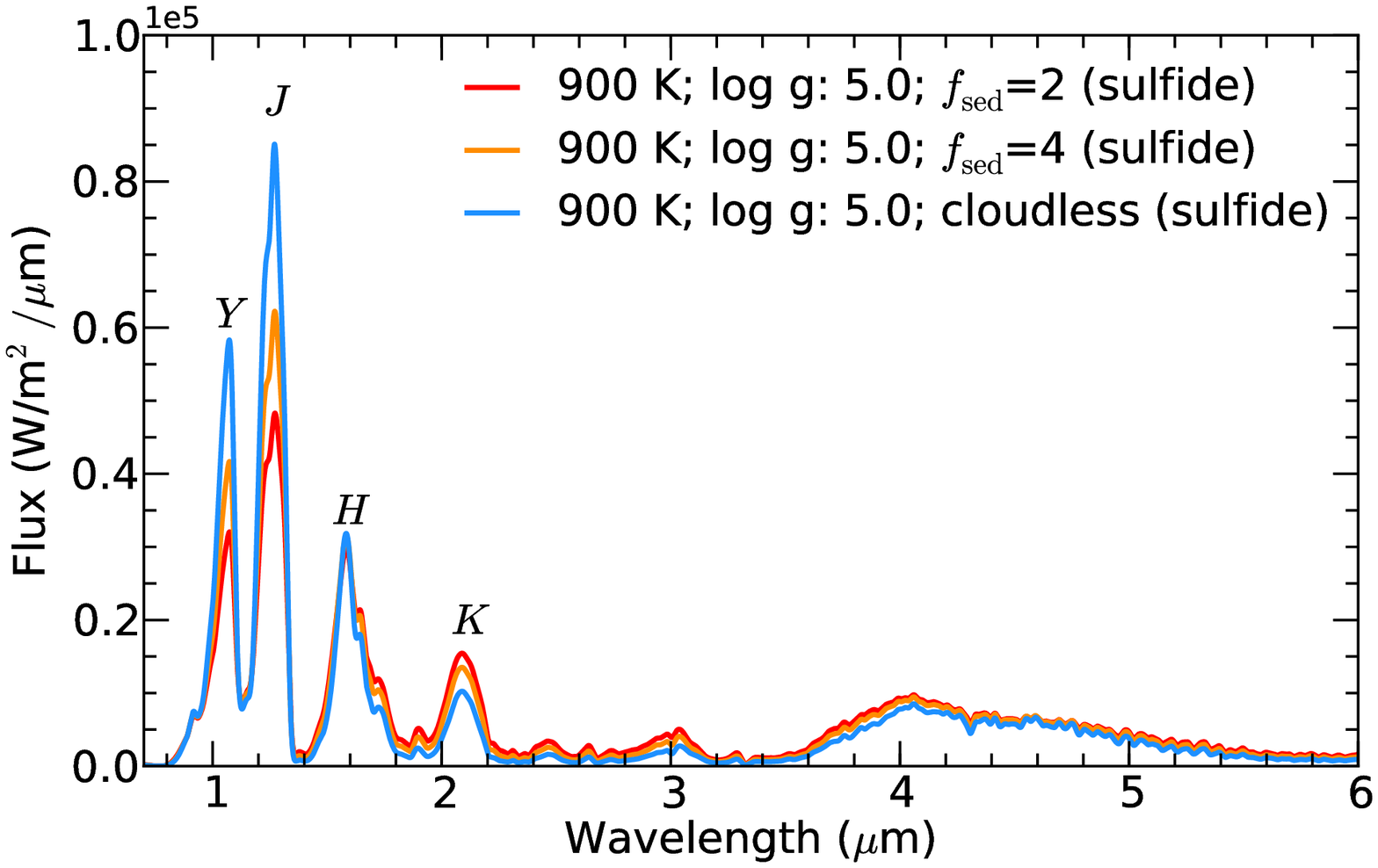}
\vspace{-6mm}
\end{minipage}
\begin{minipage}[b]{\linewidth}
\includegraphics[width=3.5in]{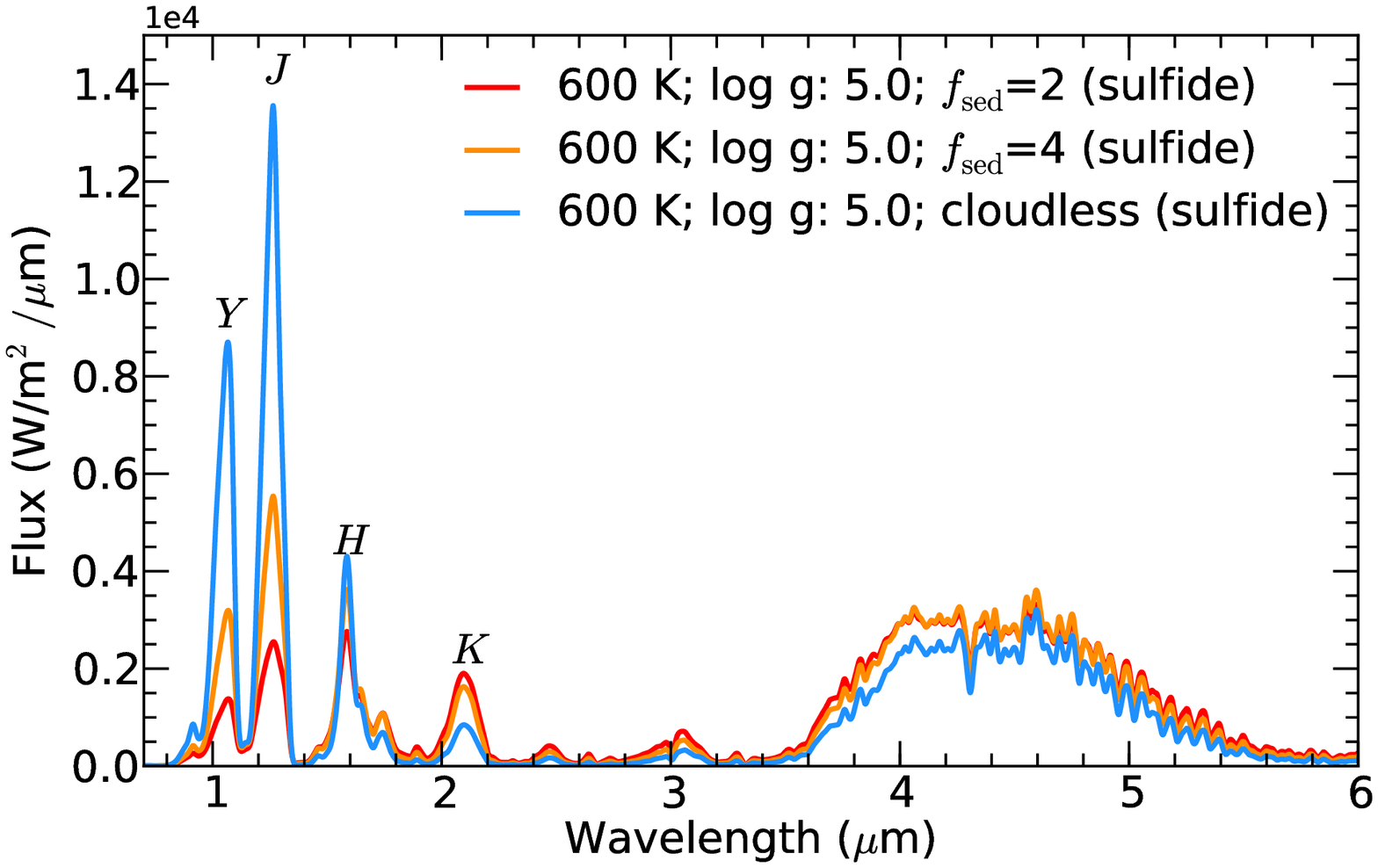}
\vspace{-6mm}
\end{minipage}
\begin{minipage}[b]{\linewidth}
\includegraphics[width=3.5in]{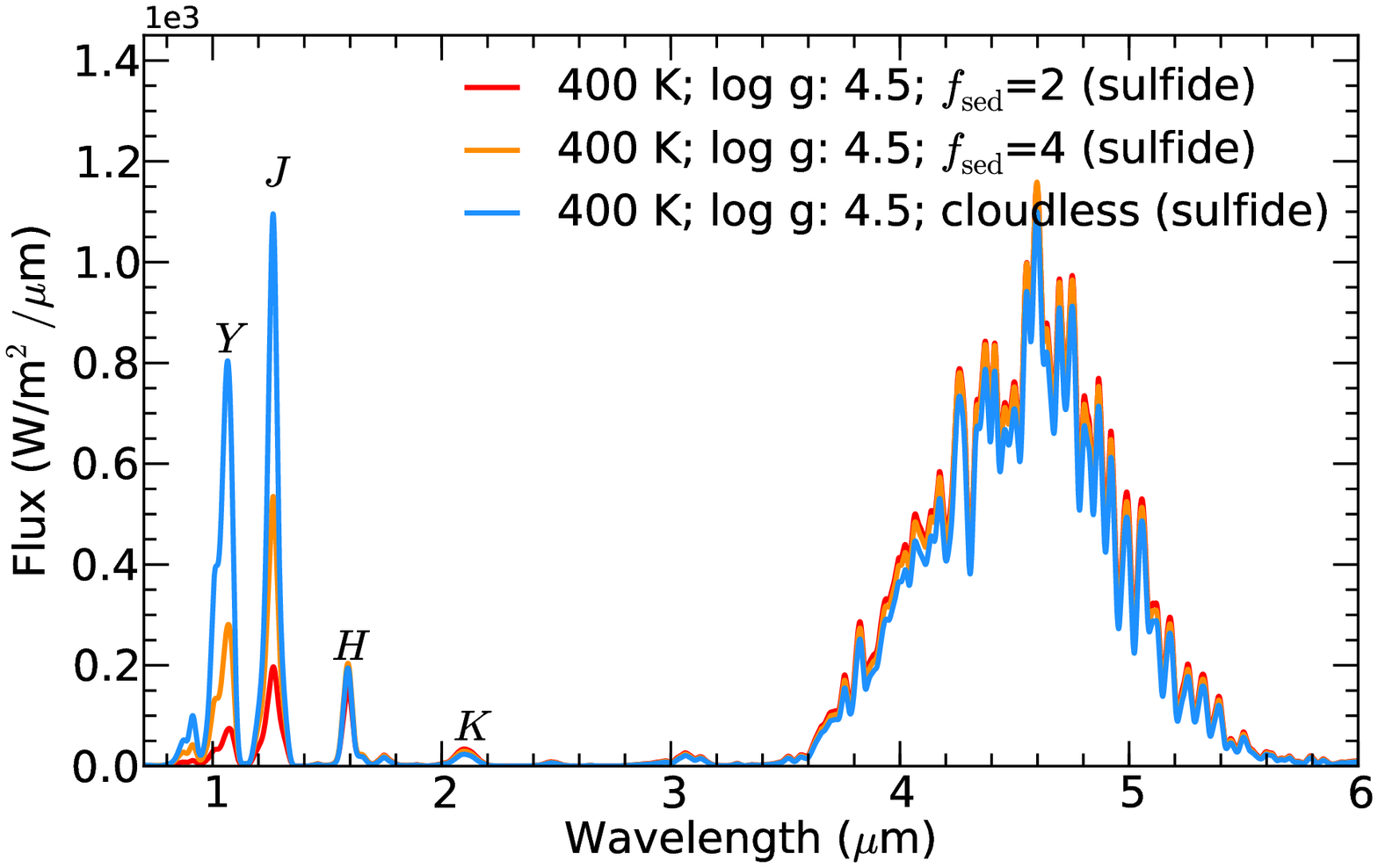}
\vspace{-6mm}
\end{minipage}
\caption{Model spectra. From top to bottom, \teff=1300 K, log $g$=5.0; \teff=900 K, log $g$=5.0; \teff=600 K, log $g$=5.0; \teff=400 K, log $g$=4.5. We show cloudy models with \fsed=2 and 4 which include the condensates Cr, MnS, \nas, ZnS, and KCl and cloudless models for comparison. Note that for the \teff=400 K, \teff=600 K and \teff=900 K models, the cloudy models are progressively fainter in $Y$ and $J$ bands and brighter in $K$ band as the sedimentation efficiency decreases. In contrast, for the \teff=1300 K case, the clouds do not significantly change the spectrum. } 
\label{specs-results}
\end{center}
\end{figure}

\begin{figure}[]
\begin{center}
\begin{minipage}[b]{\linewidth}
\includegraphics[width=3.5in]{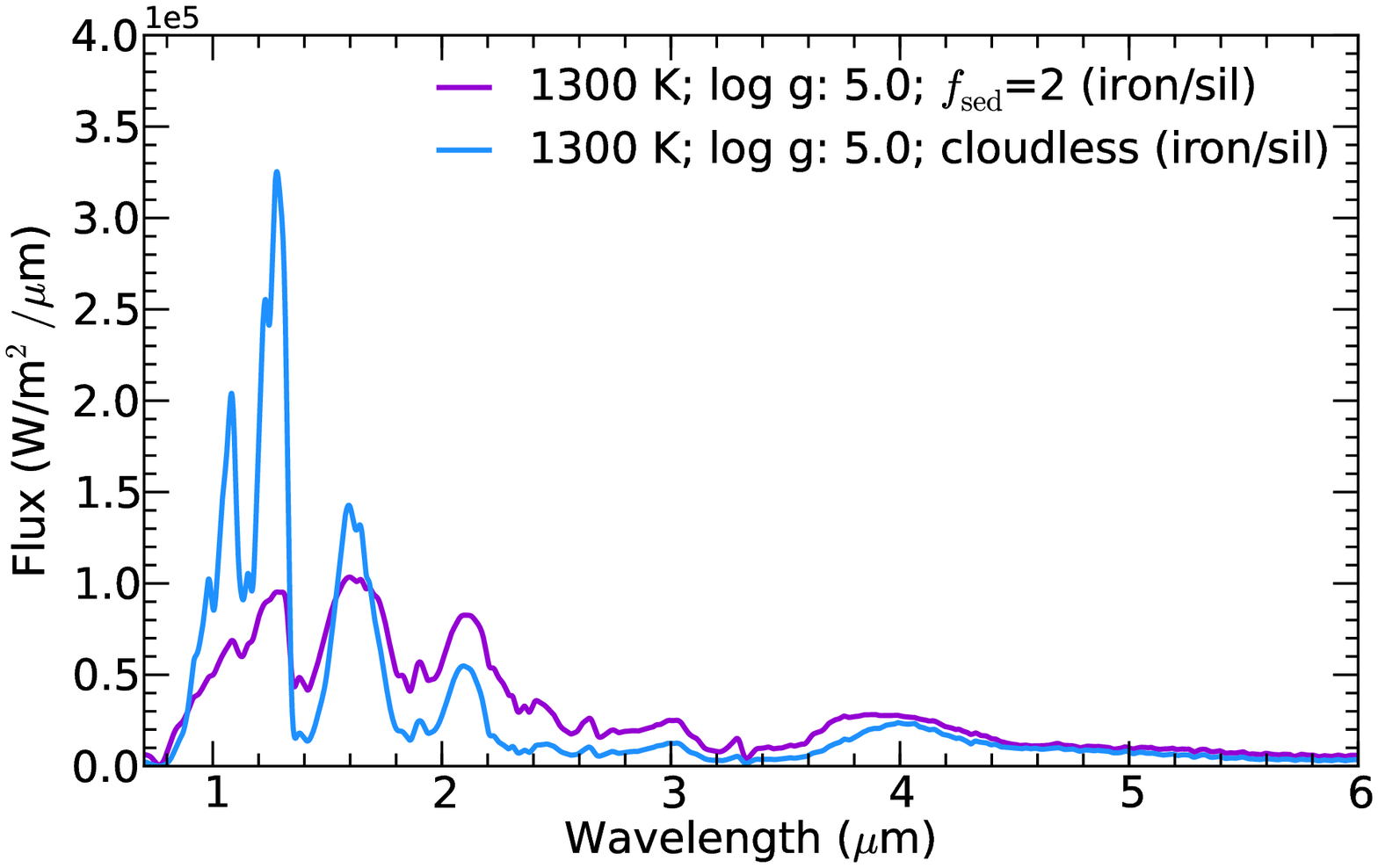}
\vspace{-6mm}
\end{minipage}
\begin{minipage}[b]{\linewidth}
\includegraphics[width=3.5in]{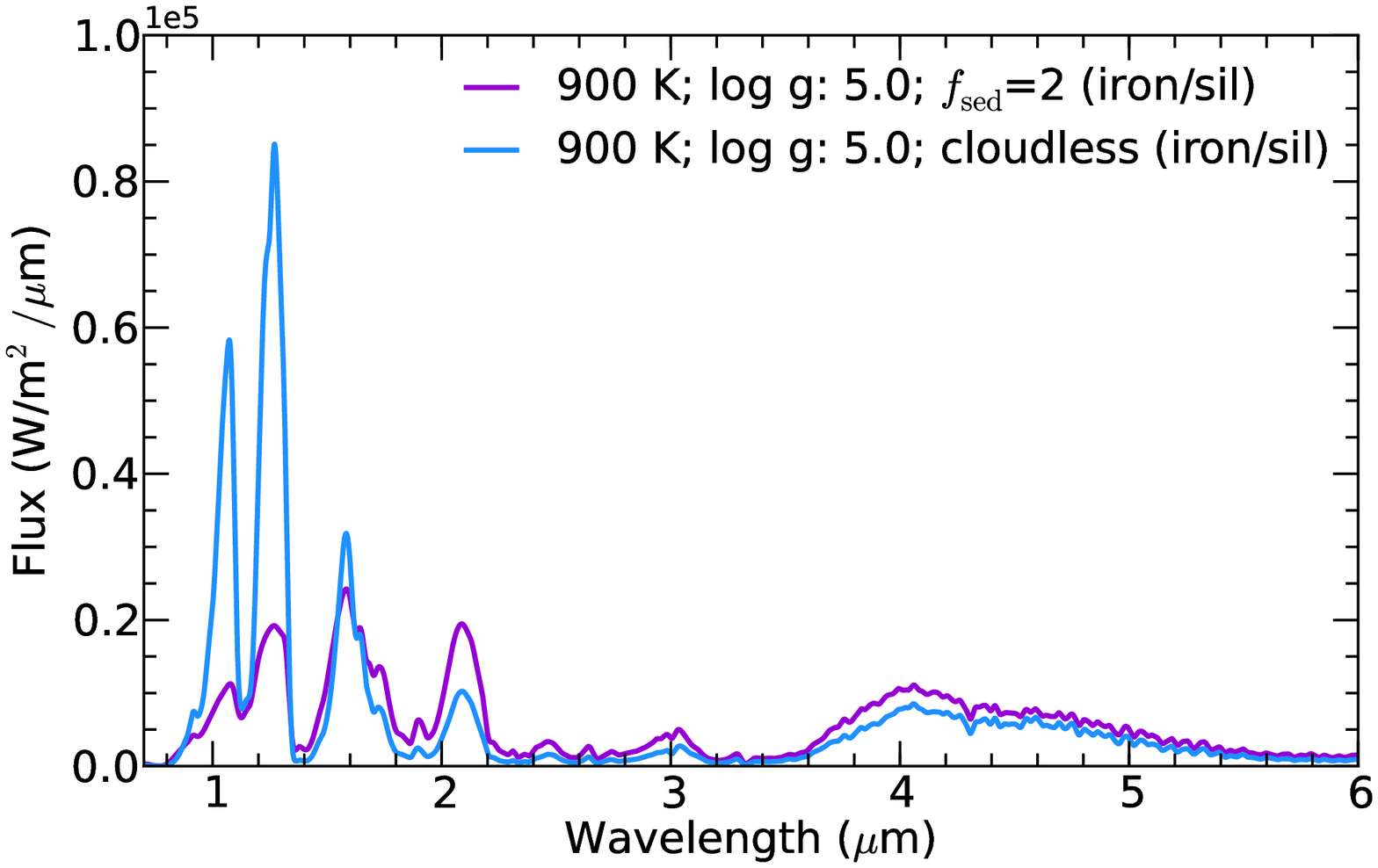}
\vspace{-6mm}
\end{minipage}
\begin{minipage}[b]{\linewidth}
\includegraphics[width=3.5in]{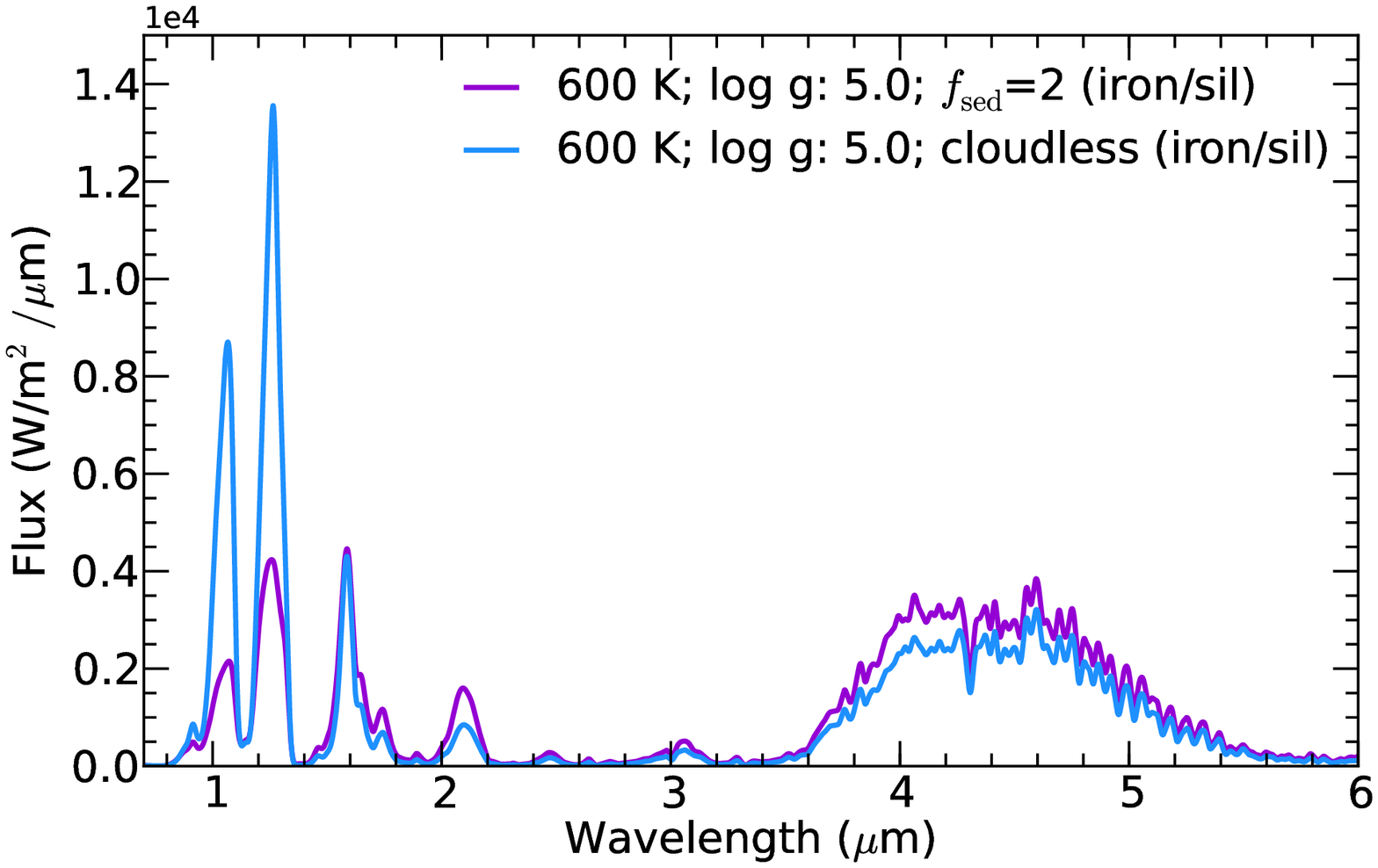}
\vspace{-6mm}
\end{minipage}
\begin{minipage}[b]{\linewidth}
\includegraphics[width=3.5in]{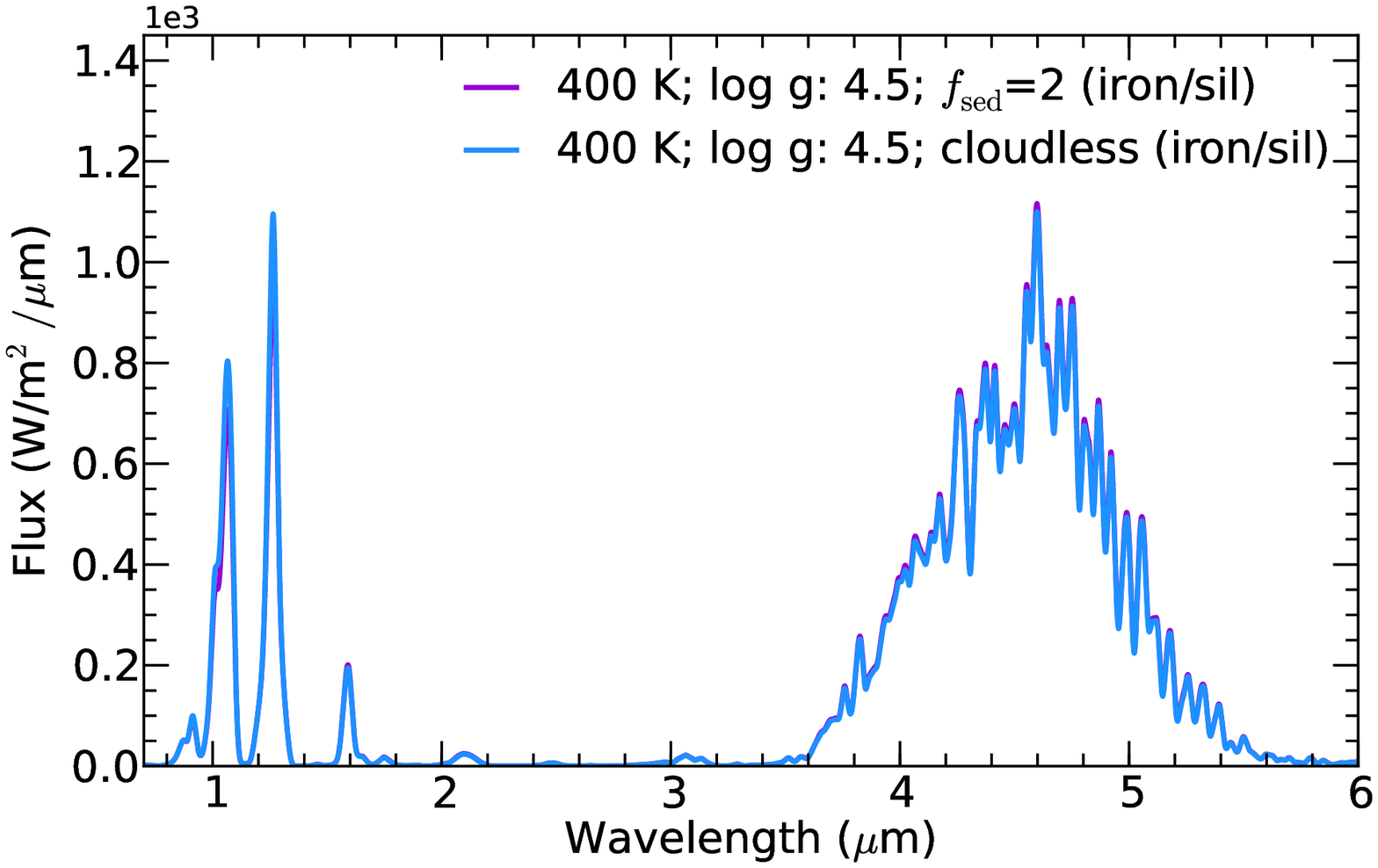}
\vspace{-6mm}
\end{minipage}
\caption{Model spectra with iron/silicate clouds. As in Figure \ref{specs-results}, from top to bottom, \teff=1300 K, log $g$=5.0; \teff=900 K, log $g$=5.0; \teff=600 K, log $g$=5.0; \teff=400 K, log $g$=4.5. We show cloudy models with iron/silicate/corundum clouds (no sulfide clouds) with \fsed=2 and cloudless models for comparison. Note that these clouds, unlike the sulfide clouds in Figure \ref{specs-results}, significantly change the shape of the 1300 K model. } 
\label{specs-silicate}
\end{center}
\end{figure}

In Figures \ref{BT} and \ref{specs-results}, we show the spectra of the same example models at 1300, 900, 600, and 400 K: Figure \ref{BT} shows the wavelength-dependent brightness temperatures from these models, while \ref{specs-results} shows the model fluxes computed from the top of the atmosphere. The brightness temperature gives some insight into the depth into the atmosphere probed at each wavelength. Flux from 0.8 to 1.3 \micron\ comes from the deepest, hottest layers of the atmosphere. Clouds change the flux in this wavelength range by limiting the depth from which flux emerges; conversely, the clouds do not change the depth probed between $\sim$2 and 5 \micron\ because the clouds form below the layers from which most of the flux is emerging. However, the hotter atmospheric temperatures at a given pressure (see Figure 3) lead to slightly higher fluxes at these wavelengths.  Though not plotted here, flux in the mid-infrared also comes from above the cloud layer and is slightly higher because the entire pressure-temperature profile is hotter.  

For the hottest of these models (\teff=1300 K), the cloudy spectra look almost identical to the cloudless spectrum. This model is too hot to have much mass of these condensed species form in the photosphere. For a cooler model (\teff=900 K), the cloudy spectra look different from the cloudless spectrum. As we decrease the sedimentation efficiency \fsed\ in the model, the flux in $Y$ and $J$ bands decreases and the flux in $K$ band increases. 

For the coldest two models shown (\teff=400, 600 K), the cloudy spectra look dramatically different from the cloudless spectrum in the near-infrared; even the thinnest cloud considered here (\fsed=5) causes the flux in $Y$ and $J$ to decrease by 50\% and the flux in $K$ to correspondingly increase. Decreasing the sedimentation efficiency enhances this effect.  

Figure \ref{specs-silicate} shows the effect of the iron and silicate clouds on the spectra of models with the same effective temperatures (1300 K, 900 K, 600, and 400 K) and surface gravity. Note that unlike the sulfide clouds, these iron and silicate clouds substantially change the shape of the 1300 K and 900 K models by suppressing the flux in $Y$, $J$, and $H$ and increasing the flux in $K$ band and the mid-infrared. This strong effect at higher temperatures is due to the large amount of iron and silicate condensed in the visible atmosphere at those temperatures.

\subsection{Cloud Structure in Model Atmospheres} \label{cloudstructures}

\begin{figure}[]
  \begin{minipage}[b]{\linewidth}
    \includegraphics[width=3.33in]{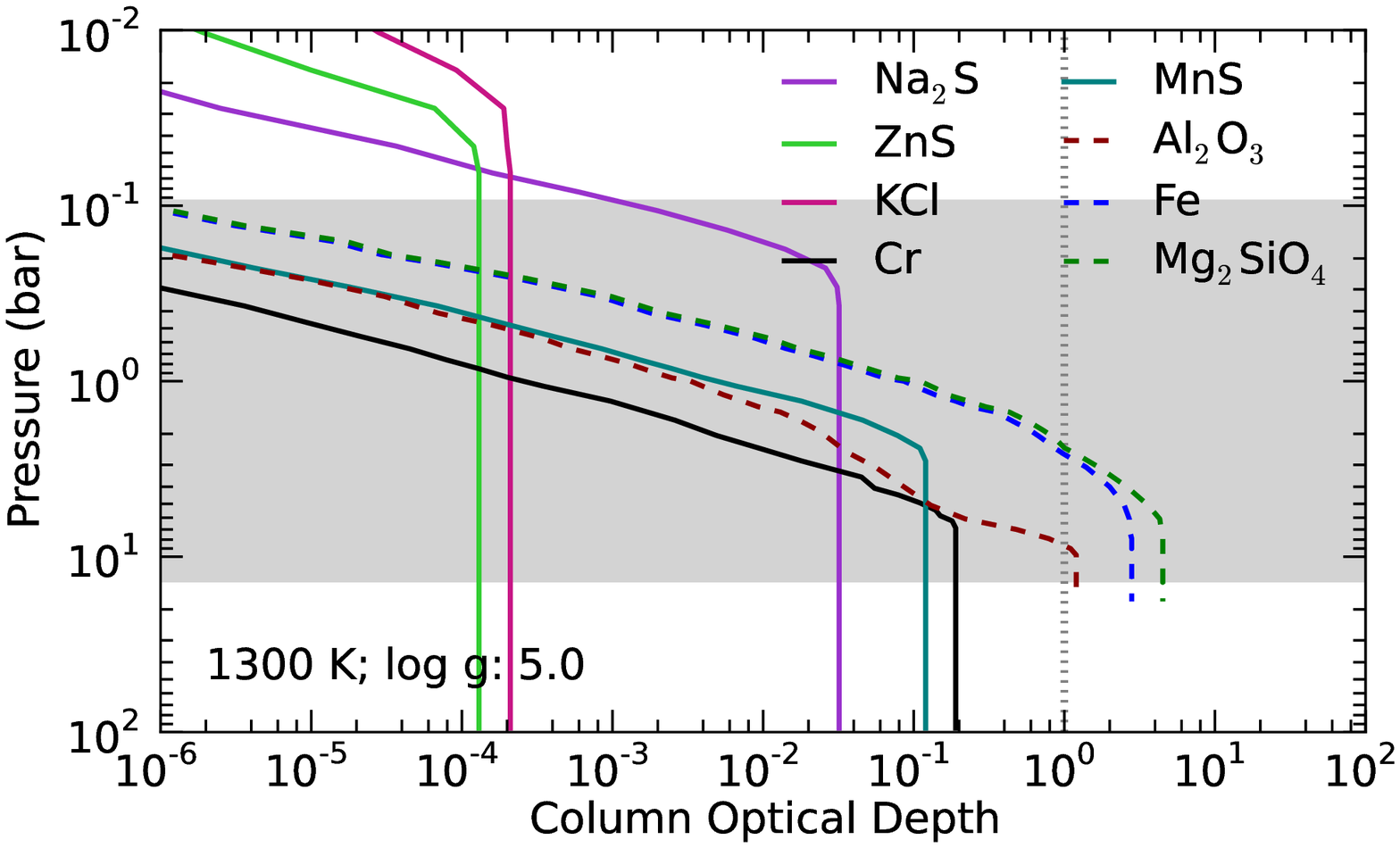}
    \vspace{-3mm}
  \end{minipage}
  \begin{minipage}[b]{\linewidth}
    \includegraphics[width=3.33in]{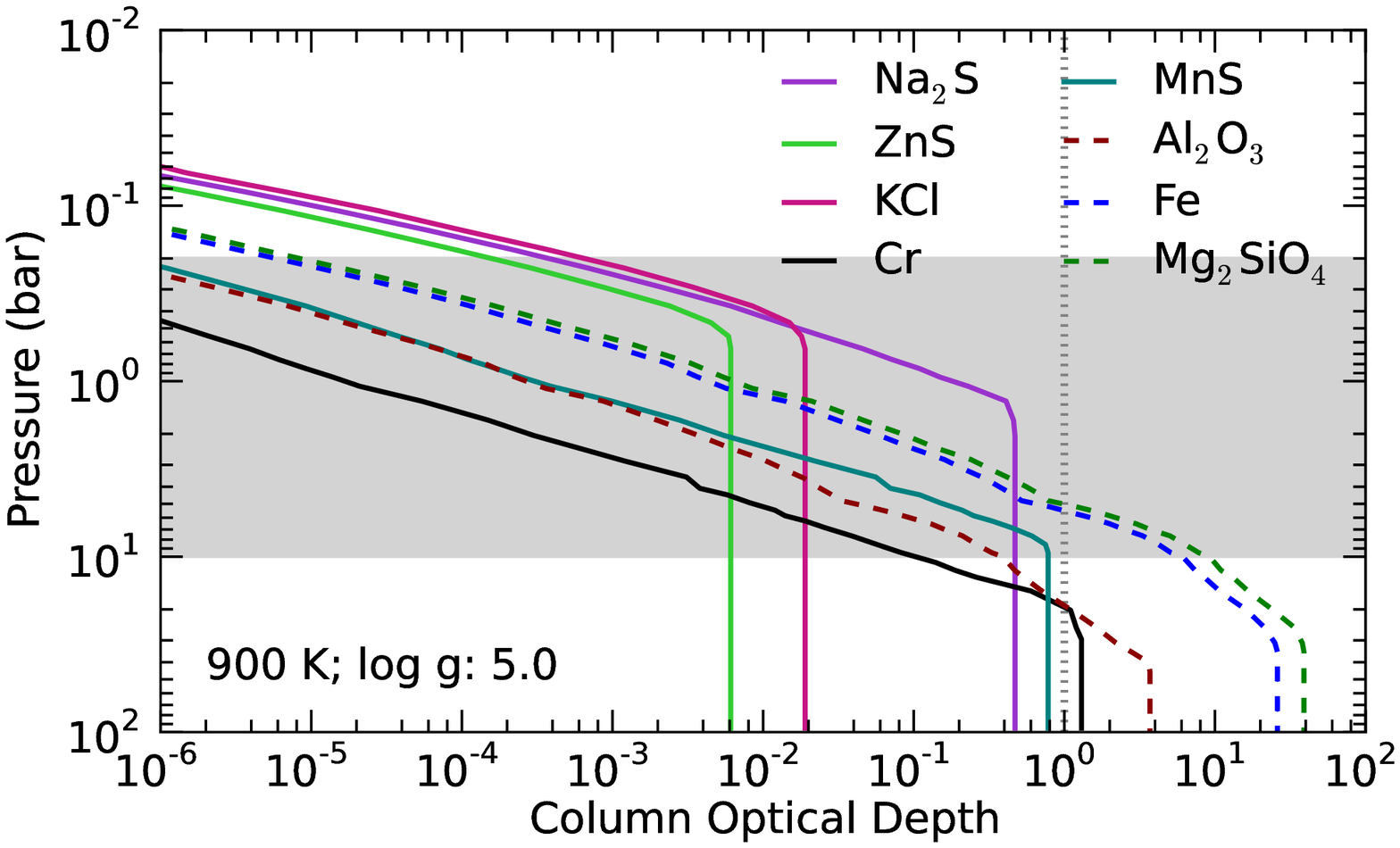}
    \vspace{-3mm}
  \end{minipage}
    \begin{minipage}[b]{\linewidth}
    \includegraphics[width=3.33in]{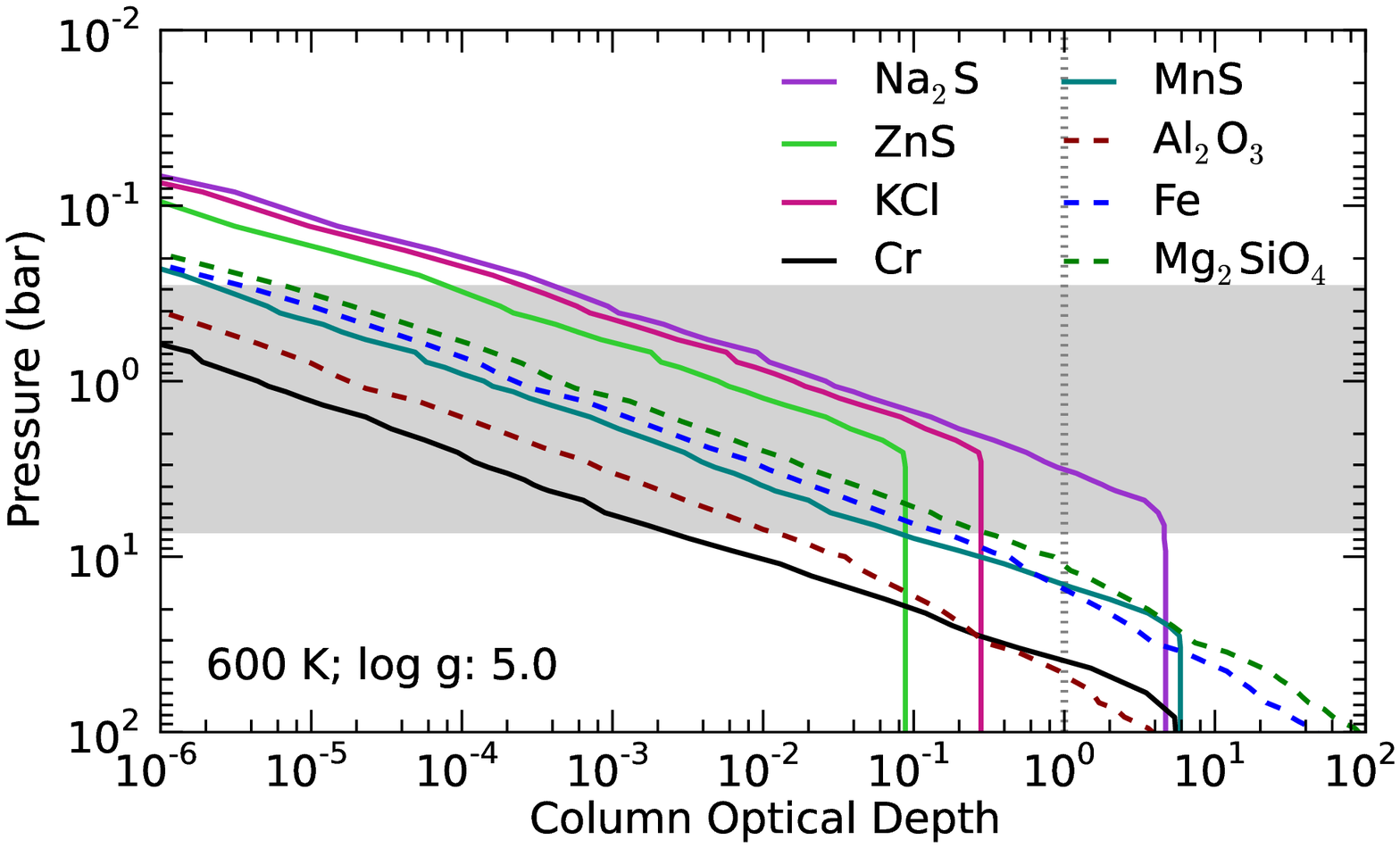}
    \vspace{-3mm}
  \end{minipage}
    \begin{minipage}[b]{\linewidth}
    \includegraphics[width=3.33in]{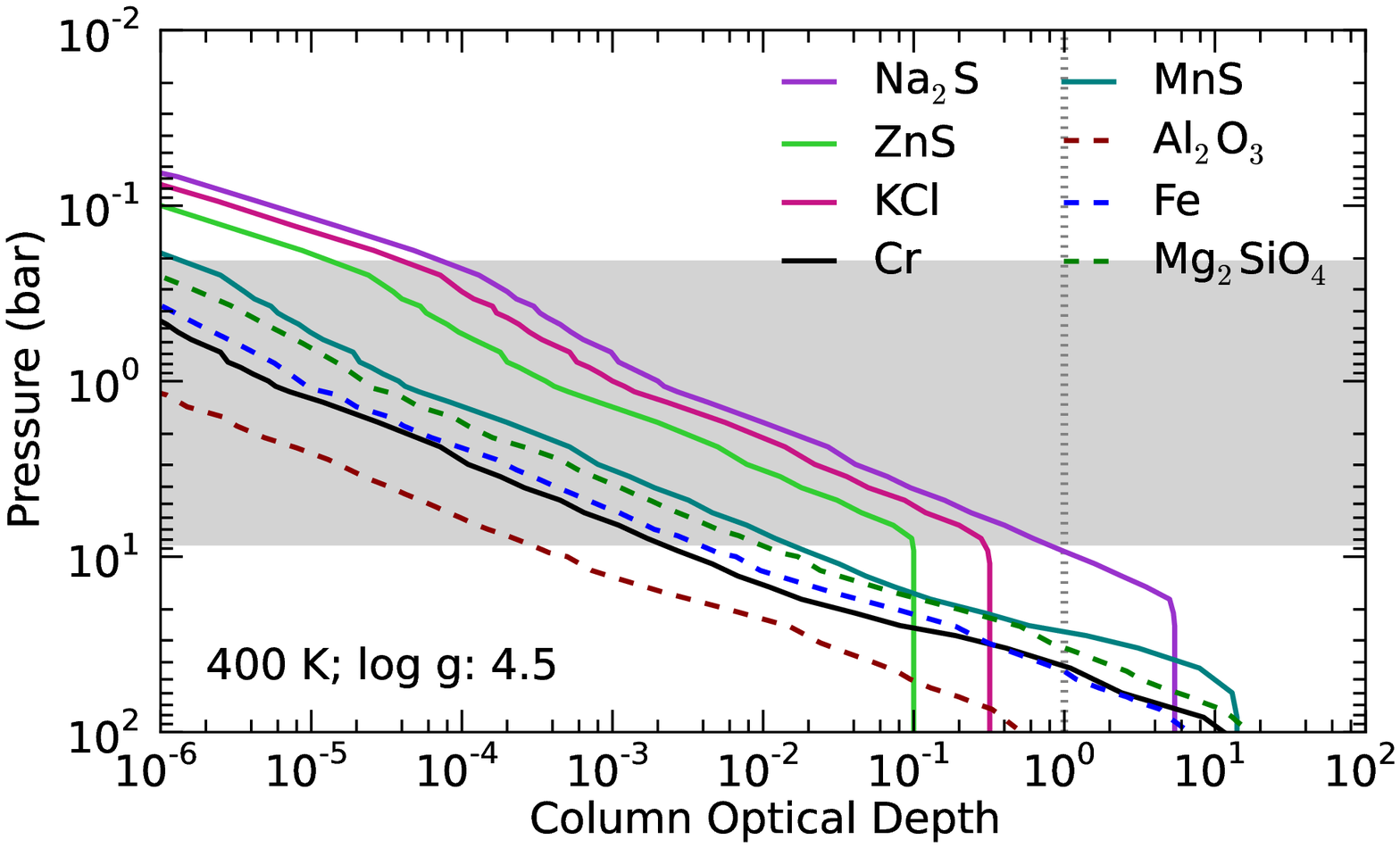}
    \vspace{-4mm}
  \end{minipage}
\caption{Pressure vs.~column optical depth. The column optical depth of each cloud species is plotted. The solid lines denote the clouds examined in this study: \nas, ZnS, KCl, Cr, and MnS. The dashed lines  show the column optical depths of models that include only the more refractory clouds corundum (Al$_2$O$_3$) iron (Fe), and forsterite (Mg$_2$SiO$_4$) to show where those clouds would form in comparison to the sulfide clouds. All models use \fsed=2. The shaded grey area shows the region of the atmosphere which lies within the $\lambda=$ 1 to 6 \micron\ photosphere. Note that the \nas\ cloud is by far the most important of the added clouds for the 600 K model in the near-infrared. Also note that if the Al$_2$O$_3$, Fe, and Mg$_2$SiO$_4$ persisted to effective temperatures of 900-1300 K, they would be quite visible, which would not match observations. }
\label{colopd}

\end{figure}

Figure \ref{colopd} shows the distribution of clouds in the model atmospheres for the three example cases. The locations of iron, silicate, and corundum clouds, using models that only include those clouds---the standard \ct{Saumon08} cloud configuration---are plotted for reference.  The column optical depth is given by Equation 16 in \ct{AM01}, which calculates the cumulative geometric scattering optical depth by cloud particles through the atmosphere. 

For the 1300 K model, all of the sulfide clouds have tiny optical depths in the photosphere and do not affect the emergent spectra. The silicate and iron clouds would have significant optical depth ($\tau$=2-3) and would substantially change the emergent spectra.

\begin{figure}[]
\includegraphics[width=3.5in]{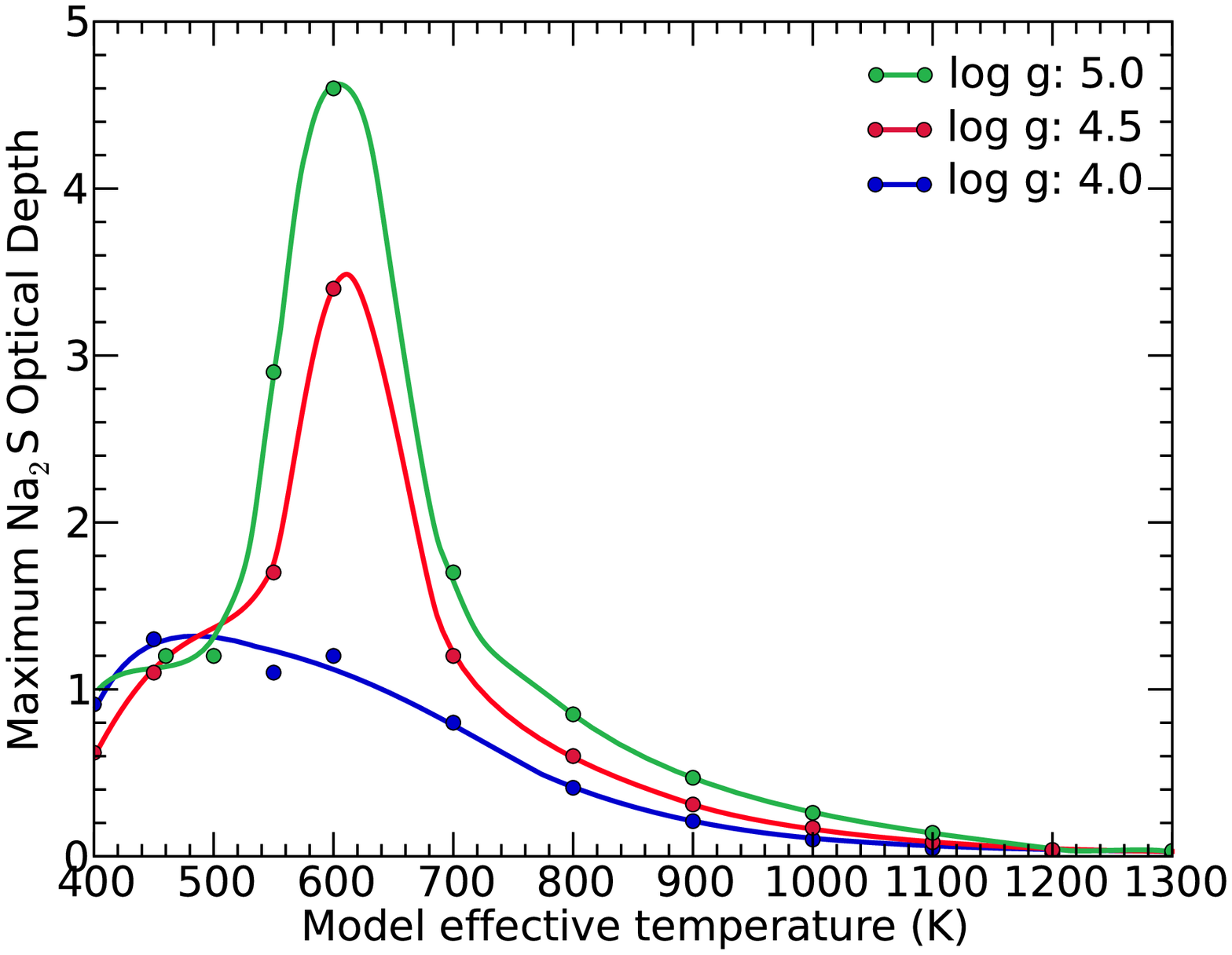}
\caption{The column optical depth of the \nas\ cloud above the bottom of the 1-6 \micron\ photosphere in each model is plotted as a function of model effective temperature.  The curves connecting the points are there to guide the eye. Three different surface gravities are shown and all models use \fsed=2. The column optical depth peaks at temperatures of about 600 K, and models with higher surface gravity have a greater \nas\ column optical depth. }
\label{maxtau}
\end{figure}

For the 900 K model, all of the sulfide clouds have a column optical depth smaller than 1 in the photosphere. KCl and ZnS have tiny optical depth ($\tau<2\times10^{-2}$) and will not create an observable change in the spectrum. \nas\ and MnS have optical depth between 0.1 and 1 and will change the model spectra slightly. 

For the 600 K model, \nas\ is the most important condensate opacity source. KCl has a small optical depth and ZnS has a negligible optical depth. This result is expected, based on the abundances of each species (see Table \ref{cloudstable}). The other two clouds, MnS and Cr, are below the near-infrared photosphere, so also do not change the spectrum. The silicate and iron clouds would also be below the photosphere. 

Using our full grid of models, we can examine the importance of the \nas\ cloud as a function of \teff\ and surface gravity.  Figure \ref{maxtau} shows how the column optical depth of this cloud varies from 400-1300 K and log $g$ from 4.0-5.0 for a constant value of the parameter \fsed.  Moving to \teff\ values below 1300 K, the \nas\ cloud grows in importance as it forms progressively deeper in the atmosphere, so that there is larger mass of condensate in the cloud.  The maximum optical depth of the \nas\ cloud at pressure levels above the bottom of the 1-6 \micron\ photosphere is largest for models with \teff\ of 600 K.  At lower \teff, much of the cloud opacity is below the visible atmosphere.  The optical depth within the photosphere is significant ($\tau\gtrsim$ 1) for models between 400 and 700 K.

Figure \ref{maxtau} indicates that the cloud optical depth within the photosphere is largest for higher surface gravity atmospheres for a constant value of \fsed. However, this does not necessarily predict that higher gravity brown dwarfs will have thicker clouds than lower gravity brown dwarfs because the parameter \fsed\ is not necessarily independent of gravity. 

\section{Comparison with Observations}
\subsection{Color-Magnitude Diagrams} \label{generaltrends}

\begin{figure*}[]
\begin{minipage}[b]{0.33\linewidth}
\vspace{-2mm}
\includegraphics[scale=0.42]{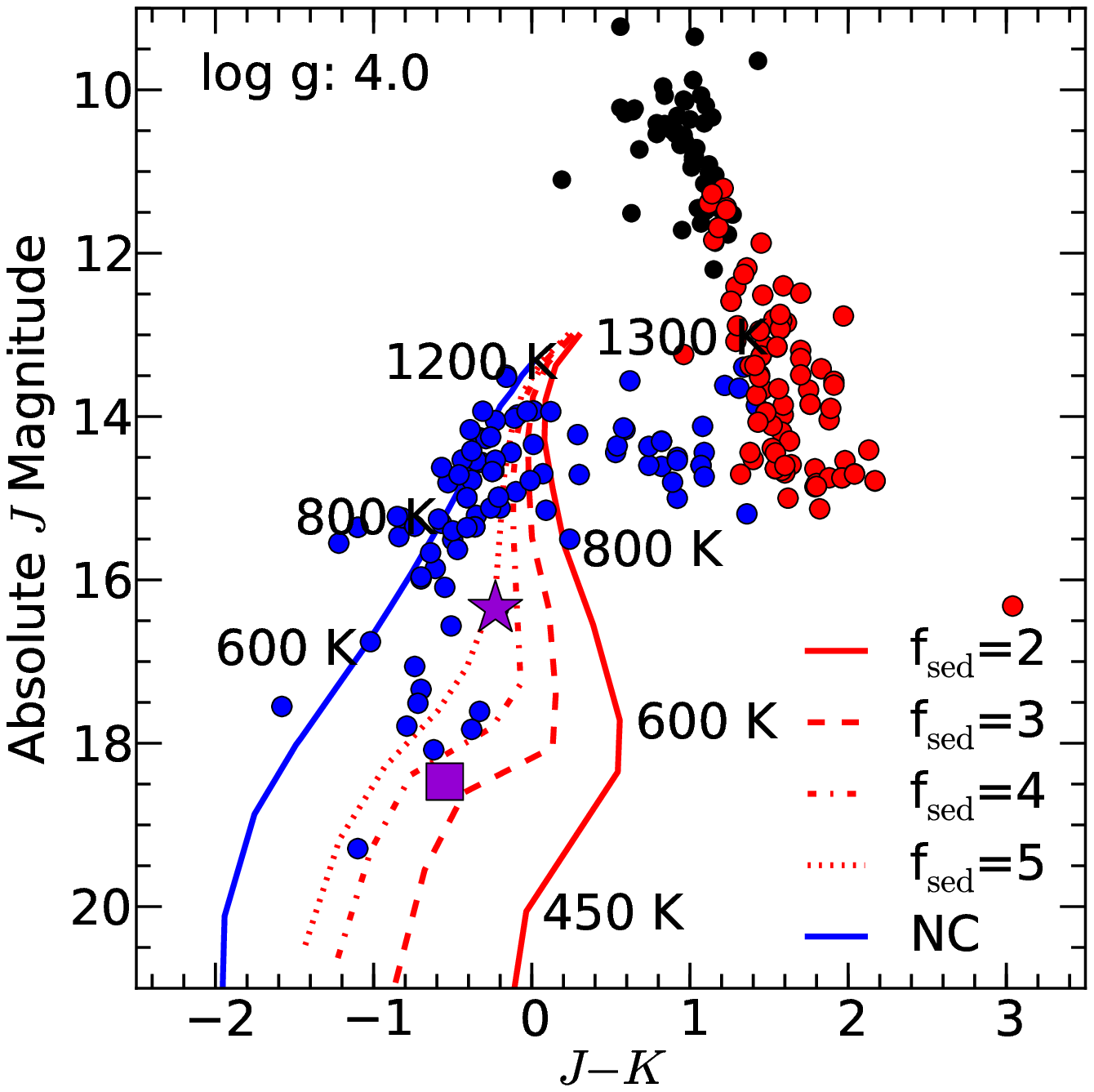}
\vspace{-2mm}
\end{minipage}
\begin{minipage}[b]{0.33\linewidth}
\vspace{-2mm}
\includegraphics[scale=0.42]{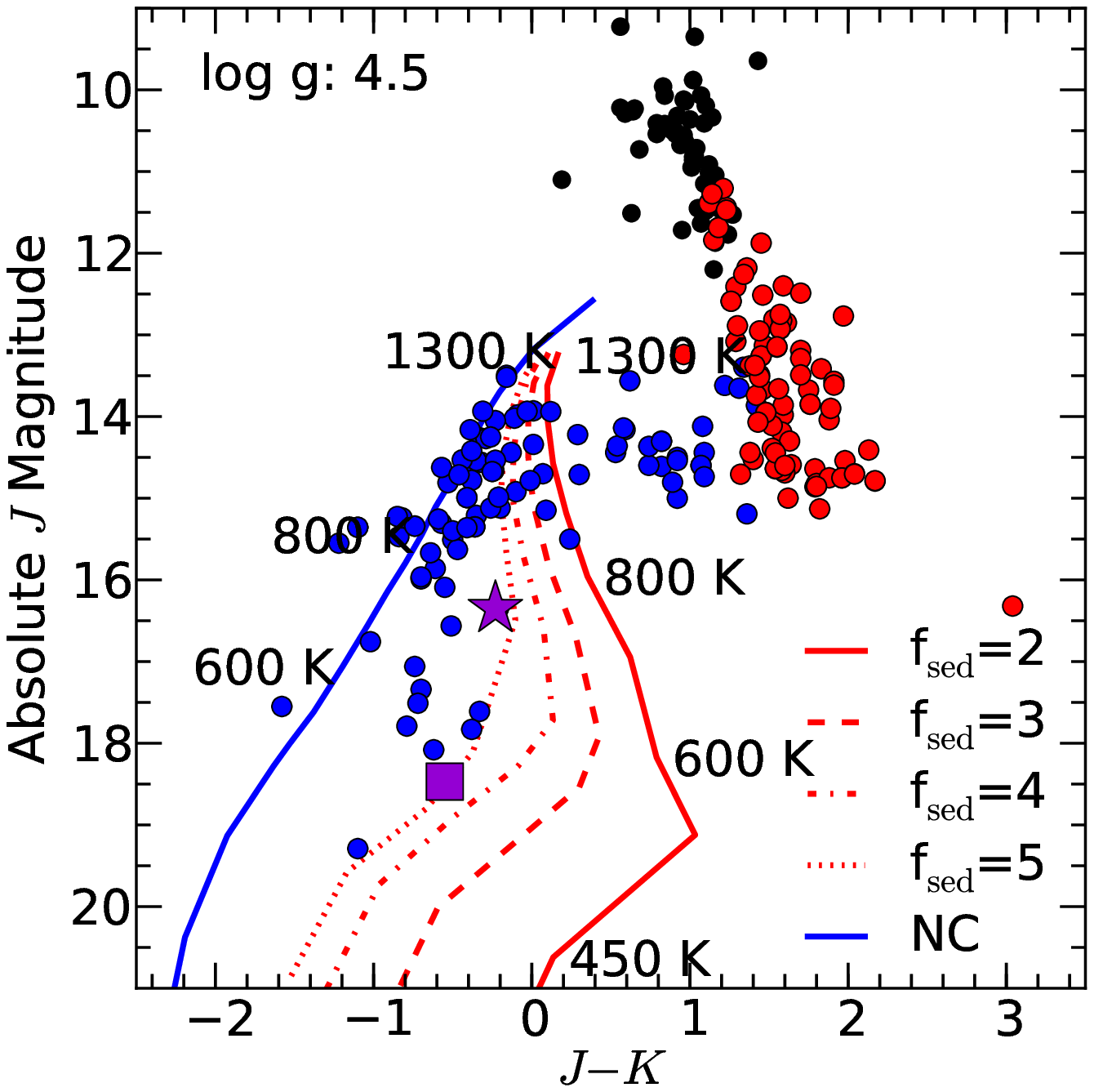}
\vspace{-2mm}
\end{minipage}
\begin{minipage}[b]{0.33\linewidth}
\vspace{-2mm}
\includegraphics[scale=0.42]{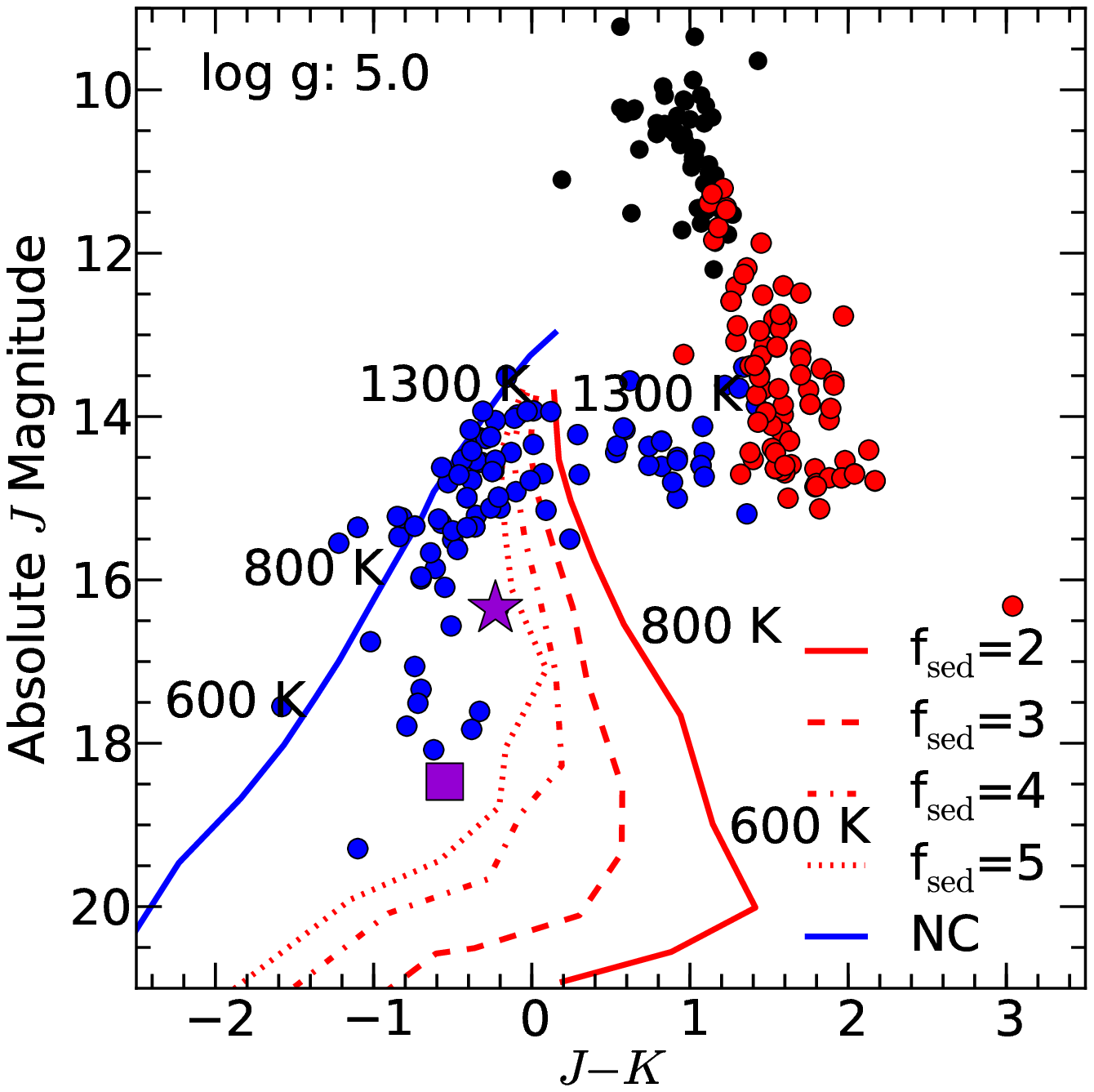}
\vspace{-2mm}
\end{minipage}

\begin{minipage}[b]{0.33\linewidth}
\vspace{-2mm}
\includegraphics[scale=0.42]{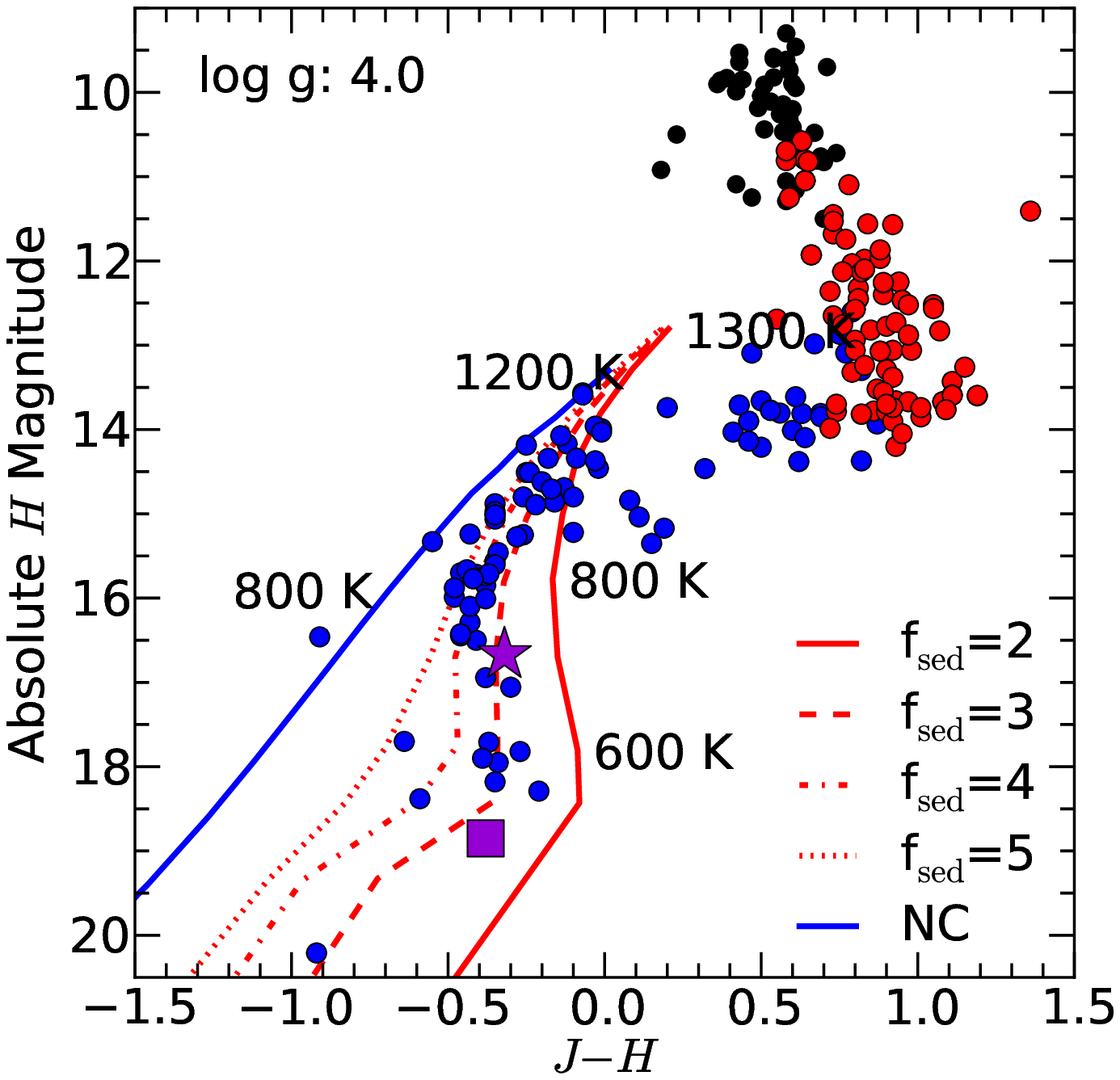}
\vspace{-2mm}
\end{minipage}
\begin{minipage}[b]{0.33\linewidth}
\vspace{-2mm}
\includegraphics[scale=0.42]{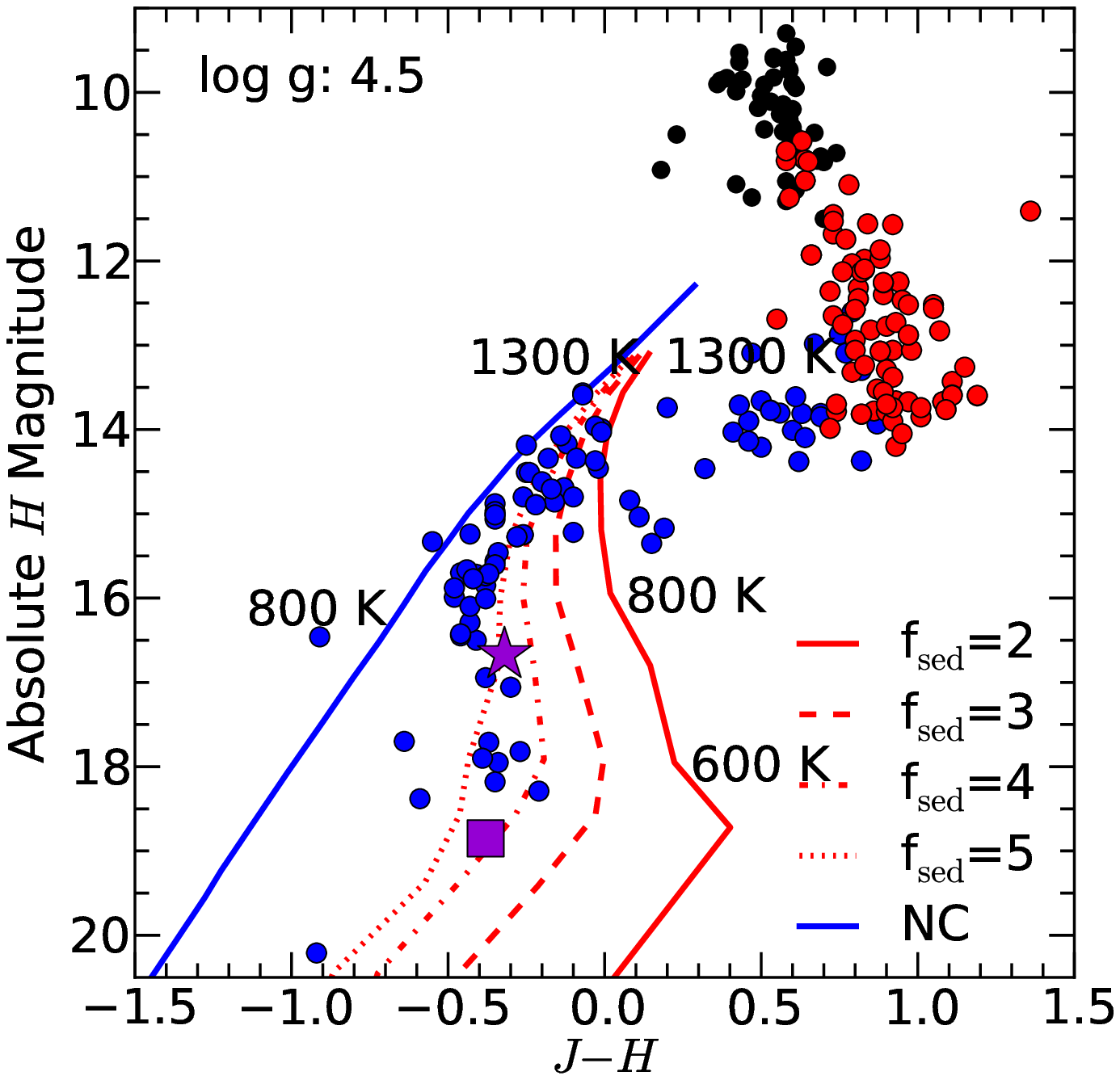}
\vspace{-2mm}
\end{minipage}
\begin{minipage}[b]{0.33\linewidth}
\vspace{-2mm}
\includegraphics[scale=0.42]{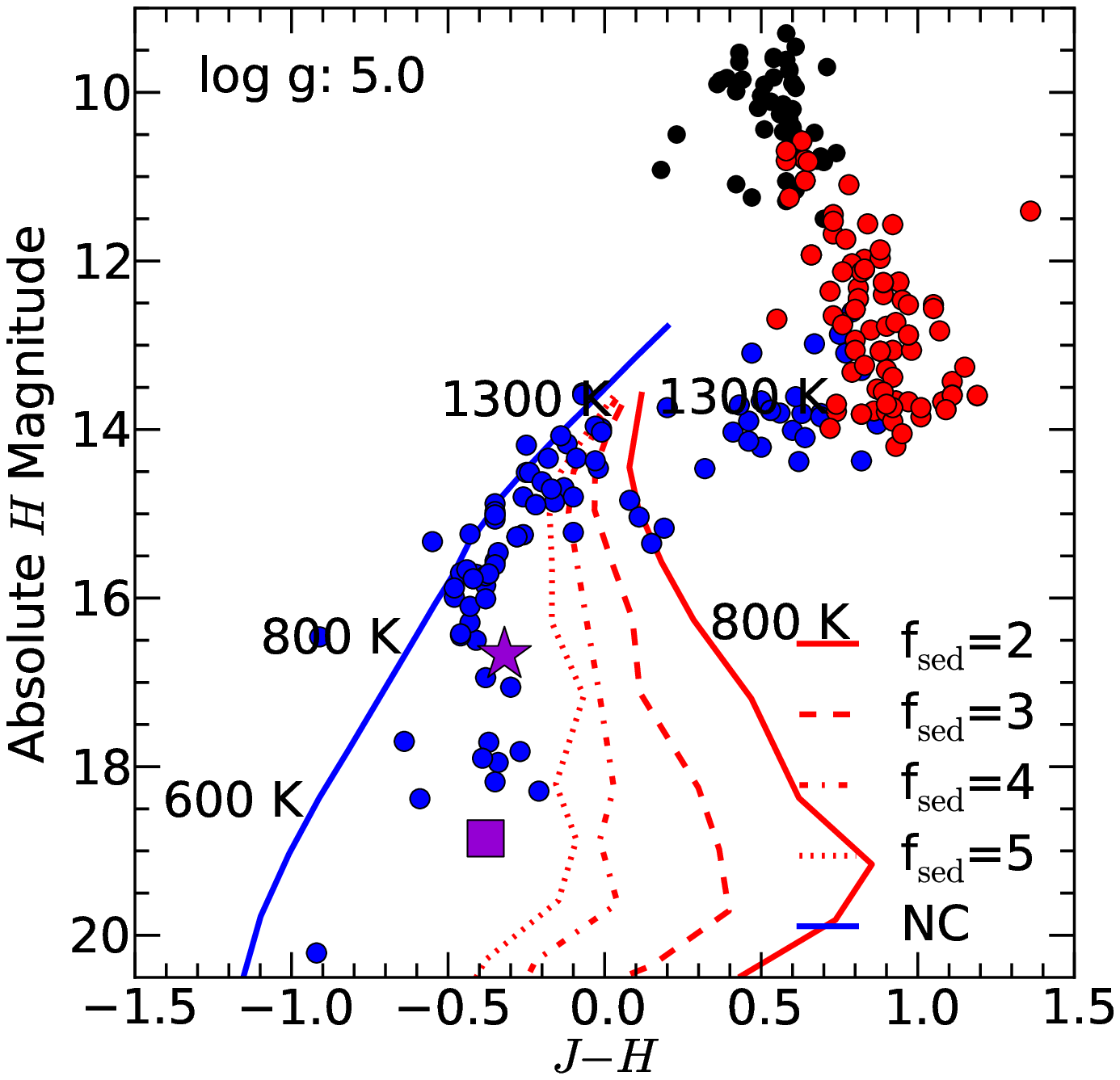}
\vspace{-2mm}
\end{minipage}
\caption{Color-magnitude diagrams for M, L, and T dwarfs. As in Figure \ref{intro-colormag}, observed ultracool dwarf color is plotted against the absolute magnitude for all known brown dwarfs with measured parallax. In the top 3 plots, $J-K$ color is plotted against absolute J magnitude; in the bottom 3 plots, $J-H$ color is plotted against absolute $H$ magnitude. All photometry is in the MKO system. M dwarfs are plotted as black circles, L dwarfs as red circles, and T dwarfs as blue circles. Observational data are from \ct{Dupuy12, Faherty12}. The locations of the brown dwarfs Ross 458C and UGPS 0722-05, the objects to which we compare model spectra to observations in Figures 11 and 12, are shown with a purple star and square symbol, respectively. \emph{Models.} Models are plotted as lines. Each labeled temperature marks the approximate locations of the model with that effective temperature.  Three representative gravities are plotted: from left plot to right plot, log g=4.0, 4.5, and 5.0. Blue lines are cloudless models and red lines are cloudy models (\fsed=5, 4, 3, and 2 from left line to right line in each plot) that include the opacity of only the newly added clouds---\nas, Cr, MnS, ZnS, and KCl. }

\label{colormag-results}
\end{figure*}

In Figure \ref{colormag-results}, we plot the photometric colors in the near-infrared of all brown dwarfs with measured parallaxes and apparent $J$ and $K$ magnitude errors smaller than 0.2 magnitudes \cp{Dupuy12, Faherty12}. We also plot the calculated photometric colors of our suite of cloudless and cloudy models from 400-1300 K with several representative surface gravities. In Section \ref{generaltrends} we discuss the general trends of the photometric colors of models at various effective temperatures and cloud sedimentation efficiencies. In Section \ref{comparison} we compare our model results to the photometric observations. 

As discussed in Sections \ref{modelspecs} and \ref{cloudstructures}, clouds in T dwarf atmospheres tend to suppress the flux in $Y$ and $J$ bands and increase the flux in $K$ band. The flux shift from $J$ to $K$ gives cloudy models larger (redder) $J-K$ and $J-H$ colors than cloudless models. 

In Figure \ref{colormag-results}, the hottest cloudy models have nearly the same near-infrared colors as cloudless models. As we decrease the effective temperature of a cloudy model, more cloud material condenses; the model has a redder photometric color than a cloudless model with the same \teff. If we reduce the sedimentation efficiency ($f_{\rm sed}$) of the cloud, the cloud becomes optically thicker, and the model has a redder photometric color.

The upper panels, which show $J-K$ photometric colors, show that our sulfide cloud models can easily reach the colors of red T dwarfs, with \fsed\ values of 4-5. The bulk of the T dwarf population is bluer than the \fsed=5 model.  However, the cooler T dwarfs are generally well-matched by the models. In $J-H$, the color directly affected by cloud opacity limiting the depth to which one sees, the model colors are an excellent match to the data. The cloudy models are a much better match than the corresponding cloud-free models.

\subsection{Comparison to Observed T Dwarfs} \label{comparison}

\subsubsection{Expected Surface Gravity of T dwarfs}

Based on \ct{Saumon08} evolution models and assuming that observed brown dwarfs will have ages less than 10 Gyr, we expect that the coldest objects modeled, between 400-600 K, will have surface gravities less than log $g$=5.2. Hotter objects, between 1000 and 1300 K, will have surface gravities less than log $g$=5.5. 

\subsubsection{Cloud Sedimentation Efficiency}

For the L dwarfs, we are generally able to match photometric colors by including silicate, iron, and corundum clouds with a sedimentation efficiency parameter of \fsed=2$\pm$1 \cp{Stephens09, Saumon08}. However, for these cooler T dwarfs, models with these sulfide clouds with \fsed=2 are redder than observed brown dwarfs. If we assume \fsed\ is larger---around 4 or 5---we are able to match observed colors.  The cloud model does not explicitly suggest any physical mechanism for why \fsed\ would be different. However, since these objects are about 1000 K cooler than L dwarfs, it would not be surprising if these objects populate a different physical regime, and would have substantially different rates of atmospheric mixing and cloud condensation.  Indeed, a large increase in \fsed\ with lower \teff\ values is one way to quickly clear away the silicate and iron clouds \cp{Knapp04} 

\subsubsection{WISE Color-Color Diagrams}

\ct{Cushing11} announced the discovery of six proposed Y dwarfs found by the WISE mission. Objects around the T-to-Y transition are cold enough to have NH$_3$ absorption features in the near infrared.

We have obtained near-infrared photometry for two proposed Y0 dwarfs discovered using the Wide-field Infrared Survey Explorer (WISE) by Cushing et al. (2011).  $YJH$ was obtained for WISEP J140518.40$+$553421.5 and $YJ$ for J154151.65$-$225025.2, using the near-infrared imager NIRI \cp{Hodapp03} on the Gemini-North telescope on Mauna Kea, Hawaii. The photometry is on the Mauna Kea Observatories system \cp{Tokunaga05}.The data were obtained via the Gemini queue program GN-2012A-Q-106 on 2012 February 10; the program aims to supplement and improve on the photometry presented in the discovery paper. Individual exposure times of 60~s were used at $Y$ and $J$, and 30~s at $H$; a 9-position dither pattern with 10" offsets was repeated as necessary for sufficient signal to noise. The total exposure time for WISEP J140518.40+553421.5 was 9 minutes at $Y$ and $J$ and 58.5 minutes at $H$; for  WISEP J154151.65-225025.2 the total exposure time was 18 minutes at each of $Y$ and $J$. Data were reduced in a standard fashion and flatfielded with calibration lamps on the telescope. The UKIRT faint standards FS 133 and 136 were used for photometric calibration; $J$ and $H$ were taken from  \ct{Leggett06}, and $Y$ from the UKIRT online catalog (http://www.jach.hawaii.edu/UKIRT/astronomy/calib/phot cal/fs ZY MKO wfcam.dat).
\begin{figure*}[]
  \begin{minipage}[b]{0.5\linewidth}
    \includegraphics[width=3.5in]{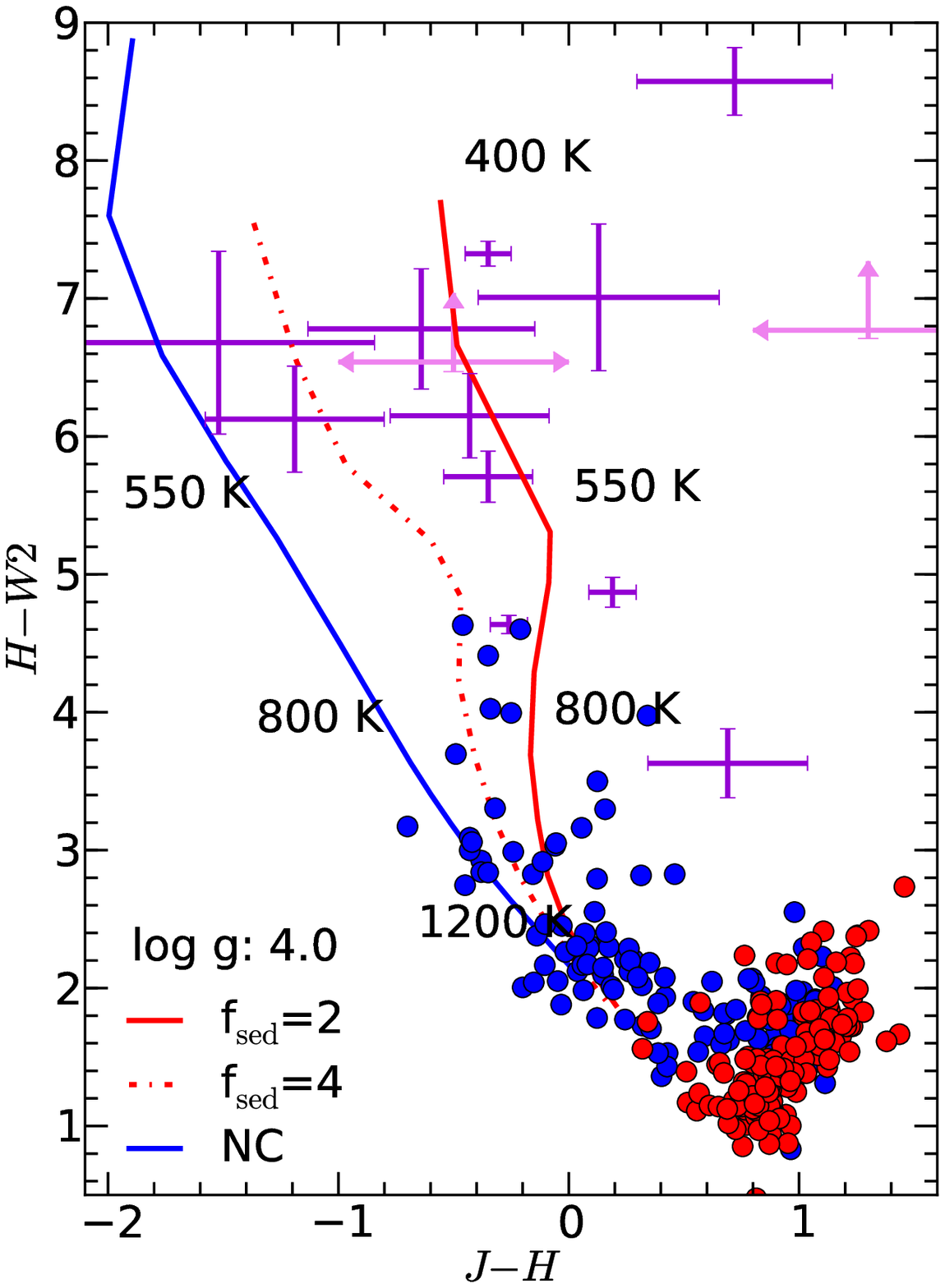}
  \end{minipage}
  \begin{minipage}[b]{0.5\linewidth}
    \includegraphics[width=3.5in]{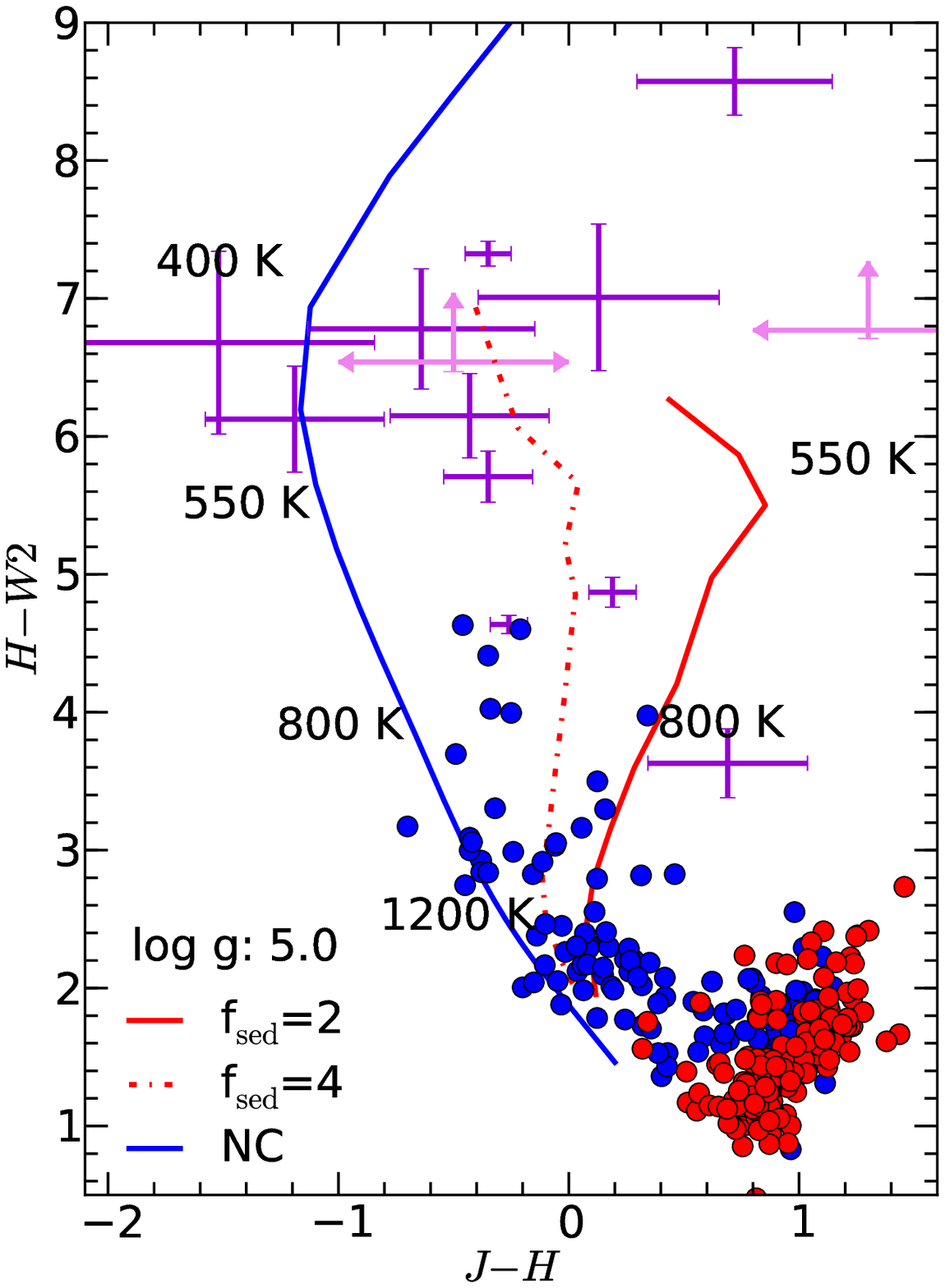}
     \end{minipage}
  \vspace{-4mm}

   \caption{Color-color diagrams using WISE and near infrared data. Observed $J-H$ versus $H-W2$ colors of L and T dwarfs \cp{Kirkpatrick11} and proposed WISE Y dwarfs \cp{Cushing11, Kirkpatrick12} are plotted. For $J$ and $H$ bands we use MKO photometry. L and T dwarfs are plotted as red and blue dots, respectively. WISE Y dwarfs are plotted as purple error bars; Y dwarfs with magnitude upper limits are shown in pink. Model photometric colors are shown as solid and dashed lines; the blue line shows a cloudless model and the red lines show two cloudy models (from left to right, \fsed=4 and \fsed=2). Each labeled temperature marks the approximate location of the models with that effective temperature. Many of these cold brown dwarfs have photometric colors closer to the cloudy models than the cloud-free model. The left plot shows log $g$=4.0 and the right plot shows log $g$=5.0.}
  \label{wise}
\end{figure*}

The final reduced magnitudes are:   $Y = 21.41 \pm 0.10$,   $J = 21.06 \pm 0.06$ and  $H = 21.41 \pm 0.08$ for WISEP J140518.40+553421.5; $Y = 21.63 \pm 0.13$ and  $J = 21.12\pm 0.06$ for WISEP J154151.65-225025.2.

 Figure \ref{wise} shows how clouds will effect these cold objects.  $H-W2$ is a useful temperature indicator for these objects, while $J-H$ is sensitive to both the cloud structure and gravity. As the \teff\ of the non cloudy models decreases from 800 to 500 K, the models become progressively bluer in $J-H$ color. However, most of the proposed WISE Y dwarfs are redder than the cloudless model. The models that include the sulfide clouds match their colors better.  This result tentatively suggests that for objects colder than T dwarfs, the sulfide clouds remain important. Of course, for objects with effective temperatures of $\sim$350 K, water will condense; at that point, H$_2$O clouds should contribute to the spectra \cp[e.g.][]{Burrows03}.

\subsection{Comparison to Observed T Dwarf Spectra}

We now compare model spectra to two relatively cold, red T dwarfs, Ross 458C and UGPS 0722--05. The near-infrared spectra of these two objects are not well-matched by cloudless T dwarf spectra; by including our neglected clouds, which for these cool objects are dominated by the \nas\ cloud, we match their spectra more accurately. 

We compare models to both near-infrared spectra and near- and mid-infrared photometry. As in previous studies of brown dwarfs \cp{Cushing08, Stephens09}, we find that in different bands, the observations are best fit by models of different parameters. In this study, we focus on finding models that fit the shape of the spectra in the near-infrared where clouds play a significant role. 

\subsubsection{Ross 458C} \label{ross458}

Ross 458C is a late-type T dwarf (T7--9) which is separated by over 1100 AU from a pair of M star primaries. It has anomalously red near-infrared colors ($J-K=-0.21\pm0.06$). \ct{Burgasser10} obtained spectroscopic observations with the FIRE spectrograph \cp{Simcoe08, Simcoe10} on the Magellan Baade 6.5 meter telescope at Las Campanas Observatory. They fit the spectrum using cloudless and cloudy models (which include only the opacity of the iron, silicate, and corundum clouds) and find that cloudy models fit significantly better than cloudless models. \ct{Burgasser10} conclude that cloud opacity is necessary to reproduce the spectral data and invoke a reemergence of the iron and silicate clouds. We instead assume that the iron and silicate clouds are depleted, as we observe generally in other T dwarfs, and investigate the effect of the sulfide clouds. 

\begin{figure*}[ht]
  \begin{minipage}[b]{0.5 \linewidth}
    \includegraphics[width=3.6in]{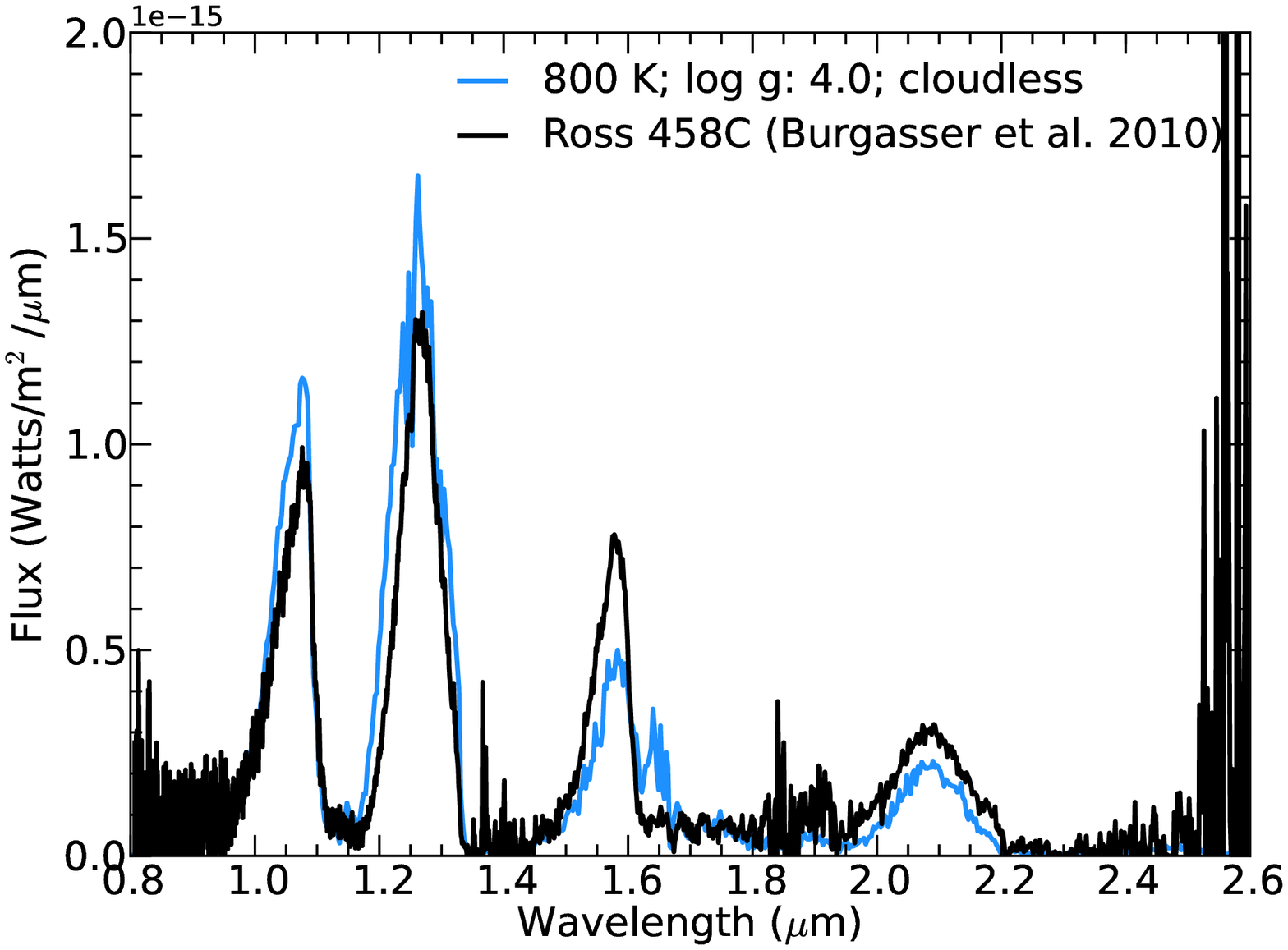}
    \vspace{-4mm}
  \end{minipage}
  \begin{minipage}[b]{0.5 \linewidth}
    \includegraphics[width=3.6in]{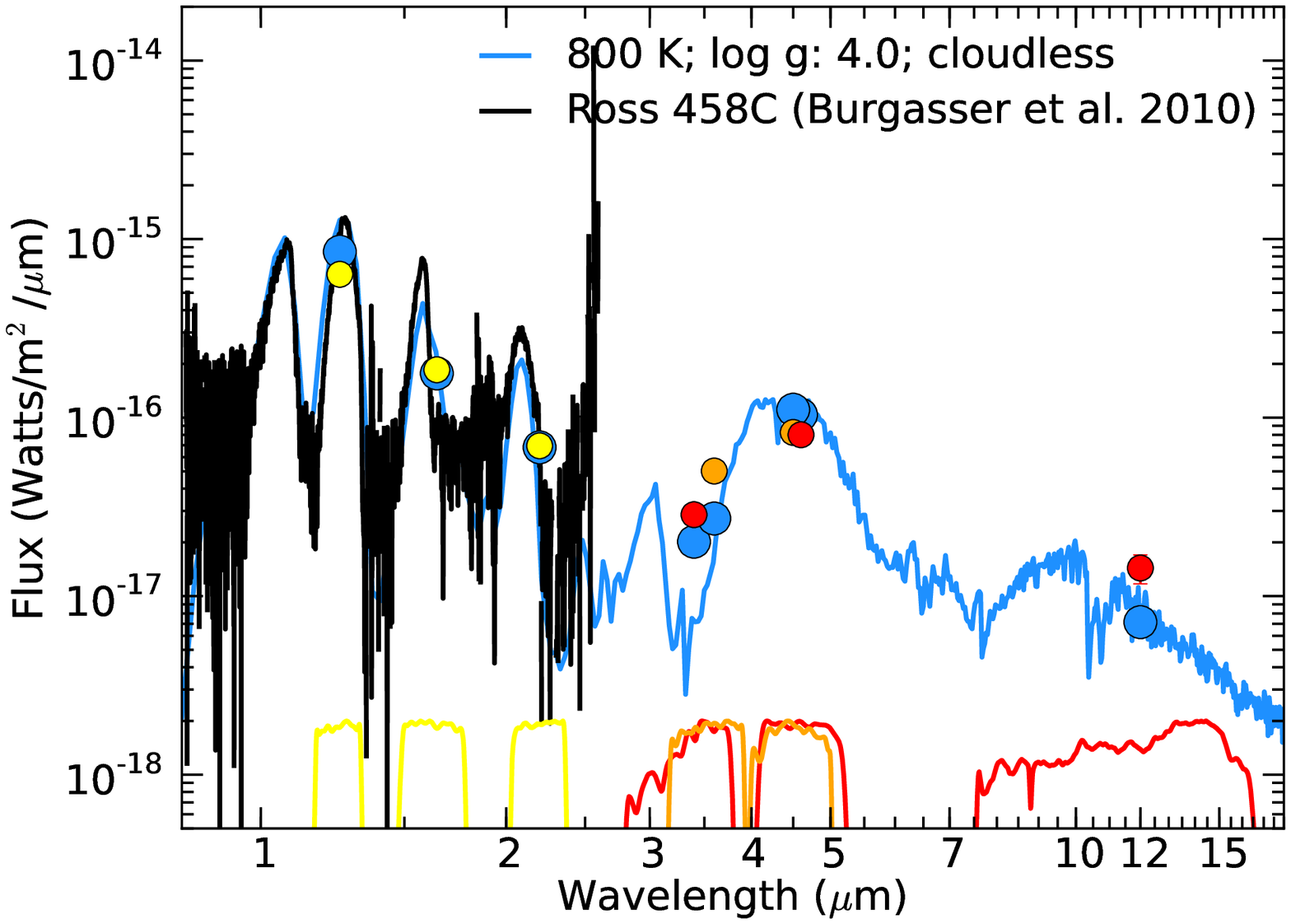}
    \vspace{-4mm}
  \end{minipage}
   \begin{minipage}[b]{0.5 \linewidth}
    \includegraphics[width=3.6in]{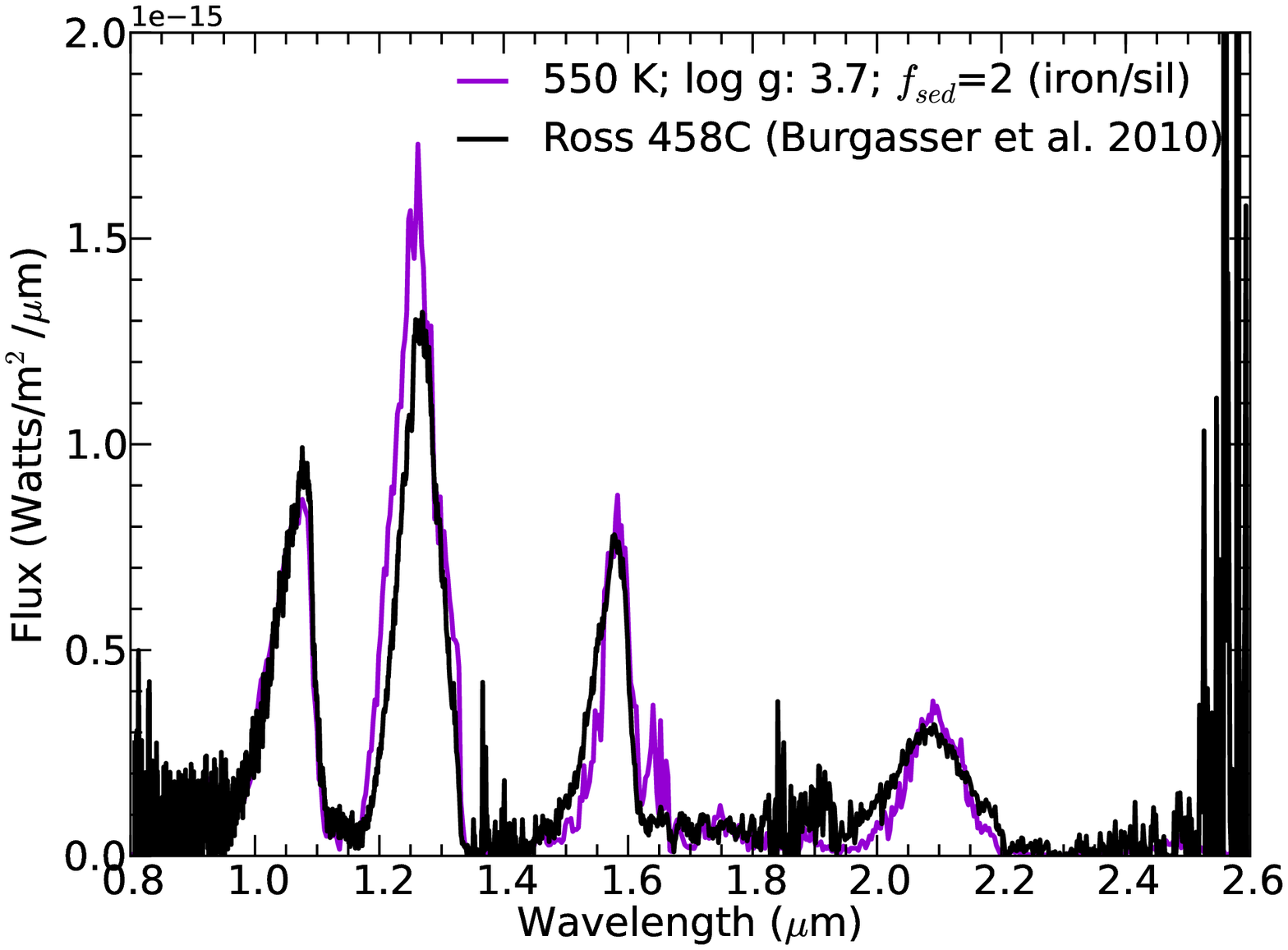}
    \vspace{-4mm}
  \end{minipage}
  \begin{minipage}[b]{0.5 \linewidth}
    \includegraphics[width=3.6in]{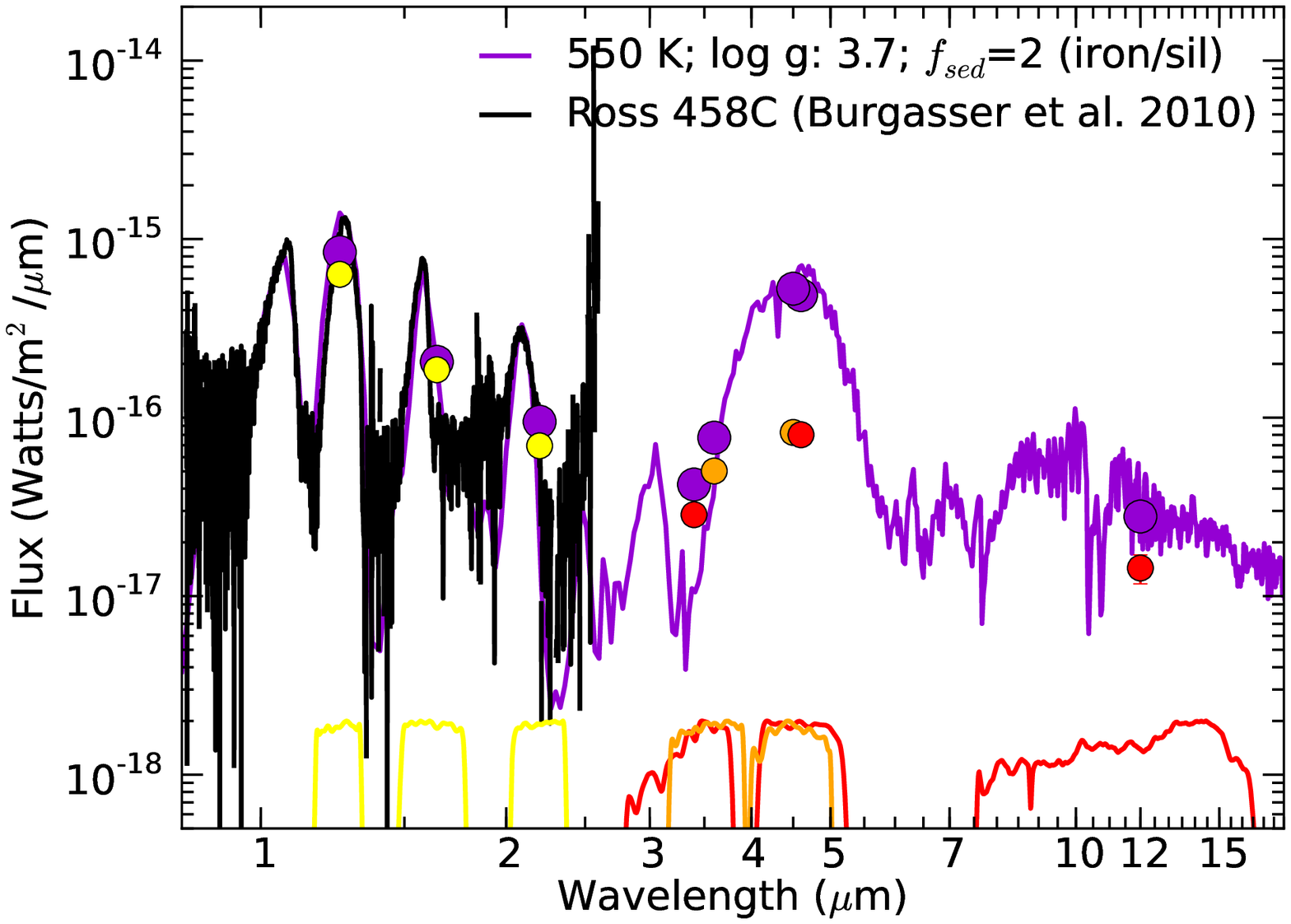}
    \vspace{-4mm}
  \end{minipage}
  \begin{minipage}[b]{0.5 \linewidth}
    \includegraphics[width=3.6in]{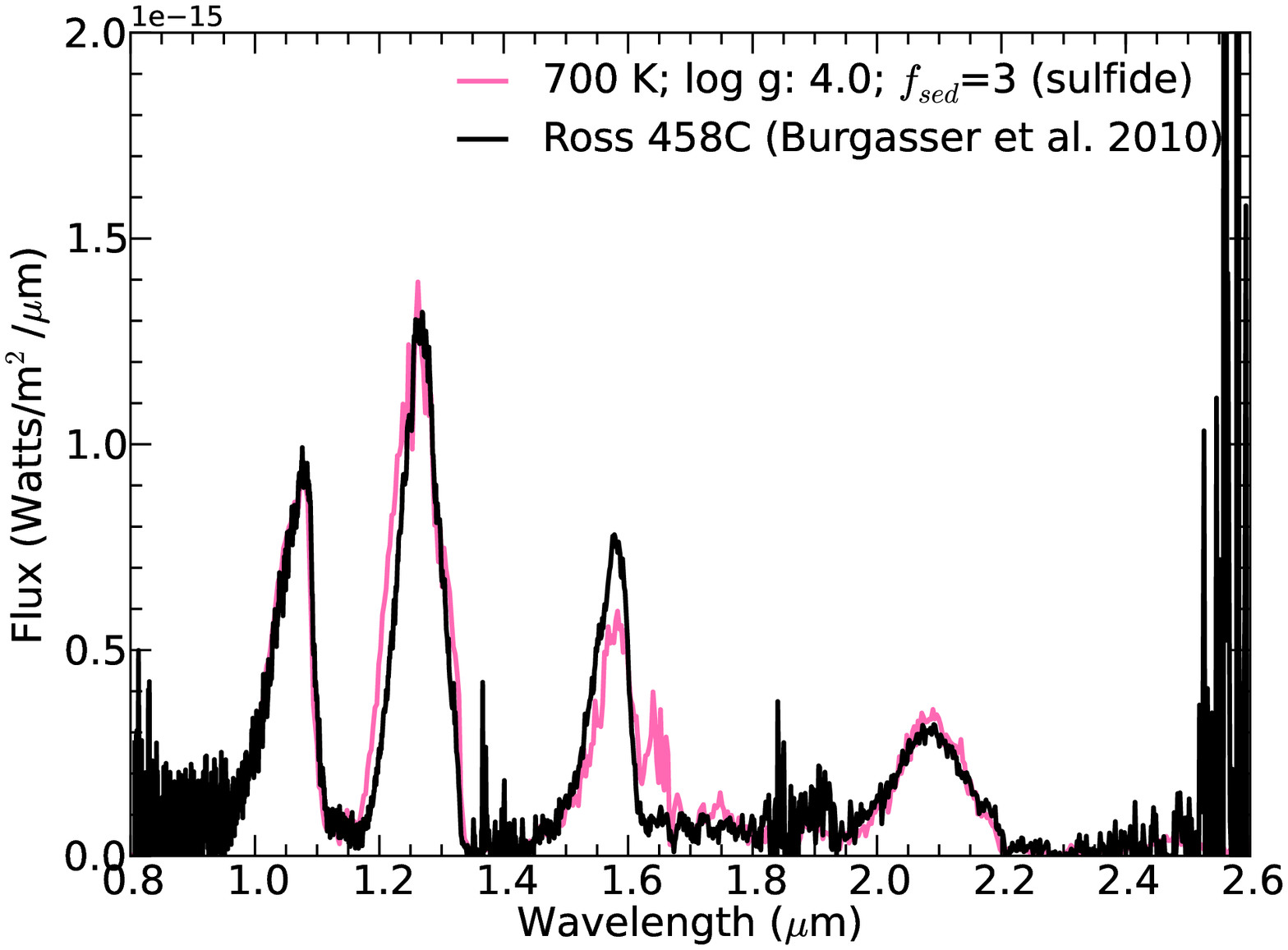}
    \vspace{-4mm}
  \end{minipage}
  \begin{minipage}[b]{0.5\linewidth}
    \includegraphics[width=3.6in]{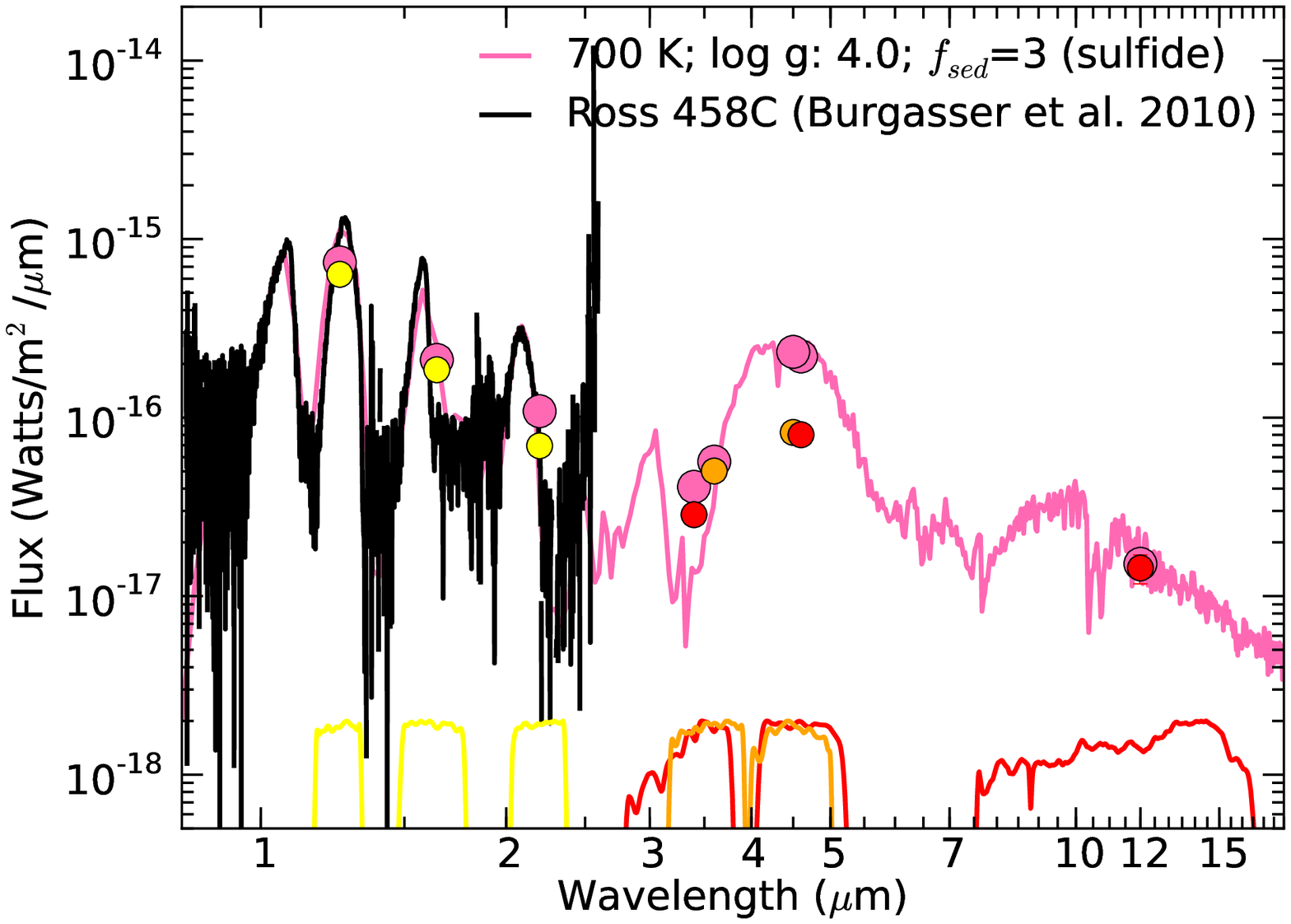}
    \vspace{-4mm}
  \end{minipage}
\caption{Ross 458C near-infrared spectrum comparison between data and models. Three different models are compared to the observed spectrum and photometry of Ross 458C from \ct{Burgasser10}. The left panels show the near-infrared spectra; the right panels show the same spectra and models with near- and mid-infrared photometry. Yellow points show $J$, $H$, and $K$ photometry; orange show Spitzer [3.6] and [4.5] photometry; red show WISE $W1$, $W2$, and $W3$ photometry. The filters for the photometric bandpasses are shown with corresponding colors along the bottom. The top row shows a cloudless model spectrum that best matches the data, which has an effective temperature of 800 K and surface gravity log $g$=4.0. The middle row shows the best matching cloudy spectrum using iron and silicate clouds. The bottom row shows the best matching cloudy spectrum using sulfide clouds. Both cloudy models have significantly lower effective temperature (100-250 K cooler) than the cloudless best-matching model, but have similar (low) surface gravity. For this object, we can largely match the overall features of the spectrum using either iron/silicate clouds or sulfide clouds.}
\label{ross458-spec}
\end{figure*}

In Figure \ref{ross458-spec} we show the FIRE spectrum and the best fitting cloudy and cloudless models. We also show photometry in $J$, $H$, and $K$ \cp{Dupuy12}, WISE photometry \cp{Kirkpatrick12}, and Spitzer photometry \cp{Burningham11}.  The spectra used to generate these results differ somewhat from those in \ct{Burgasser10} because we use models that include recent improvements to the opacity database \cp{Saumon12}  for both ammonia and the pressure-induced opacity of H$_2$ collisions. All models are fit by eye to the observations. 

Like \ct{Burgasser10}, we find that clouds are essential to match the spectrum of Ross 458C. Figure \ref{ross458-spec} shows the best fitting cloudless model and the two best fitting cloudy models (one including the iron and silicate clouds and the other including the sulfide clouds). The cloudless model is a poor representation of the spectral data; the flux in $Y$ and $J$ is too high and the flux in $H$ and $K$ is too low. The cloudy models are better representations of the relative flux in each band. 

\ct{Burgasser10} found that the surface gravity of Ross 458C must be low (log $g$=4.0) for models to match the observed spectrum. Likewise, we find that our best-fitting models have surface gravities of 4.0 (cloudless), 3.7 (silicate clouds), and 4.0 (sulfide clouds). 

We conclude that we do not need to invoke a reemergence of iron and silicate clouds into the photosphere of Ross 458C to reproduce the observed spectrum. Instead, we are able to reproduce the spectrum using the sulfide clouds which are naturally expected to form in the photospheres of cool T dwarfs. Section \ref{sulfsil} contains additional discussion on which cloud species we expect to be important. 

The very red slope to the L band spectrum of Ross 458C---much redder than all the models---is reminiscent of the behavior of some cloudy L dwarfs, including 2MASS2224 and  DE 0255 (L3.5 and L9 respectively) and may be a signature of very small dust grains \cp{Stephens09}.

The discrepancies at 4.5 \micron\ are likely to be a result of non-equilibrium chemistry, which is not included in these models. This effect is discussed in more detail in Section \ref{results-non-eq}.

\subsubsection{UGPS J072227.51-054031.2} \label{0722}

\begin{figure*}[]
  \begin{minipage}[b]{0.5\linewidth}
	\includegraphics[width=3.7in]{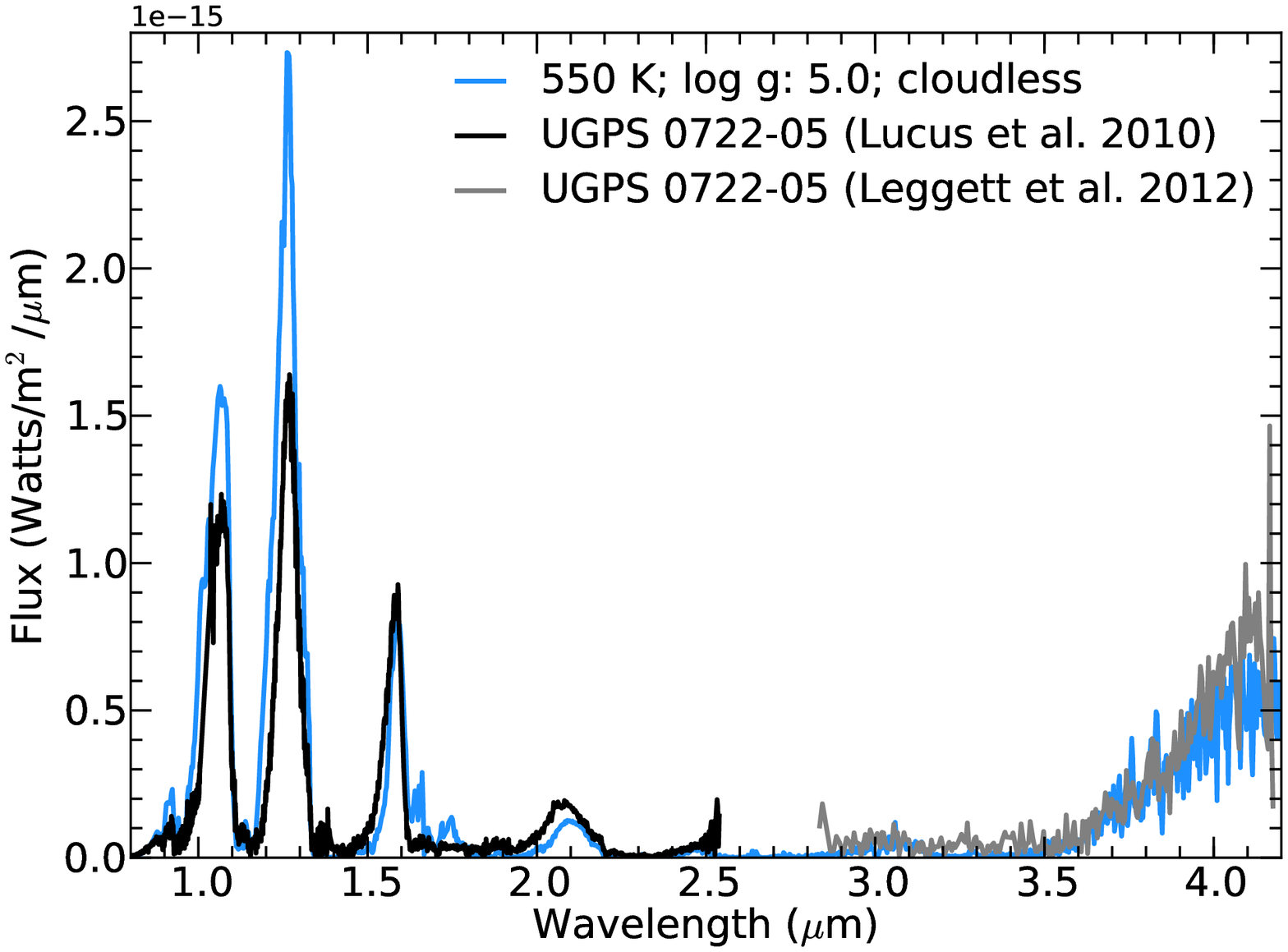}
     \vspace{-4mm}
 \end{minipage}
  \begin{minipage}[b]{0.5\linewidth}
	\includegraphics[width=3.7in]{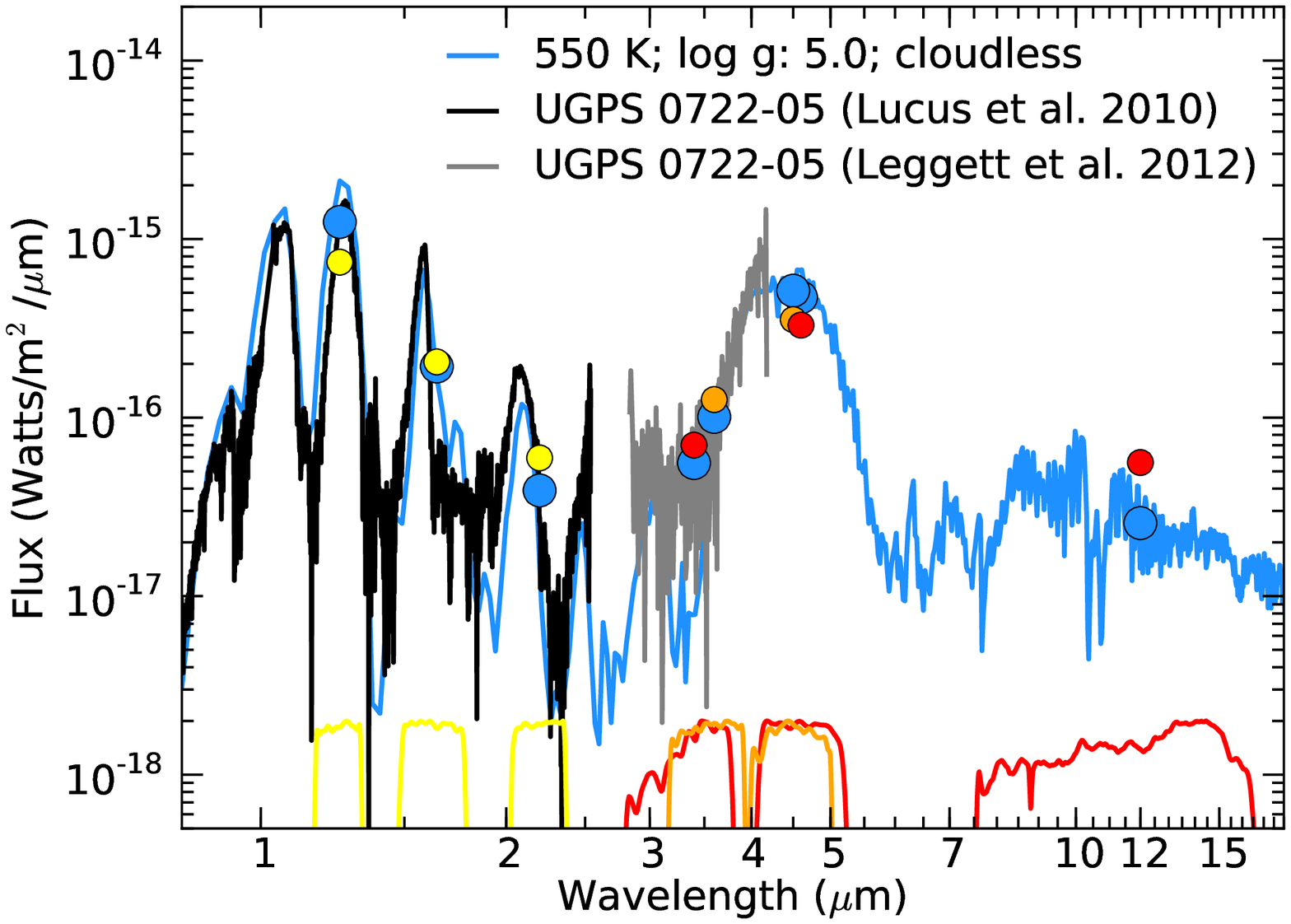}
     \vspace{-4mm}
 \end{minipage}
   \begin{minipage}[b]{0.5\linewidth}
	\includegraphics[width=3.7in]{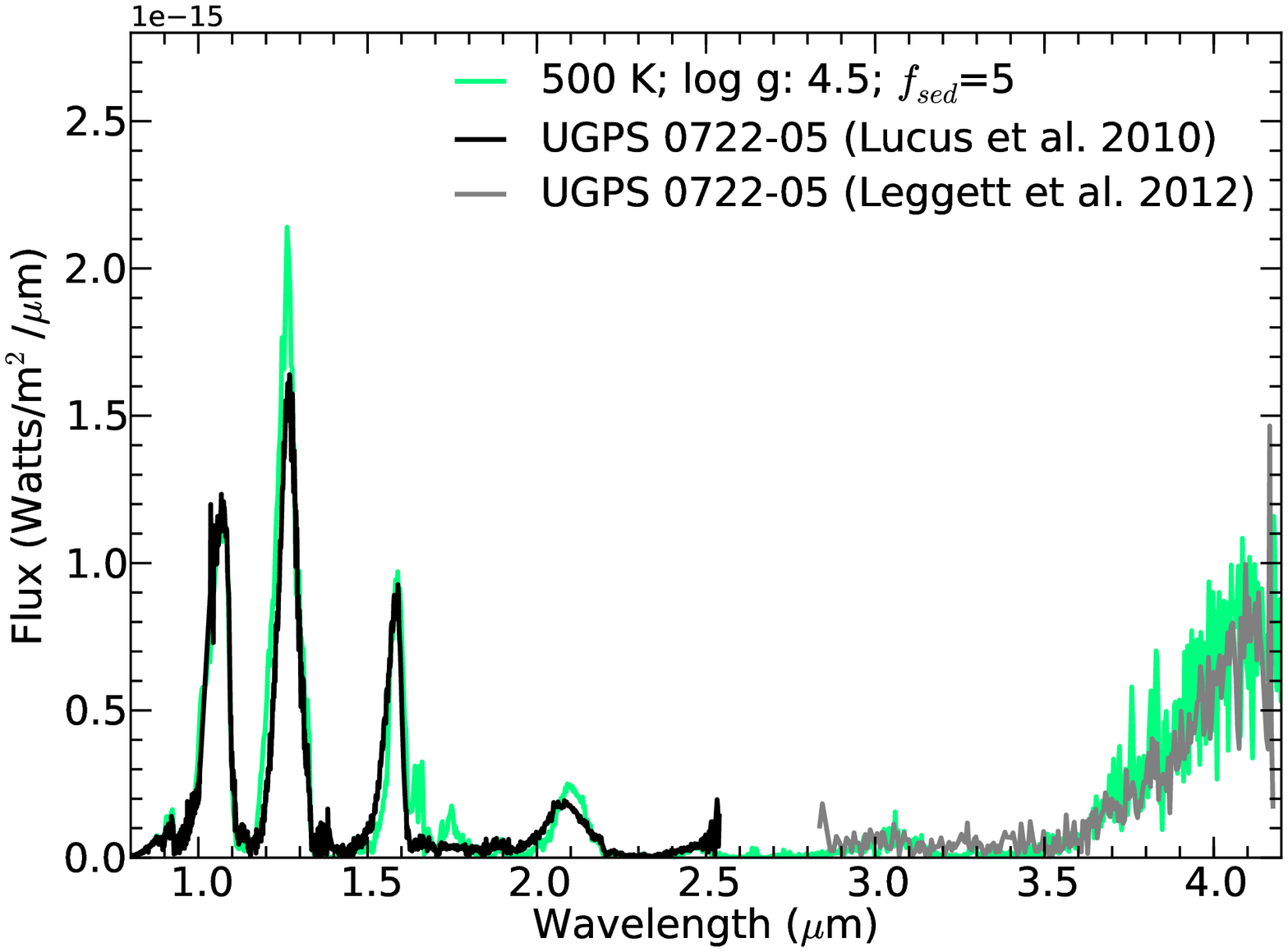}
     \vspace{-4mm}
 \end{minipage}
  \begin{minipage}[b]{0.5\linewidth}
	\includegraphics[width=3.7in]{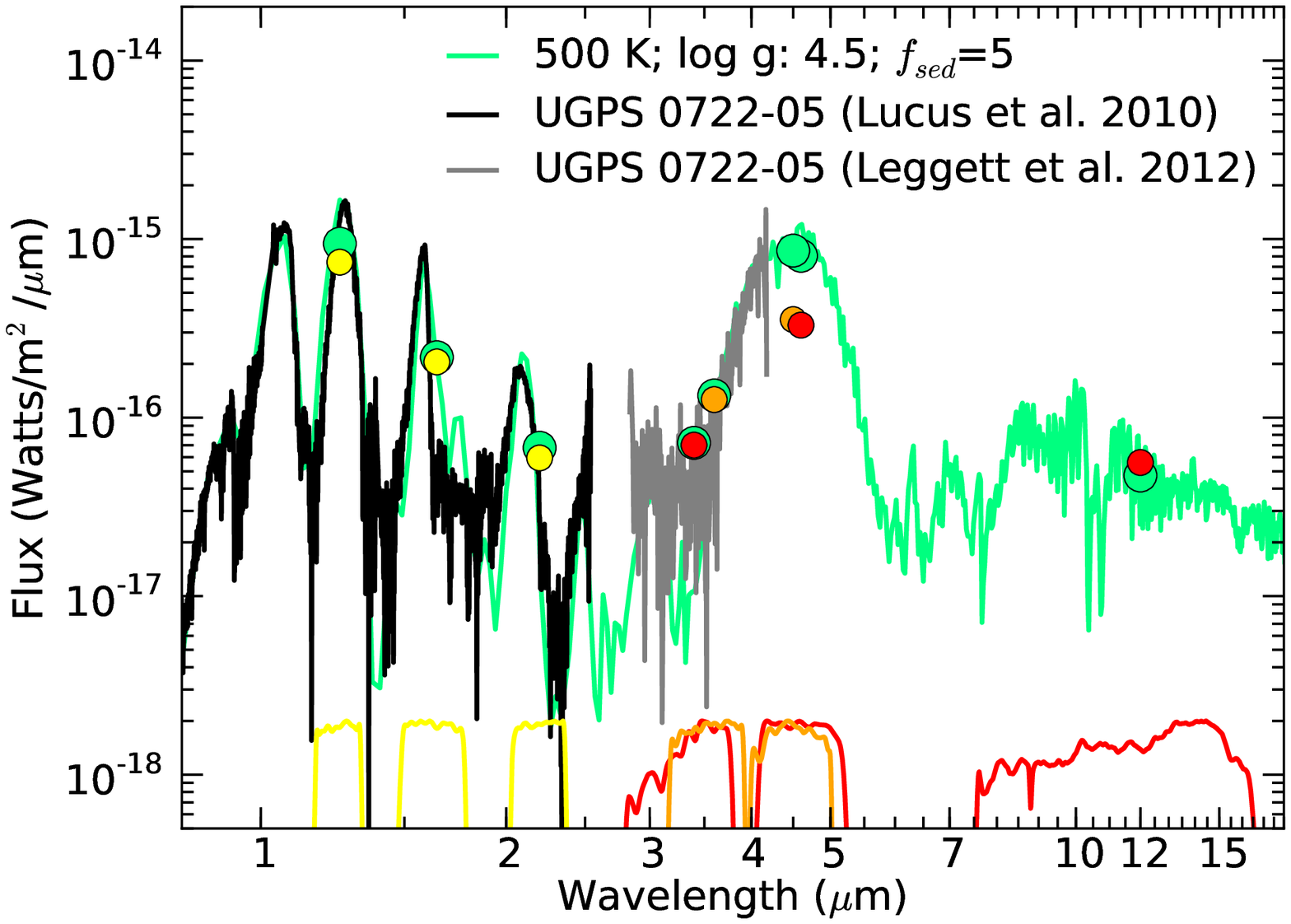}
     \vspace{-4mm}
 \end{minipage}
\caption{UGPS 0722--05 near-infrared spectrum comparison. Two different models are compared to the observed spectrum of UGPS 0722--05 from \ct{Lucas10}. As in Figure \ref{ross458-spec}, the left panels show the near-infrared spectra; the right panels show the same spectra and models with near- and mid-infrared photometry. Yellow points show $J$, $H$, and $K$ photometry; orange show Spitzer [3.6] and [4.5] photometry; red show WISE $W1$, $W2$, and $W3$ photometry. The filters for the photometric bandpasses are shown with corresponding colors along the bottom. The top plot shows a cloudless model spectrum that best matches the data, which has an effective temperature of 550 K and surface gravity log $g$=5.0. The bottom plot shows the best matching cloudy spectrum using sulfide clouds; it has an effective temperature of 500 K, log $g$=4.5, and \fsed=5. }
\label{0722-spec}
\end{figure*}

UGPS J072227.51-054031.2 (hereafter UGPS 0722--05) is a T9 or T10 dwarf with an effective temperature of approximately 500 K, discovered by \ct{Lucas10}. Previous spectral analysis with cloudless models has been unsuccessful at modeling the flux in the near-infrared in $Y$ and $J$ bands \cp{Leggett12}. 

In Figure \ref{0722-spec} we plot the near-infrared spectra published in \ct{Lucas10} and \ct{Leggett12} with the cloudy and cloudless models that are fit by eye to be the closest representations of the data.  We also show $J$, $H$, $K$ and Spitzer photometry \cp{Lucas10} and WISE photometry \cp{Kirkpatrick12}. These models have similar temperatures and gravities to previous studies; \ct{Leggett12} presented fits with \teff\ between 492 and 550 K and log g=3.52 to 5.0, whereas our fits have \teff\ of 600 K (cloudless) and 500 K (with sulfide clouds) and both have log $g$=4.5.

Note that the flux in $Y$ and $J$ in cloudless models is systematically too high. The opacity of the sulfide clouds provides a natural mechanism to decrease the flux in $Y$ and $J$ and increase the flux in $H$ and $K$ bands. The match to the models is still not perfect. This may be due in part to incomplete line lists for methane; these cold objects have a significant amount of CH$_4$ in the atmosphere, which absorbs strongly in the near infrared. The discrepancy at 4.5 \micron\ is most likely due to absorption of CO as a result of non-equilibrium chemistry (see Section \ref{results-non-eq}). Outstanding issues in T dwarf modeling are discussed in Section \ref{outstandingissues}.

\subsection{Non-Equilibrium Chemistry} \label{results-non-eq}

We note that both of our preferred sulfide cloud models predict brighter $M$ band (4.5 \micron) photometry than is observed.  This is likely a consequence of our neglect of non-equilibrium mixing of CO in this study.  As first predicted by \ct{Fegley96}, such mixing is an important process in the atmospheres of brown dwarfs \cp{Noll97, Marley02, Saumon06, Stephens09} as it is in solar system giants \cp{Barshay78} and the relative impact of the mixing increases with decreasing gravity \cp{Hubeny07, Barman11}.  Absorption by excess atmospheric CO decreases the thermal emission in M band and is likely responsible for the mismatches seen in Figures \ref{ross458-spec} and \ref{0722-spec}, particularly for the lower gravity Ross 458C.  

The formation of the clouds considered in this study does not involve the species most affected by non-equilibrium chemistry such as CO and CH$_4$. The cloud models will therefore be only minimally affected by the changes in the pressure-temperature profile of the atmosphere due to the changes in gas opacity. However, the overall spectra of models will look quite different in regions where CO absorbs strongly, such as the 4.5 \micron\ feature, so future, more comprehensive fits of sulphide cloud models to observations will have to include non-equilibrium models.

\section{Discussion}

\subsection{Formation of Clouds}

Clouds must form in brown dwarf atmospheres as they cool; there is no way to avoid the condensation of different species as the atmosphere reaches lower effective temperatures.  In these models, we parameterize the opacity of clouds by creating a distribution of cloud material in the atmosphere which has a distribution of cloud particle sizes. Within the models, we can change those distributions. A cloud that sediments into a small number of large particles will settle into a thin layer and will not significantly change the emergent spectrum; the same cloud material organized into an extended cloud with small particle sizes will have a dramatic effect on the model spectrum. 

For these reasons, we require a model of cloud particle sizes and distribution as well as the underlying chemistry. When we consider models that include new or different clouds, we do not change any of the underlying chemistry of condensation; we change the opacity of the condensate particles and in doing so modify the effect that the cloud formation has on emergent flux.

\subsection{Sulfide or Silicate Clouds?} \label{sulfsil}

\ct{Burgasser10} invoked the reemergence of silicate clouds to explain the spectrum of Ross 458C. We suggest instead that the initial emergence of sulfide clouds would have a similar effect on the spectrum and provide a more natural explanation for the results. 

From observations, it is clear that the range in T dwarf colors just following the L/T transition is small; spectra of T dwarfs show no evidence that clouds still affect the emergent flux for objects slightly cooler than this transition. This observation suggests that the iron and silicate clouds have dissipated between 1400 and 1200 K (for typical field dwarfs) and are no longer important in T dwarf atmospheres. If iron and silicate clouds were sometimes important in T dwarf atmospheres, we would expect to see a population of relatively quite red objects at effective temperatures between that of Ross 458C and the L dwarfs; no brown dwarf with these properties has been observed. 

As T dwarfs cool, the range in observed infrared colors increases; a population of red T dwarfs develops, which are redder in the near-infrared than cloudless models predict. Based on these observations, we favor a mechanism that cannot strongly alter \teff$\sim$900-1200 K T dwarf atmospheres, but naturally reddens \teff$\lesssim$800 K T dwarfs. 

The emergence of sulfide clouds provides that natural explanation for this range in T dwarf colors at low effective temperatures. Just as the iron and silicate clouds condense in the photospheres of L dwarfs and change their observed spectra, the sulfide clouds begin to condense in the photospheres of T dwarfs with temperatures cooler than 900 K, changing their observed spectra. 

We have not yet investigated whether the sulfide clouds will have identifiable spectral features that would confirm their presence in T dwarf atmospheres, but given the features in the sulfide indices of refraction (see Figure \ref{na2s_n}) these features would likely be in the mid-infrared.

\subsection{Outstanding Issues In T Dwarf Models} \label{outstandingissues}

There are several challenges in modeling T dwarfs that have not yet been addressed in these calculations. 
Because of the high densities in brown dwarf atmospheres, sodium and potassium bands at optical wavelengths are extremely pressure-broadened in T dwarf spectra \cp{Tsuji99, Burrows00, Allard05, Allard07}. The wings of these broadened bands extend into the near-infrared in $Y$ and $J$ bands, creating additional opacity at those wavelengths. For these calculations, we use the line broadening treatment outlined in \ct{Burrows00}, which is somewhat \emph{ad hoc} and potentially creates some inaccuracies in the model flux in $Y$ and $J$ bands. A calculation of the molecular potentials for potassium and sodium in these high pressure environments, as is carried out in \ct{Allard05, Allard07}, would improve the accuracy of these models. 

Another challenge in modeling T dwarf spectra is the inadequacies of methane opacity calculations; methane is the only important gas-phase absorber with inadequate opacity measurements or calculations. Uncertainties in methane absorption bands create inaccuracies in T dwarf models, especially in $H$ band where it is a very strong absorber \cp{Saumon12}.

\subsection{Breakup of \nas\ Cloud}

Sulfide clouds could form partial cloud layers with patchy clouds. One way to infer patchy cloud cover is to observe variability in photometric colors; variability can indicate high-contrast cloud features rotating in and out of view. \ct{Radigan12} studied objects at the L/T transition and inferred that the iron and silicate clouds could be in the process of breaking up and forming patchy clouds in those atmospheres. A similar study of the variability of cool T dwarfs could reveal a similar physical process in sulfide clouds. 

\subsection{Constraining Cloud Models with More Data}

A larger number of high fidelity spectra of the coldest T dwarfs would help to constrain these cloudy models. Currently there are a few objects with effective temperatures cooler than 700 K with moderate resolution spectra. A larger sample of objects would give us better statistics on the overall population of T dwarfs, with different surface gravities, metallicities, and cloud structures.

\subsection{Water Clouds}

At cooler effective temperatures, water clouds will condense in brown dwarf atmospheres \cp{Burrows03}.  Oxygen is 300 times more abundant than sodium in a solar-composition gas and the silicate clouds only use 20\% of the total oxygen in the atmosphere, so water clouds will be much more massive and  important in shaping the emergent flux. As missions like WISE find colder objects \cp{Kirkpatrick11, Cushing11} and these objects are observed spectroscopically, future models of brown dwarfs will need to include the condensation of these more volatile clouds.

Before the water clouds condense, \ct{Lodders99, Lodders06} predict that RbCl and CsCl will condense; however, the abundances of Cs and Rb are very low \cp{Lodders03} so these clouds will be optically thin. If equilibrium conditions prevail, NH$_4$H$_2$PO$_4$ would also condense \cp{Fegley94, Visscher06} with a mass similar to that of the \nas\ cloud.  Whether NH$_4$H$_2$PO$_4$ condenses or P remains in the gas phase as PH$_3$ (as on Jupiter and Saturn) deserves further study to examine potential effects on the spectra of the coolest brown dwarfs.

\subsection{Application to Exoplanet Atmospheres}

Observations and models of T dwarfs provide a testbed to study planetary atmospheres. While brown dwarfs are more massive than planets, the atmospheres of T dwarfs have similar effective temperatures to those of young giant planets \cp{Burrows97,Fortney08b}. The study of T dwarfs provides crucial tests of cloudy atmosphere models that will be applicable to directly-imaged exoplanet atmospheres. 

Cloud models designed originally for brown dwarfs are already being applied to exoplanets. Cloud models with iron and silicate clouds were originally developed to model L dwarf atmospheres; these models have been successfully applied to observations of the only directly imaged multiple planet system, HR 8799. Several studies of the HR 8799 planets have shown that the iron and silicate clouds play a significant role in their atmospheres \cp{Marois08, Barman11, Galicher11, Bowler10, Currie11, Madhu11c, Marley12}. 

As instruments like the Gemini Planet Imager and SPHERE begin to discover new planets in the next few years, we will be able to apply brown dwarf models to colder planetary atmospheres in which clouds will likely play a key role in shaping their spectra.

\section{Summary}

Cloud formation is a natural and unavoidable phenomenon in cool substellar atmospheres.  At temperatures cooler than those of L dwarfs, chemistry dictates that additional condensates, beyond the silicates and iron, must form.  We have examined the effect of including the most abundant of these lower-temperature condensates, Cr, MnS, \nas, ZnS, and KCl, in brown dwarf model atmospheres.  Within the framework of the \ct{AM01} cloud model, we have  investigated the opacity of these clouds over a wide range of parameter space, across the relevant range of T dwarfs, to the warmest Y dwarfs.  From our suite of models from 400 to 1300 K, log $g$=4.0 to 5.5, \fsed=2 to 5, we have shown the likely role that these low-\teff\ clouds, dominated by sulfides, play in these cool atmospheres.

Model spectra were compared to two T dwarfs, Ross 458C and UGPS 0722--05. These two objects have red near-infrared colors and are not well-matched by cloudless models.  Model spectra that include the sulfide clouds match the observed spectra of both objects more accurately than cloudless models. 

The photometric colors of the cloudy models were calculated and compared to the full population of brown dwarfs with parallax measurements. This analysis shows that the sulfide clouds provide a mechanism to match the near-infrared colors of observed brown dwarfs.  The agreement is particularly good in $J-H$, while in $J-K$ the models are somewhat too red.  WISE observations of the coolest T dwarfs and warmest Y dwarfs indicate the these new models fit observations better than cloud-free models.

Our results indicate that understanding the opacity of condensates in brown dwarf atmospheres of all \teff\ is necessary to accurately determine the physical characteristics of the observed objects. 

\acknowledgements

We thank Rabah Khenata and the rest of his team in Algeria for providing calculations of optical properties for sodium sulfide. We thank Adam Burgasser for providing the spectrum for Ross 458C. We thank Katharina Lodders for providing the condensation chemistry tables used in the model calculations as well as other very helpful advice and suggestions. We also thank the anonymous referee for his or her suggestions. Based on observations obtained at the Gemini Observatory, which is operated by the Association of Universities for Research in Astronomy, Inc., under a cooperative agreement with the NSF on behalf of the Gemini partnership: the National Science Foundation (United States), the Science and Technology Facilities Council (United Kingdom), the National Research Council (Canada), CONICYT (Chile), the Australian Research Council Australia),Minist\'{e}rio da Ci\^{e}ncia, Tecnologia e Inova\c{c}\~{a}o (Brazil) and Ministerio de Ciencia, Tecnolog\'{i}a e Innovaci\'{o}n Productiva (Argentina).


\end{document}